\documentclass[12pt,twoside]{article}
\usepackage{fleqn,espcrc1}
\usepackage{epsf,psfig}
\usepackage{epsfig}
\usepackage{graphicx}

\begin{document}

{\large {\bf Last Call for RHIC Predictions}}\\[2ex]

S.A. Bass$^1$,  M. Bleicher$^2$, \ W. Cassing$^3$, \ A. Dumitru$^4$, \ 
H.J. Drescher$^5$, \  K.J. Eskola$^6$, M. Gyulassy$^7$, D. Kharzeev$^8$,
Y.V. Kovchegov$^9$, Z. Lin$^{10}$,  D. Molnar$^{^7}$,  J.Y. Ollitrault$^{11}$, 
S. Pratt$^{12}$, J. Rafelski$^{13}$, R. Rapp$^{14}$, D.H. Rischke$^{15}$,
J.B. Schaffner$^{15}$, B.R. Schlei$^{16}$, 
A.M. Snigirev$^{17}$, H. Sorge$^{14}$, D.K. Srivastava$^{18}$,
J. Stachel $^{19}$, D. Teaney$^{14}$, R. Thews$^{13}$, S.E. Vance$^{7}$,
I. Vitev$^{7}$, R. Vogt$^{20}$,
 X.N. Wang$^{20}$, B. Zhang$^{10}$, J. Zim\'anyi$^{21}$
\\[3ex]  

$^{1}$ Physics Dept., Duke Univ., Durham, NC, USA,
$^{2}$ Inst. Theor. Phys., J.W. Goethe Univ., 
Frankfurt am Main, Germany,
$^{3}$ Inst. Theor. Phys., Univ. Giessen, Germany,
$^{4}$ Physics Dept., Yale Univ., New Haven, CT, USA,
$^{5}$ Subatech, Univ. Nantes, France,
$^{6}$ Jyvaskyla Univ., Finland,
$^{7}$ Phys. Dept., Columbia Univ., New York, NY, USA,
$^{8}$ Phys. Dept. and RBRC, Brookhaven National Laboratory
Upton, NY, USA,
$^{9}$ Dept. Phys., U. Minn., Minneapolis MN, USA,
$^{10}$ Phys. Dept., TAM, College Station, TX, USA,
$^{11}$ Saclay, France,
$^{12}$ Dept. Phys., MSU, East Lansing, MI, USA, 
$^{13}$ Dept. Phys., Univ. Arizona, Tucson, NM, USA,
$^{14}$ Dept. Phys. and Astr., SUNY at Stony Brook, NY, USA,
$^{15}$ RIKEN-BNL Research Center, Brookhaven National Laboratory
Upton, NY, USA,
$^{16}$ Theor.  Div., Los Alamos National Laboratory, 
         Los Alamos, NM, USA,
$^{17}$ Moscow State Univ., Nuclear Physics Institute, Moscow, Russia,
$^{18}$ Variable Energy Cyclotron Centre, Bidhan Nagar, Calcutta, India,
$^{19}$ Dept. Phys., Heidelberg Univ., Heidelberg, Germany,
$^{20}$ Nucl. Sci. Div., LBNL, Berkeley, CA, USA,
$^{21}$ RMKI, Budapest, Hungary
  
\section*{INTRODUCTION}

Post-diction has become an art form at AGS and SPS energies.  The timing of
QM99, just as the RHIC facility comes on-line after nearly two decades of
planning and construction, provides a unique opportunity to break out of that
mold.  The one day Quark Matter 1999 session, organized by M. Gyulassy, had the
goal of updating and documenting the {\em PRE}-dictions for the upcoming RHIC
experiments. Of course, sticking one's neck out just as the experimental
"guillotine" starts its descent is risky.  The brave participants in this
session presented  a wide range of predictions summarized
in this collection of contributions. 

The contributions are organized into five section: (1) Global
and Inclusive Observables that probe  entropy production and
initial conditions in AA, (2) Hadron Flavor Observables that probe
hadro-chemical equilibration and baryon number transport,
(3) HBT and Collective Flow Observables that probe the space-time 
volume of the freeze-out surface 
of the reaction as well as the equation of state of ultra-dense matter,
(4) Jets and Penetrating Rare Probes which are sensitive to the
highest density phase of the reaction,
and (5) Exotic Possibilities that illustrate some of the 
more speculative novel phenomena that may be explored at RHIC.

\section{GLOBAL AND INCLUSIVE OBSERVABLES}
% 1
\newpage
%\subsection{K.J. Eskola, K,    Minijets and pQCD}
\subsection{K.J. Eskola: Lower limits for $\langle E_{\rm T}\rangle$ and 
$\langle N_{\rm ch}\rangle$ from pQCD \& hydrodynamics at the central 
rapidity unit in central Au-Au collisions at RHIC \cite{KJE1,KJE2}}

In heavy ion collisions at very high cms-energies the initial, very
early particle and transverse energy production at central rapidities
is expected to be dominated by multiple production of minijets,
i.e. gluons, quarks and antiquarks with transverse momenta $p_T\sim
1...2\, {\rm GeV} \gg \Lambda_{\rm QCD}$ \cite{BM}.  Assuming
independent multiple semi-hard parton-parton collisions, the average
transverse energy carried by the minijets produced with $p_{\rm T}\ge p_0$
at the central rapidity window $\Delta y=1$ in a central (${\bf b=0}$) 
$AA$-collision can be computed in the lowest order (LO) perturbative QCD (pQCD) as
\cite{EKL}
%\begin{equation}
%\bar N_{AA}(\sqrt s, p_T\ge p_0,\Delta y,{\bf 0}) = T_{AA}({\bf 0})
%\int_{{p_0, \Delta y}} dp_Tdy_1dy_2 \frac{d\sigma}{dp_Tdy_1dy_2}.
%\end{equation}
%and the average transverse energy carried by these partons as \cite{EKL}
\begin{equation}
 E_{\rm T,pQCD}^{AA}(\sqrt s, p_T\ge p_0,\Delta y,{\bf 0}) 
= T_{AA}({\bf 0})
\int_{{p_0, \Delta y}} dp_Tdy_1dy_2 \frac{d\sigma}{dp_Tdy_1dy_2}p_T.
\end{equation}
The differential cross section above is that of each binary (LO)
parton collision, 
\begin{equation}
\frac{d\sigma}{dp_T^2dy_1dy_2} = 
        \sum_{{ijkl=}\atop{q,\bar q,g}}
        x_1f_{i/A}(x_1,Q) \, x_2f_{j/A}(x_2,Q)
        {d\hat\sigma\over d\hat t}^{ij\rightarrow kl} 
\end{equation}
where the rapidities of the outgoing partons are $y_1$ and $y_2$.  The
nuclear collision geometry is accounted for by the standard nuclear
overlap function $T_{AA}({\bf 0})\sim A^2/\pi R_A^2$. In order to
have at least one of the two outgoing partons in each minijet
collision in the rapidity window $\Delta y$, appropriate kinematical
cuts have to be made, see \cite{EKL,EKR,EK} for details. We use the
GRV94-LO parton distributions \cite{GRVLO}, and, to consistently
include both the $x$- {\it and} the $Q$-dependence of the nuclear
effects in the parton distributions (~$f_{i/A}(x,Q)\ne
f_{i/p}(x,Q)$~), we use the recent EKS98-parametrization \cite{EKS98}.
We emphasize that the results presented here are obtained by using
merely the lowest order pQCD but in order to approximate
the effects of the next-to-leading-order terms in the minijet cross
sections, we include $K$-factors 2.0 (1.5) for RHIC (LHC), together
with the scale choice $Q=p_{\rm T}$.

{} From saturation arguments \cite{EK}, and from requiring agreement
with inclusive pion spectra at central rapidities in $p\bar p$
collisions at $\sqrt s = 200$ GeV, we expect  $p_0\sim 1...2 $ GeV
for Au-Au collisions at RHIC. The few-GeV minijets are produced within
a short proper time $\tau_i\sim1/p_0\sim 0.2 ... 0.1$ fm/$c$, so they
serve as initial conditions for further evolution of the early QGP.
In addition to the pQCD component, at RHIC we also expect a
non-negligible non-perturbative (soft) component in the initial
transverse energy production. For the simple estimates here, we take
the soft component directly from the measured average $E_{\rm T}$ at
$\eta\sim0$ at central Pb-Pb collisions at the SPS \cite{NA49}. The
{\it initial} transverse energy at $\tau=1/p_0$ thus becomes $E_{\rm
T}^i = E_{\rm T,pQCD}^{AA} + E_{\rm T}^{\rm NA49}$, as shown by the
dashed lines in Fig. 1a.  Dividing by the initial volume $V_i= \pi
R_A^2\tau_i\Delta y$, we get a Bjorken-estimate of the initial energy density:
$\epsilon_i = E_{\rm T}^i/V_i$. In a fully thermalized, 1+1
dimensional boost-invariant hydrodynamic system, the $pdV$ work causes
the energy density to decrease as $\epsilon\sim \tau^{-4/3}$
\cite{BJ} and, especially, {\it the transverse energy to decrease}
as $E_{\rm T}\sim \tau ^{-1/3}$. We do not attempt to follow
the system through a phase transition here but simply decouple the
system at $\epsilon_f = 1.5$ GeVfm$^{-3}$. The resulting $E_{\rm T}^{\rm
final}$ represents a lower limit of $\langle E_{\rm T}\rangle$ (mod the
decrease in the mixed/hadronic phase), and is plotted by the solid curves 
in Fig.~1a.  It should also be noted that when transverse expansion is 
included, the loss of $E_{\rm T}$ is less than in the 1+1 dimensional case
considered above.

\begin{figure}[h]
\vspace{4cm}
\centerline{\hspace*{0.5cm} 
\epsfxsize=8.5cm\epsfbox{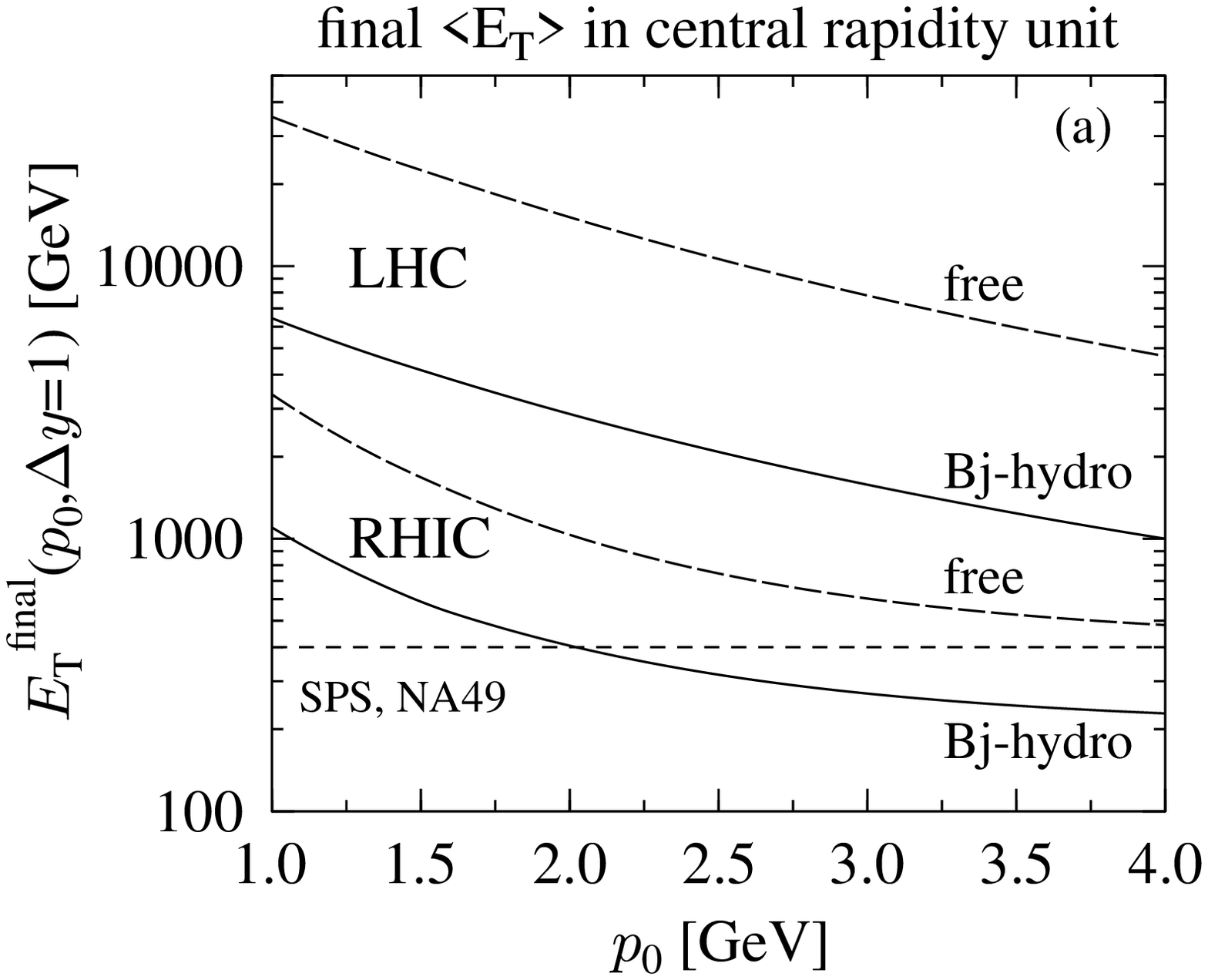}
\epsfxsize=8.5cm\epsfbox{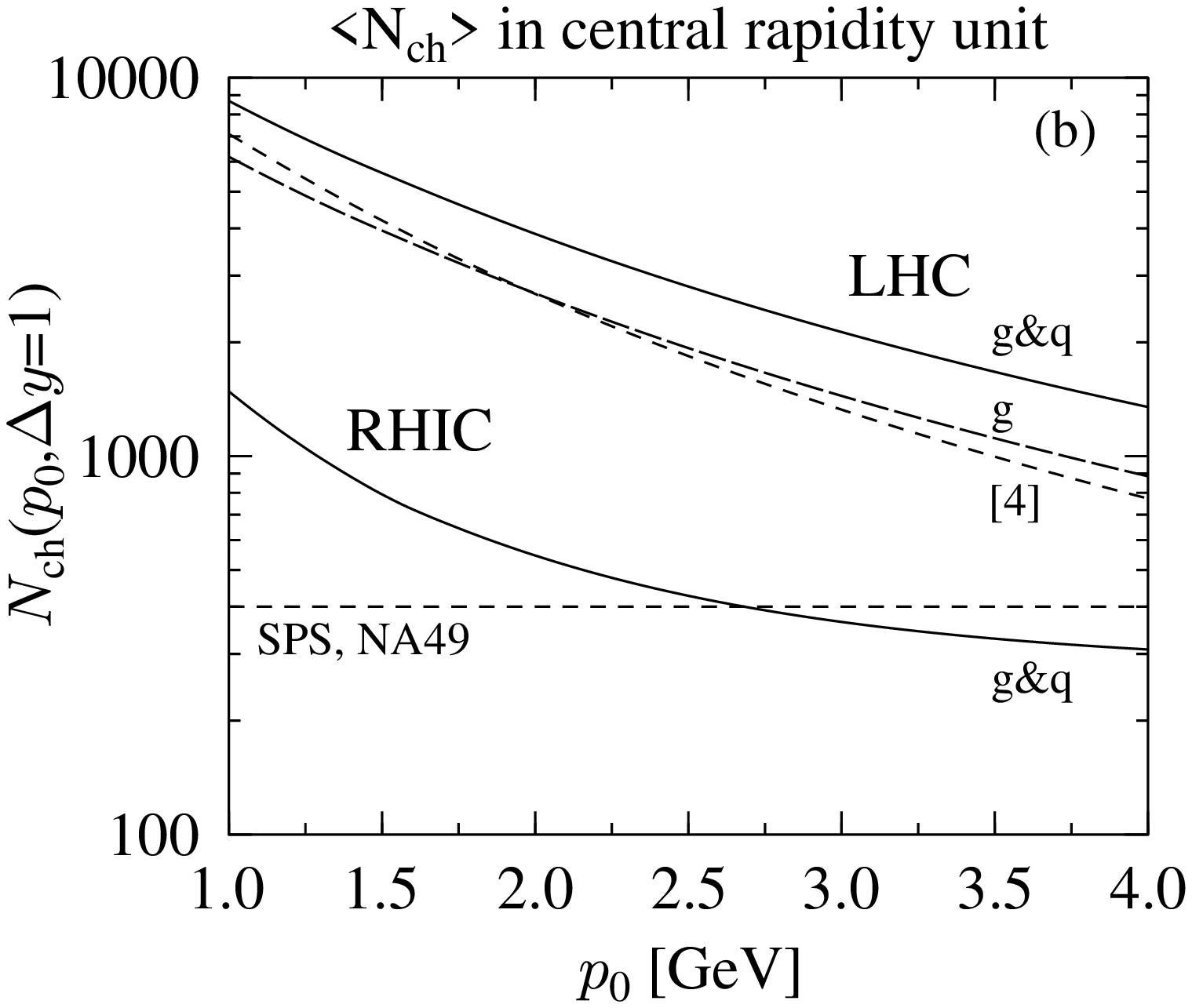}
}
\vspace{-6cm}
\caption{ Final $\langle E_T \rangle$  and $\langle N_{ch} \rangle$ 
in the central rapidity unit.}
\label{DATA}
\end{figure}

To get a corresponding estimate of the final state charged pion
multiplicity (Fig.~1b), we convert the initial energy density
$\epsilon_i$ into a temperature $T_i$, from which the initial rapidity
density of entropy $S_i$ can be computed.  For simplicity (in spite of
gluon dominance), let us assume full thermalization of gluons and
quarks here. In an isentropic boost-invariant 1+1 dimensional flow,
the rapidity density of entropy is conserved, so $S_i=S_f\approx
4\frac{3}{2}N_{\rm ch}$ \cite{EKR}, where $N_{\rm ch}$ is the final
state charged pion multiplicity in the central rapidity unit. In other
words, we obtain a lower limit of $N_{\rm ch}$ from entropy of the 
initial state.

To show the dependence on the transverse momentum cut-off, we plot the
lower limits for $\langle E_{\rm T}\rangle$ and $N_{\rm ch}$ as
functions of $p_0$ (the solid curves). The upper limit of $E_{\rm
T}^{\rm final}$ is the initial $E_{\rm T}^i$, so $p_0$ can be fixed
from some other arguments, the measured $\langle E_{\rm T}\rangle$
serves in principle also as a measure of thermalization. Our favorite
estimates for the lower limits can be read off from the figures at
$p_0\sim 1.5$ for RHIC and at $p_0\sim 2$ GeV for the LHC.

% 2

\newpage
\subsection{Yu.V. Kovchegov:  Nuclei and  Classical Yang Mills}
%3
%\input{Kovchegov/kovin.tex}
 This contribution is based on the paper
\cite{mM} in collaboration
with  A. H. Mueller. The transparencies can be viewed at the web
address given in \cite{qm}.

\begin{figure}[b]
\begin{center}
\epsfxsize=5cm
\leavevmode
\hbox{ \epsffile{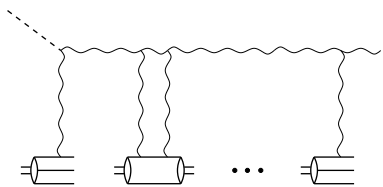}}
\end{center}
\caption{Gluon production in the current--nucleus interaction as 
envisioned in the text.}
\label{cng}
\end{figure}

First let us consider scattering of a gauge invariant current $j = -
\frac{1}{4} F_{\mu \nu}^a F_{\mu \nu}^a$ on a nucleus. The single
gluon inclusive production amplitude for this process viewed in
covariant gauge $\partial \cdot A = 0$ is depicted in
Fig. \ref{cng}.
 The current (dashed line) interacts with a nucleon in
the nucleus producing a gluon, which then rescatters on the other
nucleons in the nucleus. The interactions of the propagating gluon
with nucleons are pictured in the quasi--classical approximation
\cite{mM,k,qm}. That limits the interactions to one
(inelastic) or two (elastic) gluon exchanges (see Fig. \ref{cng}) in
the eikonal approximation. Interactions involving more gluons are
suppressed by the powers of $\alpha_s$. We consider the case of not
very high center of mass energy and not very large transverse momentum
of the produced particles, so that there is no large logarithms to
enhance those extra powers of $\alpha_s$. Our expansion is equivalent
to resummation in the powers of the effective parameter $\alpha_s^2
A^{1/3}$ \cite{mM,k}, with $A$ the atomic number of the nucleus.

 A calculation outlined in \cite{mM} yields for the number of gluons
 produced at the transverse position $x_\perp$:
\begin{eqnarray}\label{N}
\tilde{N} (x_\perp^2) = \int d^2 b \, \frac{N_c^2 - 1}{\pi^2 \alpha
N_c x_\perp^2 } \left[ 1 - \exp \left( - \frac{2 \pi^2 \sqrt{R^2 - b^2}
x_\perp^2 \alpha N_c}{N_c^2 - 1} \rho xG(x, 1/x_\perp^2 ) \right)
\right],
\end{eqnarray}
where we assumed that the nucleus is a sphere of radius $R$ and has a
constant density $\rho$. The gluon distribution of each nucleon is
given by $xG(x, Q^2 )$ and is taken at the two-gluon level
\cite{mM}. We note that in covariant gauge everything is
given by the final state interactions, which result in a simple
Glauber expression of Eq. (\ref{N}). If one views the process in the
light cone gauge $A_+ = 0$, where the nucleus is moving in the
``plus'' direction, everything can be pictured in terms of initial
state interactions. The nuclear wave function, given by non--Abelian
Weizs\"{a}cker--Williams field \cite{k}, provides us with a gluon
which already has all the information about the multiple rescatterings
that broaden its transverse momentum.  This gluon simply interacts
with the current producing a final state on-shell gluon \cite{mM}. The
resulting production cross section is, of course, the same as given by
Eq. (\ref{N}), since it is a gauge invariant object.

Similar effects happen in the proton--nucleus (pA) collisions. We are
just going to state the answer for the total inclusive gluon
production cross section in the quasi--classical approximation
\cite{mM}:
\begin{eqnarray}\label{sig}
\frac{d \sigma}{d^2 l d y} = \frac{1}{\pi} \, \int d^2 b \, \left\{
\frac{\partial}{\partial l^2} xG(x, l^2 ) + xG(x, \left< l_\perp^2
\right> ) \, \frac{ e^{ - \frac{\underline{l}^2}{ \left< l_\perp^2
\right> }}}{ \left< l_\perp^2 \right>} \right\}.
\end{eqnarray}
Here $\left< l_\perp^2 \right>$ is the typical transverse momentum
squared of the produced gluon \cite{mM}. Formula (\ref{sig}) has a
very simple physical interpretation: the first term on the right hand
side corresponds to the gluon emission off the proton after the proton
passes through the nucleus and the second term provides us with
gluon's transverse momentum broadening corresponding to the case when
the gluon is already present in the proton's wave function before the
collision and multiply rescatters on the nucleons in the nucleus
during the collision.

\begin{figure}[b]
\begin{center}

\vspace{-1.cm}
\epsfxsize=9cm
\leavevmode
\hbox{ \epsffile{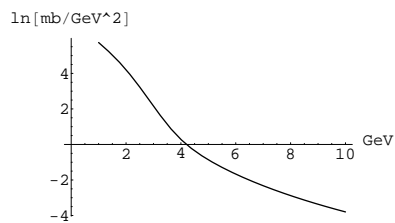}}
\end{center}
\caption{Differential gluon production cross section 
as a function of gluon's transverse momentum for p+Au collision.}
\label{pred}
\end{figure}

Eq. (\ref{sig}) is written for the case of $l^2 \ll \left< l_\perp^2
\right>$. The general formula, which is given in \cite{mM} is plotted
in Fig. \ref{pred} for the case of proton--gold collision, where we
assumed a simple superposition form of the gluon distribution function
of a nucleon $xG_{p,n} \approx 3 xG_q = 3 \frac{\alpha C_F}{\pi} \ln
(Q^2/\Lambda^2) $. In Fig. \ref{pred} we show the logarithm of the
gluon production cross section $\frac{d \sigma}{d^2 l d y}$ as a
function of the transverse momentum of the gluon $l$. We put $\left<
l_\perp^2 \right> = 2 \, GeV^2 $ for RHIC ($\left< l_\perp^2 \right>$
is, in general, energy dependent). At this lowest order the rest of
Eq. (\ref{sig}) is energy (and rapidity) independent.

It is a very interesting and challenging problem to calculate the
gluon production cross section for the case of nucleus--nucleus
collision in the quasi--classical approximation, i.e., resumming all
powers of $\alpha_s^2 A^{1/3}$ (strong field limit). Some attempts in
that direction have been made already \cite{KR}. However they
correspond to the weak field limit \cite{mM,k}. The full
nucleus--nucleus problem is still to be solved.

\newpage
\subsection{X.N. Wang:  HIJING at RHIC}
%4

In this section we will present estimates of hadron spectra in central $Au+Au$
collisions at RHIC energy from perturbative QCD parton models (for high $p_T$
spectra) and HIJING model (for $dN_{\rm ch}/dy$). Since there are many nuclear
effects (including the formation of QGP) which we still don't quantitatively
understand, the extrapolation from $p\bar{p}$ data at the RHIC energy and
$p+A$, $A+A$ data at SPS energy is inevitably rigged with uncertainties. We
will try to estimate these uncertainties by considering different scenarios of
various nuclear effects, like nuclear parton shadowing and jet quenching due to
parton energy loss in dense matter. For definiteness, central collisions are
restricted to events with impact parameters smaller than 3 fm.

%\subsubsection{High $p_T$ hadron spectra \cite{wang98}}

\begin{figure}[htb]

\begin{minipage}[t]{80mm}

%\vspace{0.5cm}
%\centerline{
\psfig{file=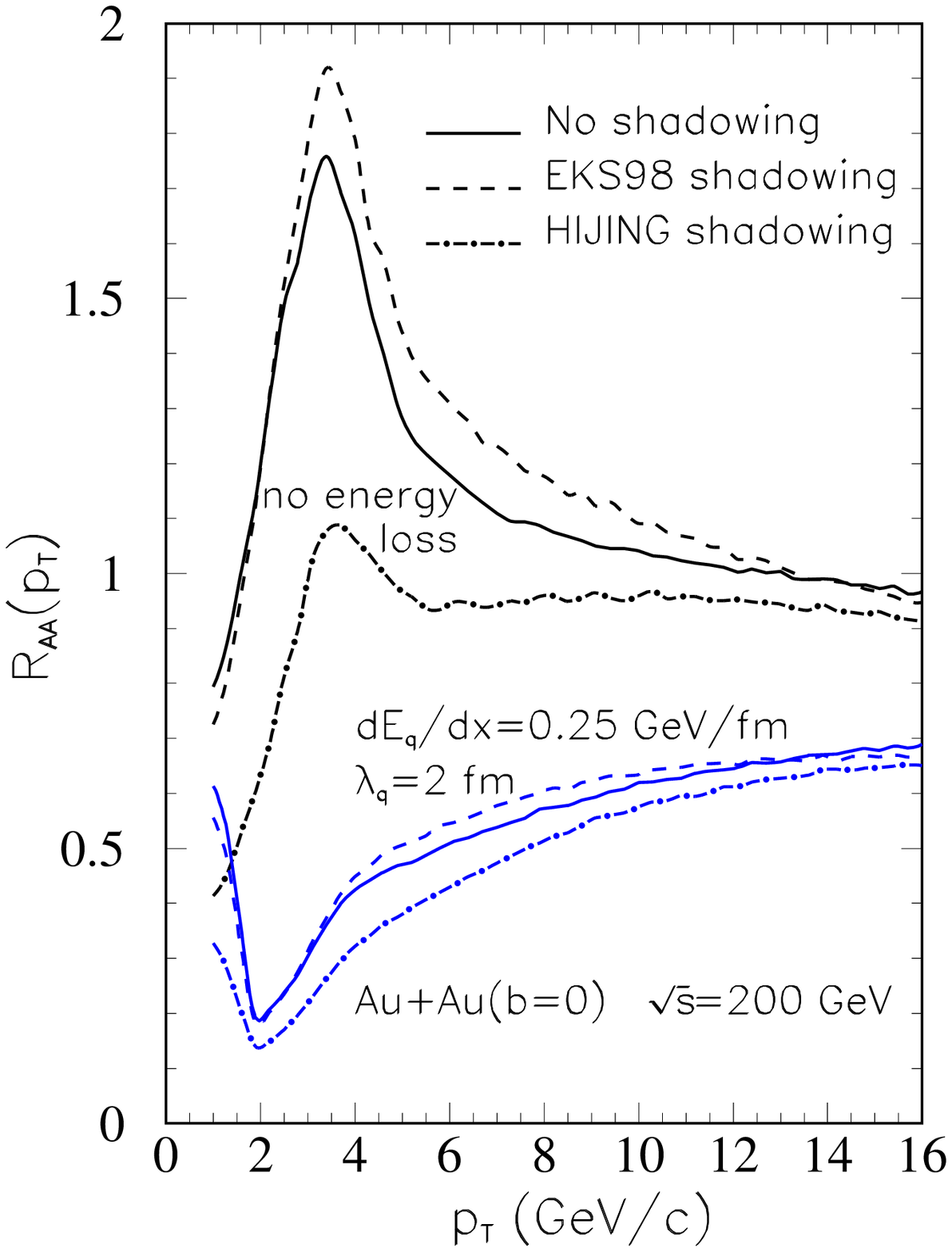,height=3.in,width=2.5in,angle=0}
%}

%\includegraphics[height=3.0in,width=2.5in]{Wang/ratio.eps}
\caption{Ratio of hadron spectra in central $Au+Au$ and $pp$ collisions as
defined in Eq.~\ref{eq:ratio}.}
\label{fig-ratio}
\end{minipage}
%
%\hspace{-2cm}
%\end{figure}
%\begin{figure}[htb]
\hspace{\fill}
\begin{minipage}[t]{75mm}
\centerline{\psfig{file=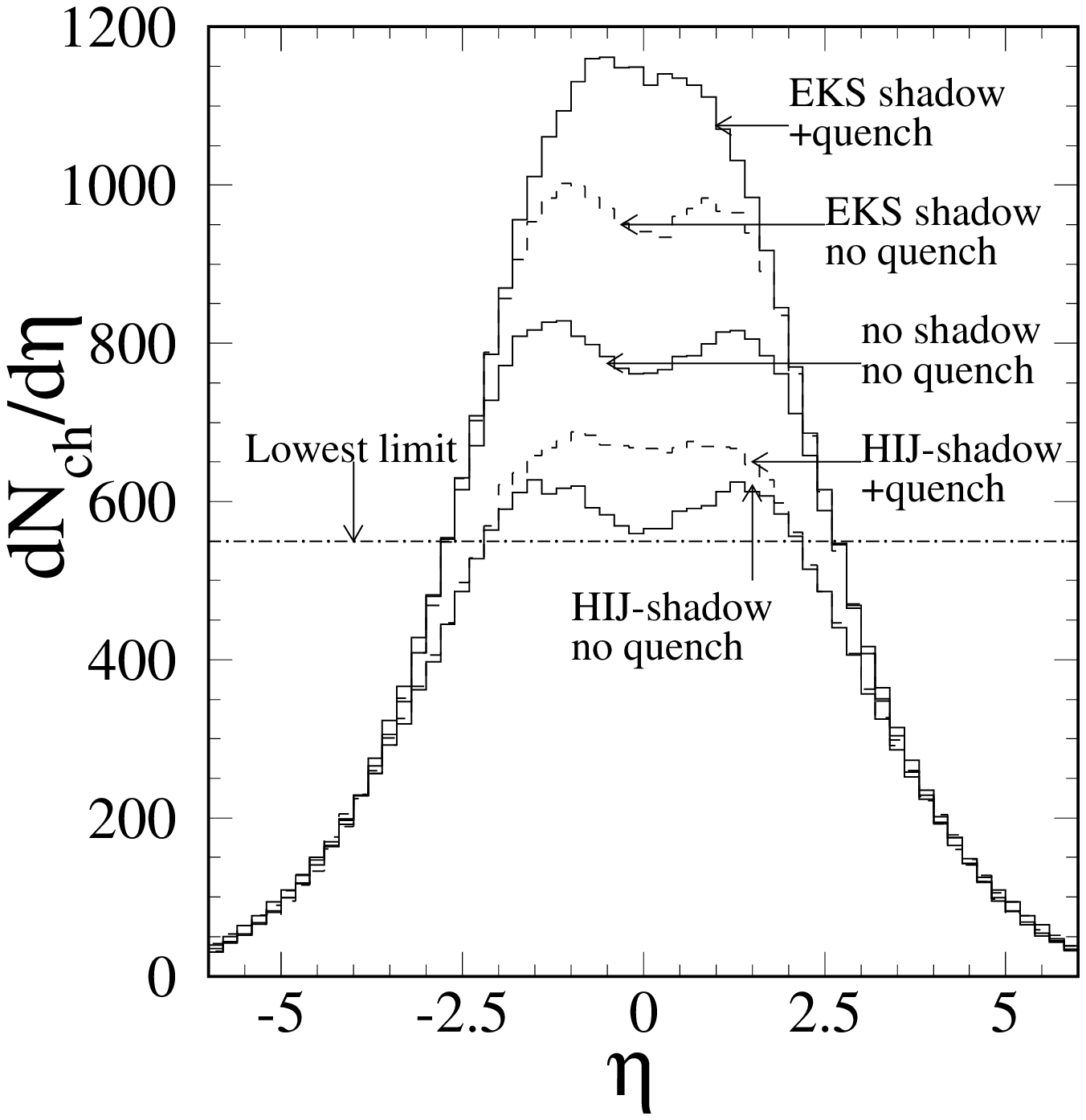,height=3.in,width=2.5in,angle=0}}
\caption{$dN_{\rm ch}/d\eta$ in central $Au+Au$ collisions. EKS \cite{eks98}
and 
HIJ \cite{hijinge} parametrizations of parton shadowing are used. $dE/dx=1$
GeV/fm is used for jet quenching scenario.}
\label{fig-dndy}
\end{minipage}
\end{figure}

Hadron spectra at large transverse momentum in $p\bar{p}$ collisions at
collider energy $\sqrt{s}=200$ GeV have been measured and are consistent with
pQCD parton model calculations \cite{wang98}. The hadron spectra for $pp$
collisions at RHIC energy are expected to be almost the same as $p\bar{p}$ in
central rapidity region. Nuclear effects like multiple parton scattering and
modification of parton distributions in nuclei will modify the hadron spectra
in $p+A$ as data have shown. These observed nuclear effects at energies
$\sqrt{s}=$20 - 40 GeV are well accounted for by including the broadening 
of initial $p_T$ due to multiple parton scattering and the measured nuclear
modification of quark distributions in the pQCD parton model \cite{wang98}.
Extrapolation to RHIC is straightforward with a maximum of uncertainty of about
25\% for $p_T$ spectra (normalized to the target nucleons $A$) at around
$p_T$=4 GeV/$c$. The uncertainty mainly comes from nuclear shadowing of gluon
distribution in nuclei which only becomes important at or above RHIC
energy. Such uncertainties are significantly larger in $Au+Au$ collisions as
shown by the upper set of curves in Fig.~\ref{fig-ratio} where the ratios
\begin{equation}
  R_{AA}(p_T)=\frac{d\sigma_{AA}/dyd^2p_T}{\langle N_{\rm binary}\rangle
      dN_{pp}/dyd^2p_T}    \label{eq:ratio}
\end{equation}
are plotted and $\langle N_{\rm binary}\rangle$ is the averaged number of
binary $NN$
collisions in $A+A$ collisions from the nuclear geometry. Two different
parametrizations\cite{hijinge,eks98} of the gluon shadowing are used in the
calculation. If the leading partons suffer medium-induced energy loss, the
spectra ratios will be suppressed as shown by the lower set of curves in
Fig.~\ref{fig-ratio}. So far, $A+A$ collisions at SPS energy have not shown any
indication of energy loss \cite{wang98}, it is difficult to predict
quantitatively the parton energy loss $dE/dx$ and that is where the biggest
uncertainty in $p_T$ spectra comes from. However, any suppression of the
spectra at large $p_T$ will be a clear indication of parton energy loss which
will be a direct evidence of early parton thermalization.

%\subsubsection{$dN_{\rm ch}/d\eta$ \cite{hijinge}}
The estimate of $dN_{\rm ch}/d\eta$ in $A+A$ collisions at the RHIC energy is much
more uncertain than the high $p_T$ spectra because of the dominance of soft
physics involved. Here we use HIJING model \cite{hijinge} which incorporate both
string-like soft production mechanism and minijet production. We include both
the effect of parton shadowing and jet quenching (parton energy loss or partial
thermalization of hard partons). The formation of QGP or the early
thermalization of soft partons and the subsequent expansion will add more
uncertainty to the estimate of the final observed $dN_{\rm ch}/d\eta$.

Shown in Fig.~\ref{fig-dndy} are estimates of $dN_{\rm ch}/d\eta$ from HIJING
with 
different parametrizations of shadowing. Uncertainties due to effect of jet
quenching are also shown with or without parton energy loss. Including all
these effects, there are an uncertainty of about a factor of 2 in the final
$dN_{\rm ch}/d\eta$. Taken the value of $dN_{\rm ch}/d\eta=2.5$ in $p\bar{p}$
collisions at the RHIC energy and assuming the $A$-scaling at RHIC energy is
the same as at SPS ($PbPb/pp \approx 250$), one has the lowest limit of
$dN_{\rm ch}/d\eta=550$ consistent with the HIJING's lowest estimate. The
clarification of jet quenching from high $p_T$ spectra and of the shadowing
from $p+A$ data will help to narrow down the uncertainty and help us to
understand the experimental value of $dN_{\rm ch}/d\eta$.

%\end{document}
\newpage
\subsection{S.A. Bass:     VNI+UrQMD}

Transport theory offers the unique possibility to cast the entire
time evolution of an ultra-relativistic  heavy-ion reaction at RHIC energies 
-- from its initial state to freeze-out -- 
into one consistent framework.
In microscopic transport models the full 
space-time evolution of all microscopic degrees of freedom -- either
all hadrons present in the system or alternatively 
(at higher beam energies) partons --
is calculated from the initial state to the final freeze-out. 
At RHIC, the description of the full reaction
dynamics from initial state to freeze-out needs to incorporate both, 
partonic and hadronic, degrees of freedom explicitly in a consistent fashion.

In this contribution \cite{transparencies}, 
a combined microscopic partonic/hadronic transport
scenario is introduced: 
the initial high density partonic phase of the heavy-ion reaction is
calculated in the framework of a parton cascade (VNI) \cite{geiger}, using 
cross sections obtained in the framework of perturbative QCD.
The partonic state is then hadronized via a 
configuration space coalescence
and cluster hadronization model and used as initial condition
for a hadronic transport calculation using the 
Ultra-Relativistic-Quantum-Molecular-Dynamics (UrQMD) \cite{urqmd} 
hadronic transport  approach, which has been extensively tested
in the AGS and CERN/SPS energy regimes. 

\begin{figure}[tbh]
\centerline{\psfig{figure=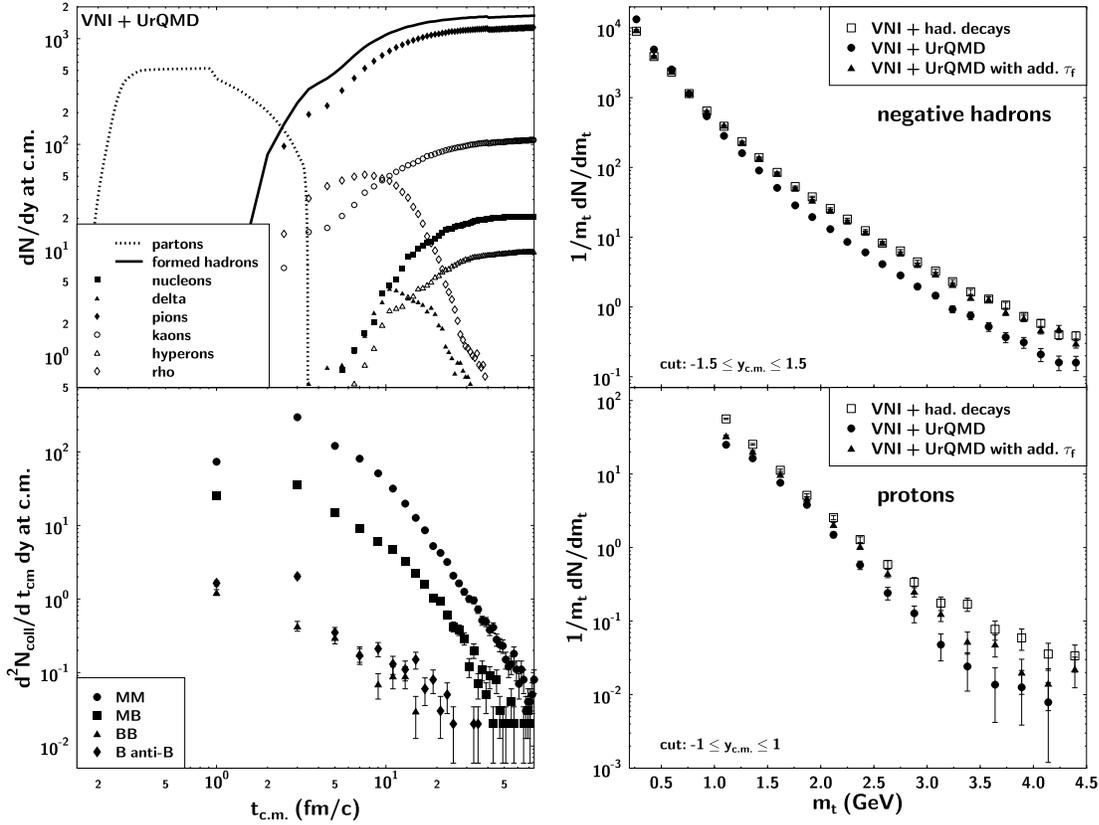,width=6in}}
\caption{\label{tevol} 
Top left: time evolution of the parton and on-shell hadron
rapidity densities at c.m. for central ($b\le 1$~fm) Au+Au collisions at
RHIC. Bottom left: Rates for hadron-hadron collisions per
rapidity at c.m. . Right: transverse
mass spectra for negatively charged hadrons (top) and protons (bottom) 
with and without hadronic rescattering. }
\end{figure}

The upper left frame of Fig.~\ref{tevol} 
shows the time evolution (in c.m. time, 
$t_{c.m.}$)
of the rapidity density dN/dy of partons (i.e. quarks and gluons) 
and on-shell hadron
multiplicities at $|y_{c.m.}|\le 0.5$.
Hadronic and partonic phases evolve in
parallel and both, parton-parton as well as hadron-hadron interactions 
occur in the same space-time volume.
The overlap between the partonic and hadronic stages of the reaction 
stretches from $t_{c.m.} \approx 1$~fm/c up to $t_{c.m.} \approx 4$~fm/c
for the mid-rapidity region. The calculation indicates 
that this overlap occurs 
not only in time but also in coordinate space -- partonic and hadronic degrees
of freedom occupy the same space-time volume during this reaction 
phase \cite{bass_vni}.
Hadronic resonances like the $\Delta(1232)$ and the $\rho(770)$  
are formed and remain populated up to 
$t_{c.m.} \approx 15 - 20$~fm/c, indicating a considerable
amount of hadronic rescattering. 

Rates for hadron-hadron collisions
per unit rapidity at $y_{c.m.}$ 
are shown in the left lower frame of Fig.~\ref{tevol}.
Meson-meson and  meson-baryon
interactions dominate the dynamics of the hadronic phase. Due to their
larger cross sections baryon-antibaryon collisions occur more
frequently than baryon-baryon interactions. 

These hadronic rescatterings influence strongly the $m_t$-spectra of final state
hadrons, e.g. negatively
charged hadrons and protons, as can be seen in the right upper and lower frames of 
Fig.~\ref{tevol}. Here calculations with (full circles) 
and without (open squares) hadronic rescattering are shown:
both, the low-$m_t$ as well as the high-$m_t$ domains of
the proton spectrum show a depletion due to hadronic rescattering. The 
proton yield is depleted by baryon-antibaryon annihilations which manifests 
itself mostly in the low-$m_t$ domain.
For $h^-$, hadronic interactions
lead to a depletion in the high-$m_t$ area of the spectrum and to an 
enhancement in the low-$m_t$ domain. This effect may significantly
affect signatures of partonic in-medium physics, like jet-quenching. 
However, the amount of hadronic rescattering crucially depends on the treatment
of hadronization: if hadrons require an {\em additional} formation time 
$\tau_f \sim \gamma/M_{had}$ 
{\em after} having coalesced from colored constituents into color-neutral 
hadrons {\it before} being able to interact, 
then rescattering effects are strongly suppressed (full triangles).

\newpage
%5
\subsection{A. Dumitru:   Hydrodynamics at RHIC}
I summarize results for Au+Au at RHIC obtained within hydrodynamics
(also combined with a
microscopic kinetic model for the late stage). 
Refs.~\cite{dumitru_webpage,DumBass} provide more figures and papers, where
also references to previous work by others can be found.

The hydrodynamical description of heavy-ion collisions is very
appealing because for a given equation of state
the (drastic) assumption of local equilibrium
uniquely determines the solution within the forward light-cone from
the specified initial conditions. Here, an
EoS with first-order phase transition between a hadronic phase
and a quark-gluon phase is employed, cf.~\cite{dumitru_webpage} for details.
A net baryon and entropy rapidity density of $dN_B/dy=25$, $dS/dy=5000$,
are employed, as might be appropriate for Au+Au at RHIC. 
\begin{figure}[htp]
\vspace*{-.5cm}
\epsfxsize=10cm
\centerline{\hspace{.8cm}
\hbox{\epsffile{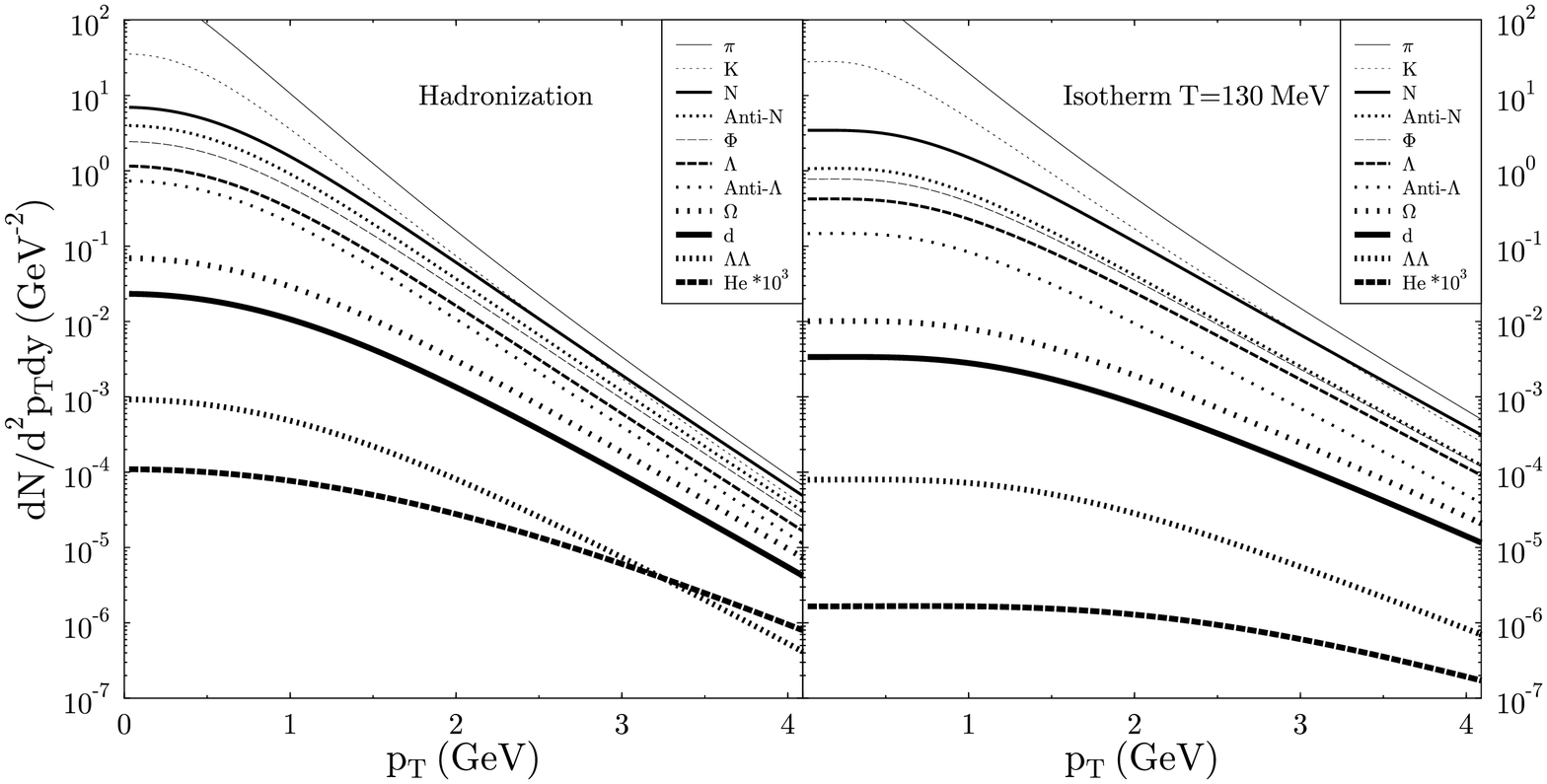}}}
\vspace*{.1cm}
\caption{$p_T$-spectra of various hadrons at hadronization (left) and at
$T=130$~MeV (right) from hydrodynamics w/o final resonance decays
(Au+Au at RHIC, $y=0$, $b=0$).}
\label{RHIC_ptspectra_fig}
\end{figure}  
The spectra of the hadrons
on a given hypersurface can be
obtained from the well-known Cooper-Frye formula. The $p_T$-spectra
at hadronization and on the
$T=130$~MeV isotherm are shown in Fig.~\ref{RHIC_ptspectra_fig}.
One observes the increasing ``stiffness'' with the particle mass, i.e.\
$\langle p_T\rangle$ increases. Also, due to the ``hadronic explosion''
after hadronization collective flow increases substantially
if ideal flow persists even
deep in the hadronic phase (which it doesn't, see below). The work performed
by the isentropic expansion (mostly by that of the QGP)
reduces the transverse energy per unit of rapidity,
$dE_T/dy$, by $40\%$ (initially it is $1.2$~TeV).

In later stages of the reaction the fluid becomes rather dilute
and ideal flow breaks down
(of course, there may be no region of validity at all).
For example, dissipation will
increase together with the mean-free paths.
Furthermore, one can not determine
the freeze-out hypersurface(s) of the various hadron species from ideal
hydrodynamics. 

A solution to these problems is to chose a suitable hypersurface were to
switch from fluid dynamics to a more detailed kinetic theory (e.g.\ the
Boltzmann equation) which explicitly accounts for the various
processes between the particles of the fluid.
One can then {\em calculate} self-consistently the freeze-out domains
of the various elementary processes as determined by the local expansion rate
and the cross-sections of those interactions, and {\em predict}
hadron momentum spectra, two-particle correlators, etc.
Since the equilibrium-limit of the microscopic transport theory is just
hydrodynamics, the switch can
be performed on any hypersurface within the region of validity of
fluid dynamics (the energy-momentum tensors and conserved currents
match on the hypersurface).

\begin{figure}[htp]
\vspace*{-.8cm}
\epsfxsize=8cm
\centerline{\hspace{.8cm}
\hbox{\epsffile{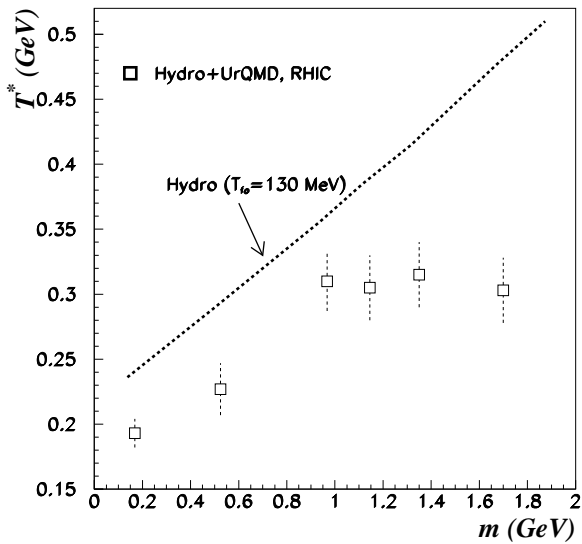}}}
\vspace*{-.5cm}
\caption{Inverse slopes of various hadron species (Au+Au at RHIC, $y=0$,
$b=0$) for pure hydrodynamics (on the $T=130$~MeV isotherm)
and for the Hydro+UrQMD model.}
\label{RHIC_islopes_fig}
\end{figure}

Calculations have been performed~\cite{DumBass} where the switch was
done on the hadronization hypersurface.
The average collective flow velocity (of $\pi$, $K$, $N$, $\Lambda$)
increases from 0.35 at hadronization to 0.4-0.45 at freeze-out.
The late hadronic ``explosion'' is found to be weaker than at SPS,
i.e.\ kinetic freeze-out at RHIC is closer to hadronization and the
system cools down less far. In particular, $\Xi$ and $\Omega$ baryons, which
have small cross sections, decouple shortly
after being hadronized and do not gain substantial flow during the
hadronic stage. Therefore, the inverse slopes of their $m_T$-spectra are
predicted to be smaller than on the $T=130$~MeV
isotherm, cf.\ Fig.~\ref{RHIC_islopes_fig}.
Another interesting observation is that
chemical reactions among the most abundant hadron species ($\pi$, $K$, $N$)
are difficult after hadronization, the respective $dN/dy$'s
change by $\le20\%$ (besides the contribution from resonance decays).
This is because the large expansion rate suppresses inelastic processes,
which typically have larger relaxation times than elastic scatterings.

% Local Variables: 
% mode: latex
% TeX-master: "y"
% End: 

%6
\newpage
\subsection{H.J. Drescher:   NEXUS at RHIC }
{\sc neXus} is a model to describe high energy interactions from 
\( e^{+}e^{-} \) annihilation up to nucleus-nucleus collisions. 
The basic physics features (like fragmentation, parton cascading, etc.)
 are tested and parameters are fixed considering the simplest reaction
 possible. This reduces possible sources of errors and uncertainties.

Basic interaction mechanism in {\sc neXus} are parton ladders, 
calculated as a hard scattering with corresponding initial and 
final state evolution of partons.
The initial state radiation is based on a forward evolution 
algorithm in order 
to treat consistently energy sharing in multiple scattering.

\begin{figure}[h]
{\par\centering{\epsfig{file=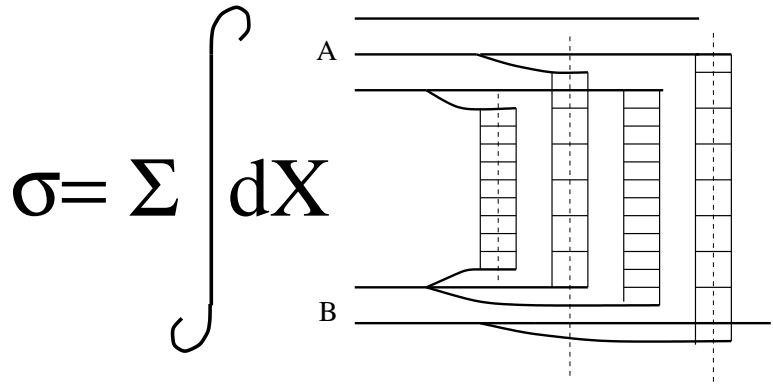,width=4.in}}\par}
\caption{Symbolic representation of the total cross section in {\sc neXus}.}
\end{figure}

It is important to notice that multiple collisions happen 
simultaneously and momentum has to be shared among all collisions 
according to the different contributions
to the total cross section. This leads to a complicated formula with a sum 
over all possible number of cut and uncut parton ladders. To choose one of 
these configurations in a Monte-Carlo framework is a highly dimensional 
problem. This is achieved with techniques coming from statistical 
physics, the Metropolis algorithm and Markov chains. At this stage the
algorithm uses some approximations, but a future version of {\sc neXus}
will treat this properly.

%Stopping is achieved via the fragmentation of excited remnants. Momentum
%transfer during the interaction can excite the nucleons according to a
%parameterized function and forms a quark-diquark string. With a certain
%probability the diquark is on the inner side, shifted towards mid-rapidity.
%The fragmentation of this string may thus shift the baryons into the inner
%region. 

After primary interactions and particle productions via string decay the 
rescattering of particles is done. For this, particles follow classical
trajectories in a Lorenz invariant frame. According to cross-sections, two
particles do an elastic or inelastic scattering. If there are more than two
particles close to each other a quark droplet is formed which evolves 
according to a Bjorken scenario. Dropping below
some critical energy density, it decays according to n-dimensional phase
space decay. This reproduces well experimentally observed behavior like 
strangeness enhancement but is based on a theoretically less firm ground.

\begin{figure}[htb]
{\par\centering{\epsfig{file=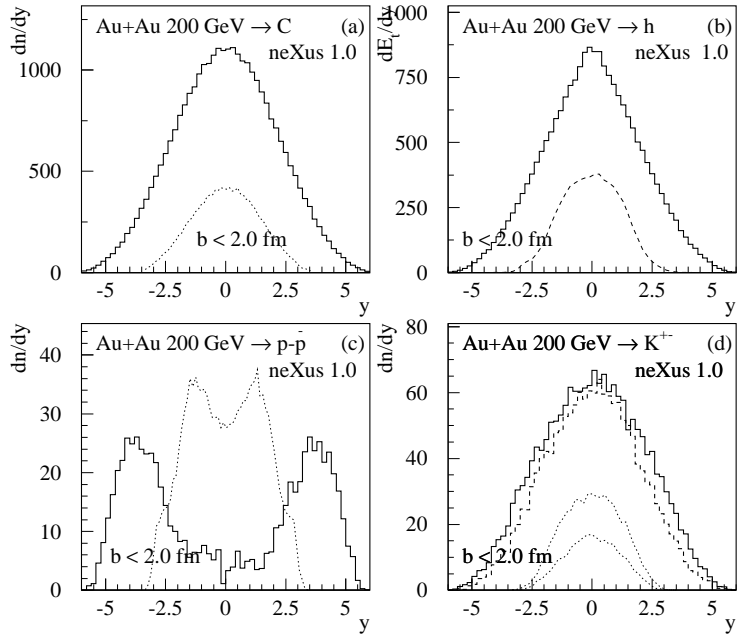,width=4.in}}\par}
\caption{\label{drescher_fig2}
{(a)} Multiplicity of charged particles (full line) 
compared to SPS lead-lead (dashed line).
{(b)} Transverse energy \( dE_{t}/dy \) of all particles (full line) 
compared to SPS (dashed). 
{(c)} Rapidity of net protons (full line) compared to SPS results 
(dashed line). 
{(d)} $K^+$ (full line) and $K^-$ (dashed) and the corresponding 
SPS results (light dashed lines). 
}
\end{figure}

{\sc neXus} calculations for gold on gold at 200 GeV center of mass
energy and an impact parameter less than 2 fm show a charged particle
multiplicity of about 1100 particles per rapidity unit in the mid-rapidity
region (Fig.~\ref{drescher_fig2}(a)). 

The transverse energy \( dE_{t}/dy|_{y=0} \) of hadrons is about \( 850
\) GeV which leads to an energy density of about \( 7\; 
{\rm GeV}/{\rm fm}^{3} \)(Figure
\ref{drescher_fig2}(b)). These two observables should have an error 
bar of about 20\% due to
approximations in the primary interactions and uncertainties in the
rescattering formalism.

More significant should be the result for the rapidity distribution of net
protons since it is quite insensitive to secondary interactions. We find a
great transparency with a multiplicity of about 5 per rapidity unit at $y=0$,
which means very little stopping power compared to the SPS results 
(Fig.~\ref{drescher_fig2}(c)).

Kaons are produced with a multiplicity of about 60 per rapidity unit 
in the central region (Fig.~\ref{drescher_fig2}(d)).

More plots can be found in \cite{drescher2}.

%7
\subsection{H. Sorge:  RQMD at RHIC (see in this proceedings)}
%8
\newpage
\subsection{M. Bleicher:  URQMD at RHIC}

As a tool for our investigation of heavy ion reactions at RHIC the 
Ultra-relativistic Quantum Molecular Dynamics model (UrQMD) is scrutinized. 
UrQMD is a microscopic transport approach based on the covariant propagation of
constituent quarks and diquarks accompanied by mesonic and baryonic 
degrees of freedom.
It includes rescattering of new and produced particles, the excitation
and fragmentation of color strings and the formation and decay of
hadronic resonances. For further details about the UrQMD model, 
the reader is referred to Ref. \cite{bl1}.
Note, that these proceedings and the used UrQMD version is based on \cite{bl2}.

The UrQMD model has been applied very successfully to
explore heavy ion reaction from below AGS energies (10 AGeV) up to the full CERN-SPS
energy (160 AGeV). This includes detailed studies of particle abundances, -spectra,
photonic and leptonic probes, $J/\Psi$'s and event-by-event fluctuations \cite{bl3}.

Let us now start to tackle the relevant questions given to us by the
upcoming of the RHIC  directly:
\begin{itemize}
\item
Is a model like UrQMD applicable to AA reactions at RHIC energies?
\item
What is the predicted amount of baryonic stopping achieved at RHIC?
\item
How many particles will be produced?
\item
How strong will the transverse expansion be?
\end{itemize}

Let me start by answering the first question: It has often been claimed
that dual string models fail to describe data above a certain center of
mass energy. Let us take a look if this region is reached at RHIC:
Comparing the UrQMD predictions to rapidity distributions in He+He at $\sqrt{s}=31 GeV$ (ISR)
leads to a very good description, if we increase to energy
further UrQMD starts to deviate from the measured multiplicity data by 35\% at $\sqrt{s}=200
GeV$. 

However the collision spectra of individual collisions shows that less
than 20\% of all reactions belong to this high energy regime ($\sqrt{s}>100
GeV$). Thus, the remaining 80\% of reactions are 'low energetic' and therefore well
treatable by string dynamics.

It has been claimed recently, that exotic mechanisms (Baryon junctions)
need to be invoked to understand the baryon number
transport at SPS and RHIC. In contrast to these approaches the UrQMD model
mainly applies quark model cross sections to the subsequent scattering
of constituent (di-)quarks in combination with a small diquark breaking component

The stopping power obtained with this approach is rather strong. The
net-proton distribution is shifted by more than one united in rapidity,
resulting in approximately 12 net-protons at midrapidity. The stopping
of hyperons is even stronger (2 units in rapidity, net-$\Lambda+\Sigma$)
leading to a large hyperon density in the central region. 

The particle production predicted by the UrQMD approach reaches more
than 1100 Pions (750 of them are charged Pions) at midrapidity with
approx. 100 additional charged Kaons at midrapidity.

This amount of charged particle abundances at midrapidity is on the 
lower bound of the expected multiplicity at $y=0$ which reaches from 600 
to 2000. It is interesting to note that the particle yield is similar
to the presented RQMD results and strongly deviates from the reported results 
obtained by pQCD dominated models.

Let us turn to the transverse expansion created in the UrQMD model.
Since the early UrQMD dynamics is based on string degrees of freedom,
the newly created quarks are not allowed to interact until they have
undergone their coalescence into hadrons (typically this needs 1fm/c in
the local rest frame of the coalescing quarks). Due to the large Lorentz 
$\gamma$-factor, this leads to a vanishing pressure in the initial reaction
zone. This behaviour is clearly visible in the distribution of the mean
transverse momentum  as a function of rapidity. 
A plateau in the
$\langle p_\perp \rangle (y)$ is observed for the newly created mesons in the
central rapidity region - the transverse momenta at $y=0$ are similar to their
$pp$ values. 
In the case of the Protons a clear dip in the the $\langle
p_\perp \rangle (y)$  at midrapidity is predicted.

Let me conclude that the semi-hadronic UrQMD model has been applied to
Au+Au reactions at RHIC. The use of such an effective model for RHIC 
may be disputable. However, rigorous  
QCD predictions (i.e. results which can be
directly compared to data to falsify or support this approach) 
are not available or are even contradicting each other.
Therefore it is of utmost importance to use a reliable phenomenological 
model, like the UrQMD, which has successfully 
described a large body of measured data at AGS and SPS to investigate
the dynamics encountered at RHIC.

\newpage
%9
\subsection{B. Zhang:   HIJING+ZPC+ART }
%10
%\documentstyle[12pt,twoside,fleqn,psfig,espcrc1]{article}

%\begin{document}

%\subsection{HIJING+ZPC+ART predictions \cite{hza1,hza2}}

To study heavy ion collisions at the
Relativistic Heavy Ion Collider (RHIC), we have developed a multi-phase
transport model that includes both initial partonic and final
hadronic interactions. Specifically, the parton cascade model ZPC
\cite{zpc} is extended to include the quark-gluon to hadronic 
matter transition and also final-state hadronic interactions based on the
ART model \cite{art}. 

Currently, the ZPC model includes only gluon-gluon elastic
scatterings with its cross section taken to be the leading
divergent cross section regulated by a medium generated screening
mass that is related to the phase space density of produced partons. 
The input of the parton phase space distribution 
to the ZPC model is obtained from the
HIJING model \cite{hijing}, which includes both hard scatterings 
via the PYTHIA routines with the nuclear shadowing effect and
soft interactions using the Lund soft momentum 
transfer model. Partons are produced from these scatterings with
a formation time determined according 
to a Lorentzian distribution with a half width given by the ratio
of its energy to the square of its transverse mass.
The positions of formed partons are then calculated by using
straight line propagation from their parent nucleon positions. 

Once partons stop interacting, they are converted into hadrons
using the HIJING fragmentation scheme after a proper
time of approximate $1.2$ fm. We consider both the
default fragmentation scheme of treating a diquark 
as a single entity and a modified one which
allows the formation of diquark-antidiquark 
pairs, that fragment into both 
$BM\bar{B}$ and $B\bar B$ with probabilities of 80\%
and 20\%, respectively.

For the hadron evolution, we have used the ART model that includes both
elastic and inelastic hadron scatterings. 
Multiparticle production is modeled through the
formation of resonances. Since the inverse double resonance
channels have smaller cross sections than those calculated directly
from the detailed balance relation, we have adjusted the double resonance
absorption cross sections to improve the model.

\begin{figure}[htb]
%\vspace{-0.15cm}
\centerline{\psfig{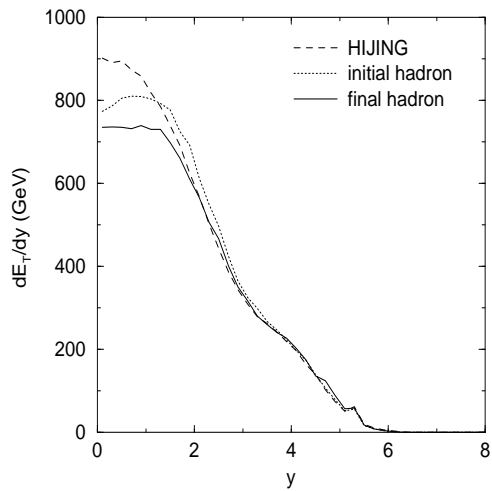}}
\caption{
Transverse energy rapidity distribution for
central (b=0) Au+Au collisions at RHIC.}
\label{fig:rhic1}
\end{figure}

\newpage
\begin{figure}[ht]
\vspace{0.2cm}
\centerline{\psfig{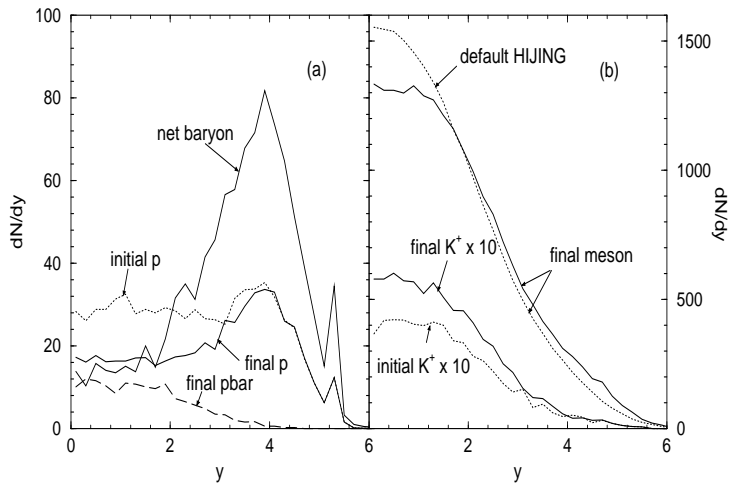}}
\caption{
Predictions of (a) baryon and (b) meson rapidity distributions for
central (b=0) Au+Au collisions at RHIC.
\label{fig:rhic2}}

\end{figure}

Hadron transverse energy rapidity distributions 
for RHIC Au+Au central (b=0) collisions are shown in
Fig.~\ref{fig:rhic1}. We see that parton evolution and fragmentation 
lower the distribution by about 100 GeV per unit rapidity while hadron 
evolution further decreases it by around 50 GeV per unit rapidity. This
shows an observable effect of final state interactions.
Fig.~\ref{fig:rhic2} gives the baryon and meson rapidity distributions.
Net baryon distribution has a large rapidity shift due to the modified
fragmentation scheme.
Many antiprotons are seen to survive the
absorption in the hadronic matter, leading to a value of about 10
at central rapidities. The final meson central rapidity distribution
shows a distinctive plateau structure. Results
using the default HIJING show a similar distribution except that
the central rapidity density is higher. Also shown in the figure
is the distribution of kaons produced from both string
fragmentation and hadronic production. The latter 
gives significant enhancement of the kaon production.

%\end{document}
\newpage
\subsection{W. Cassing:  VNI+HSD }

Intuitively one expects that the initial nonequilibrium phase of a
nucleus-nucleus collision at RHIC energies should be described by
parton degrees of freedom whereas hadrons are only formed (by
'condensation') at a later stage of the reaction which might be a
couple of fm/c from the initial contact of the heavy ions.
 Thus parton
cascade calculations \cite{Geiger1} -- including transitions rates
from perturbative QCD -- should be adequate for all initial
reactions involving a large 4-momentum transfer between the
constituents since QCD is well tested in its short distance
properties. The question, however, remains to which extent the
parton calculations can be extrapolated to low $Q^2$, where
hadronic scales become important, and under which conditions
(temperature, quark chemical potential) the expanding system of
strongly interacting partons is more appropriately described by an
expanding, but interacting hadron gas.

The practical question is, however,  if nonequilibrium partonic
and hadron/string models can be distinguished at all, i.e. do they
lead to different predictions for experimental observables? We
will thus  compare the predictions from the parton cascade VNI
\cite{Geiger1} with those from the HSD transport approach
\cite{CB99}, which involves quarks, diquarks, antiquarks,
antidiquarks, strings and hadrons as degrees of freedom.

\begin{figure}[t]
%\phantom{a}\vspace*{-12mm}
%\centerline{\psfig{file=Cassing/cassing.eps,height=10.5cm}}
\centerline{\psfig{file=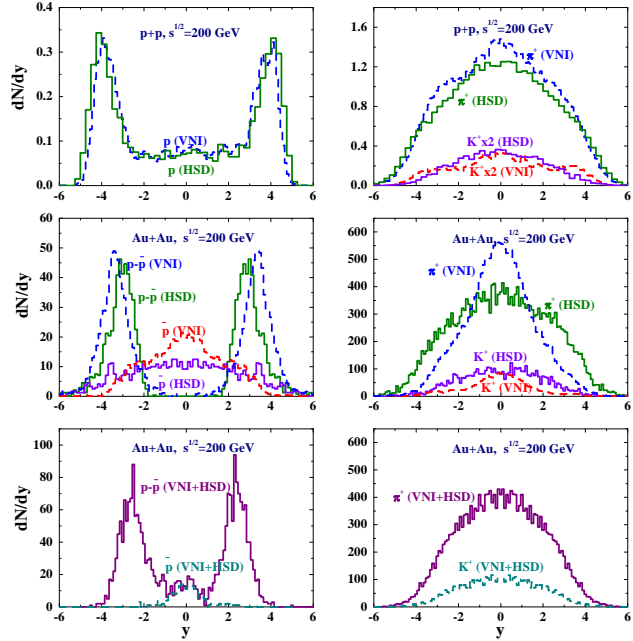,height=14.5cm}}
\phantom{a}\vspace*{-4.5cm}
\caption{The rapidity spectra of hadrons for $pp$ (upper part) and 
central collisions of $Au + Au$ at $\sqrt{s}$ = 200 GeV within 
VNI, HSD and VNI + HSD (see text).}
\label{cassing}
\end{figure}

We start with $pp$ collisions at $\sqrt{s}$ = 200 GeV.  The
calculated results for the proton, $\pi^+$ and $K^+$ rapidity
distributions in the cms are shown in Fig. \ref{cassing} (upper part)
for both models,
which are denoted individually by the labels VNI (dashed
histograms) and HSD (solid histograms).  On the level of $pp$
collisions we find no considerable differences between the two
kinetic models with respect to the rapidity distributions for $p,
\bar{p}, \pi's$ and $K, \bar{K}$. This also holds for the $p_T$ spectra.

We directly step towards central collisions ($b \leq$ 2 fm) for Au
+ Au at $\sqrt{s}$ = 200 GeV. The calculated results for the
proton (here $p-\bar{p}$), antiproton, $\pi^+$ and $K^+$ rapidity
distributions in the cms are shown in Fig. \ref{cassing}
(middle part) for both
models, which are denoted again by the labels VNI (dashed
histograms) and HSD (solid histograms). Here the hadron-string
model shows a larger stopping than the parton cascade (l.h.s.) and
a flatter distribution in rapidity of antiprotons than the
partonic cascade. The pion and kaon multiplicities turn out to be
roughly the same, but the results from the parton cascade are more
strongly peaked around midrapidity as those from the HSD approach.
  In physical terms the larger stopping of
the HSD approach and the additional production of pions and kaons
(relative to $pp$ collisions) stems from secondary and ternary
reaction channels in the hadronic rescattering phase, which are
quite abundant since the meson densities achieved are very high.
The narrow width of the antiproton and meson rapidity distribution
(as compared to $pp$ collisions) for VNI is due to an approximate
thermalization of the partonic degrees of freedom; in the
hadron-string scenario essentially 'comover' scattering occurs
with a low change of the meson rapidity distribution. Thus the
meson rapidity distributions are practically the same as for $pp$
collisions.  Also note that at midrapidity the net baryon density
$\sim N_p - N_{\bar{p}}$ is practically zero, however, even at
midrapidity at lot of baryons appear that are produced together
with antibaryons. Thus also mesons (especially $c \bar{c}$ pairs)
will encounter a lot of baryons and antibaryons on their way to
the continuum.

In order to demonstrate that the higher baryon stopping, kaon
enhancement and widening of the pion rapidity distribution is
essentially a rescattering effect, we have extended the parton
cascade VNI by the hadronic rescattering phase as described by HSD
\cite{CB99}. The results of these calculations (VNI + HSD) for
central $Au + Au$ collisions are displayed in the lower part
of Fig. \ref{cassing}. It is worth
to note that VNI+HSD almost give the same rapidity distributions
as HSD which already includes the final state interactions.

We also have performed calculations for low mass dileptons and
high mass dimuons \cite{CB99}, where especially the $J/\Psi$ and
$\psi^\prime$ peaks are of interest.  Without explicit
representation we note that in the dimuon spectra -- in the
'comover' absorption scenario \cite{BC97} -- the $\Psi^\prime$
peak vanishes almost completely whereas the $J/\Psi$ peak is
suppressed by $\approx$ 90\% in central Au + Au collisions due to
the high meson densities achieved at these energies.

\newpage
\section{HADRONIC FLAVOR OBSERVABLES} 
% 11
\subsection{J. Zim\'anyi:  Quark Coalescence}
%12
%\input{Zimanyi/zimin.tex}
%%%%%%%%% coalc.tex %%%%%%%%%%%%%%%%%%%%%%%%%%%%%%%%%%%
%
%       Lecture for QM'99 by J. Zimanyi    
There is a possibility, that at RHIC a nice
quark gluon plasma will be produced in
the collision.
This plasma will expand and cool
in the next phase. Thus it may happen,
that just before the hadronization the same type
of matter will be created as in SPS experiments.
It is assumed to consist of  constituent
quark-antiquark matter (CQM) \cite{ALCOR,ALCOR99}.

Then the hadronization may proceed via quark
coalescence process as described by the ALCOR model
just as in the case of SPS.
Thus it is worthwhile to make predictions for 
RHIC assuming coalescence mechanism.

The nonlinear ALCOR coalescence model was created for situations, 
where the subprocesses are not independent, they compete with
each other. In this model the coalescence equations relating the number 
of a given type of hadron to the product of the numbers of 
different quarks from which the hadron consists reads as: 
\begin{eqnarray}
    N_{p}  = C_p \cdot b_q \cdot b_q \cdot b_q \cdot N_q \cdot N_q \cdot N_q
&\ \ & N_{\overline p}  = C_{\overline p} \cdot 
b_{\overline q} \cdot b_{\overline q} \cdot b_{\overline q} \cdot
N_{\overline q} \cdot N_{\overline q} \cdot N_{\overline q} 
\nonumber \\
    N_{\Lambda} = C_{\Lambda} \cdot 
    b_q \cdot b_q \cdot b_s \cdot N_q \cdot N_q \cdot N_s
&\ \ & N_{\overline \Lambda}  = C_{\overline \Lambda} \cdot 
b_{\overline q} \cdot b_{\overline q} \cdot b_{\overline s} \cdot
N_{\overline q} \cdot N_{\overline q} \cdot N_{\overline s} 
\nonumber \\
    N_{\Xi}  = C_{\Xi} \cdot 
    b_q \cdot b_s \cdot b_s \cdot N_q \cdot N_s \cdot N_s
&\ \ & N_{\overline \Xi}  = C_{\overline \Xi} \cdot 
b_{\overline q} \cdot b_{\overline s} \cdot b_{\overline s} \cdot
N_{\overline q} \cdot N_{\overline s} \cdot N_{\overline s} 
\nonumber \\
    N_{\Omega}  = C_{\Omega} \cdot 
    b_s \cdot b_s \cdot b_s \cdot N_s \cdot N_s \cdot N_s
&\ \ & N_{\overline \Omega}  = C_{\overline \Omega} \cdot 
b_{\overline s} \cdot b_{\overline s} \cdot b_{\overline s} \cdot
N_{\overline s} \cdot N_{\overline s} \cdot N_{\overline s} 
\label{coal}
\end{eqnarray}
\begin{eqnarray}
    N_\pi  &=& C_\pi \cdot 
b_q \cdot b_{\overline q} \cdot N_q \cdot N_{\overline q}
\nonumber \\
    N_K  &=& C_K \cdot 
b_q \cdot b_{\overline s} \cdot N_q \cdot N_{\overline s}
\nonumber \\
    N_{\overline K}  &=& C_{\overline K} \cdot 
b_{\overline q} \cdot b_s \cdot N_{\overline q} \cdot N_s
\nonumber \\
    N_{\eta} &=& C_{\eta} \cdot 
b_{\overline s} \cdot b_s \cdot N_{\overline s} \cdot N_s
\label{coalm}
\end{eqnarray}
Here the normalization coefficients, $b_i$,  are determined uniquely
by the requirement, that { the number of the constituent
quarks do not change during the hadronization ---
which is the basic assumption for all quark counting methods}:
\begin{eqnarray}
 N_s &=& 3 \cdot N_\Omega + 2 \cdot N_\Xi + 1 \cdot N_\Lambda +
      1 \cdot N_{\overline K} + 1 \cdot N_{\eta}  \nonumber \\
 N_{\overline s}  &=& 3 \cdot N_{\overline \Omega} + 
2 \cdot N_{\overline \Xi} + 1 \cdot N_{\overline \Lambda} +
      1 \cdot N_{K} + 1 \cdot N_{\eta} \nonumber \\
 N_q &=& 3 \cdot N_p + 1 \cdot N_\Xi + 2 \cdot N_\Lambda + 1 \cdot N_{K}+
      1 \cdot N_\pi  \nonumber \\
 N_{\overline q}  &=& 3 \cdot N_{\overline p} + 
1 \cdot N_{\overline \Xi} + 2 \cdot N_{\overline \Lambda} +
      1 \cdot N_{\overline K}
    + 1 \cdot N_\pi  \ . \label{cons} 
\end{eqnarray}
In eq.~(\ref{cons}) $N_\pi $ is the number of directly produced pions.
(Most of the observed pions are created in the decay
of resonances.)
Substituting eqs.~(\ref{coal} - \ref{coalm}) into
 eq.~(\ref{cons}) one obtains  equations
for the for normalization constants. These constants
are then given in terms of  quark numbers and  $ C_i $ factors.
 However, one can predict some relations even without solving the
set of nonlinear equations, eq.~(\ref{cons}) \cite{Bial98,Zim99}. 
Thus one obtains the interesting relations for the 
antiparticle - particle ratios:
$$
\frac{\overline \Lambda}{\Lambda} = D \cdot \frac{\overline p}{p}\ \ ,
\ \ \ \ \
\frac{\overline \Xi}{\Xi} = D \cdot \frac{\overline \Lambda}{\Lambda}\ \ ,
\ \ \ \ \
\frac{\overline \Omega}{\Omega} = D \cdot \frac{\overline \Xi}{\Xi}\ \ ,
$$
where $D=K^+/K^-$ (see details in Ref.~\cite{Zim99}).
At SPS energy the measured value is $D=1.8 \pm 0.2$ \cite{EXP3}
which agrees within the experimental errors with $D=1.95$ obtained in ALCOR.
Our prediction for RHIC energy is $D=1.2$.

\begin{figure}
\vskip -0.4in
\centerline{
\psfig{figure=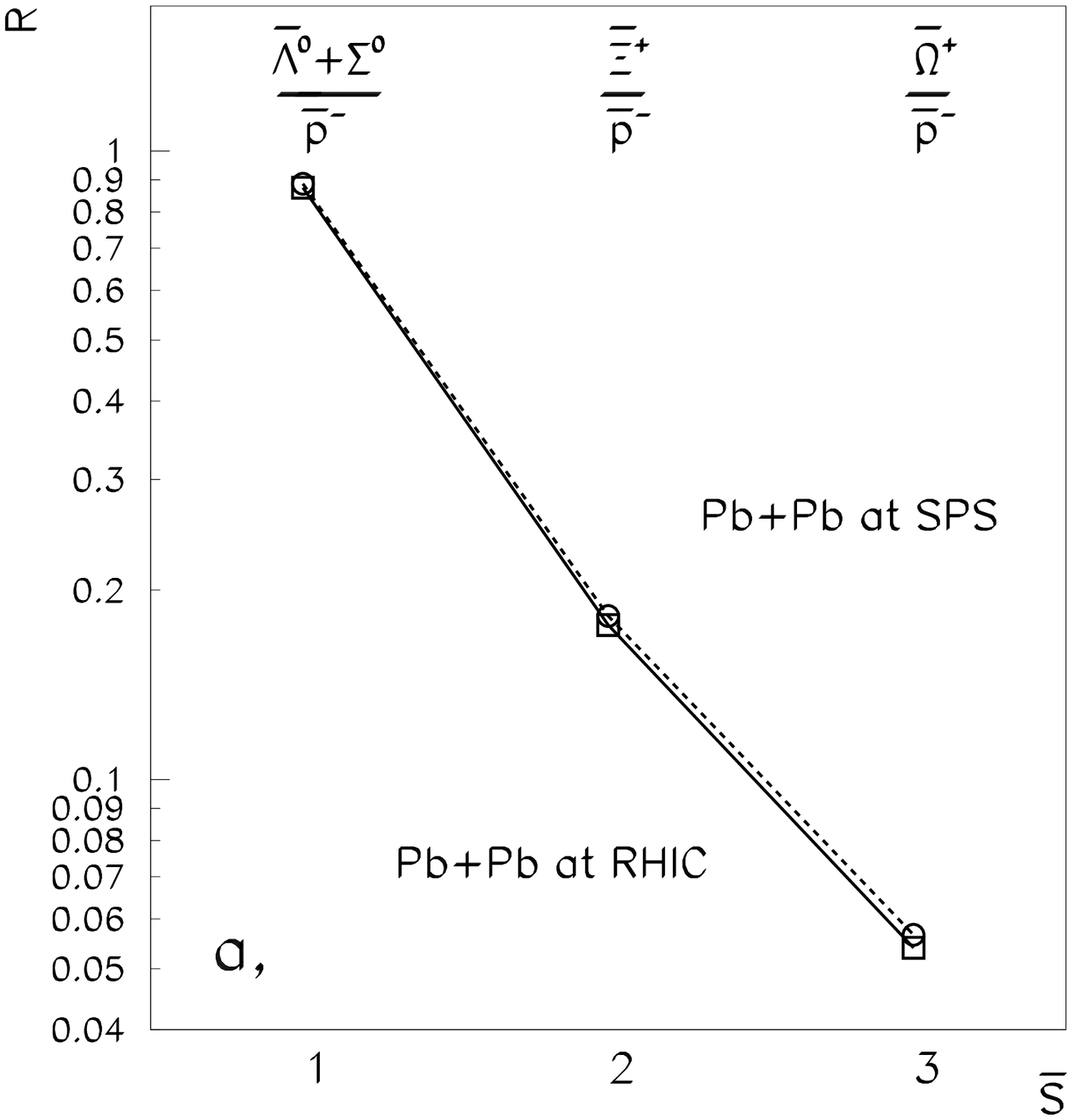,height=4.0in,width=3.0in}
\psfig{figure=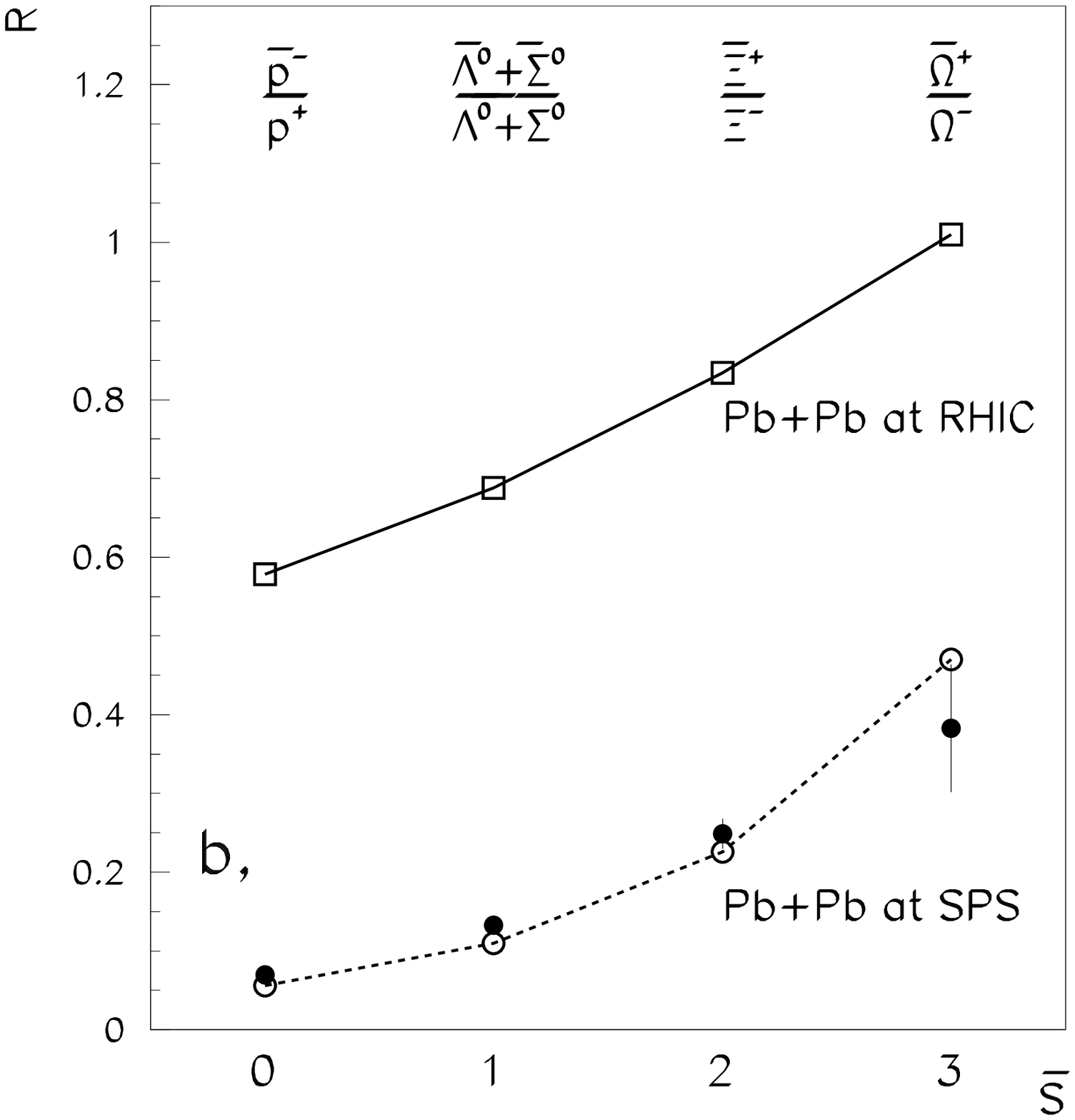,height=4.0in,width=3.0in} }
\vskip -0.05in
\caption[]{
 \label{figurezim} 
Particle ratios in Pb + Pb collisions calculated for SPS (open circle)
and predicted for RHIC (open square). Experimental data (filled circle)
are from Refs.~\cite{EXP1,EXP2}.
In Fig.~\ref{figurezim}(a) the ratios for SPS and RHIC are practically 
the same.
}
\end{figure}

Some particle ratios are shown for the
SPS and RHIC energies in Fig.~\ref{figurezim}.  
For the RHIC calculations the 
same parameters were used as for the SPS, except
the number of produced quark antiquark pairs 
($N_{q{\overline q}}=834, N_{s{\overline s}}=92$ at SPS;
$N_{q{\overline q}}=4200, N_{s{\overline s}}=462$ for RHIC).
This number 
was determined by the requirement, that the same
 $ h^- $ multiplicity should be obtained as  
from HIJING.

%%%%%%%%%%%%%%%%%%%% REFERENCES %%%%%%%%%%%%%%%%%%%%%%%

\newpage
\subsection{ J. Stachel:  Fireball Model Predictions}

Predictions for hadron yields in the central region at RHIC
\cite{col_stachel,talk_stachel} can be based
on a thermal model describing successfully the data obtained in present
fixed target experiments at the AGS \cite{thermal1} and SPS
\cite{thermal2,thermal_pbpb} for central collisions of
Si/S and Au/Pb beams with heavy targets (Au,Pb). The statistical model
treats the system as a grand canonical ensemble and has two free
parameters, a temperature $T$ and a baryon chemical potential
$\mu_B$. Interactions of the particles are taken into account in form of
an excluded volume correction corresponding to repulsion setting in for
all hadrons at a radius of 0.3 fm \cite{thermal_pbpb}. The hadron yield
ratios resulting from this model agree particularly well with the most complete
set of data obtained sofar, the one for central Pb + Pb collisions at
the SPS. Putting these results in perspective vis-a-vis the expected
phase boundary between the quark-gluon-plasma and the hadron gas the
hadrochemical freeze-out points are where one expects the phase boundary and
it it suggestive that hadron yields are frozen at the point when
hadronization is
complete \cite{phase_gerry,inpc}. This suggests for RHIC a hadrochemical
freeze-out temperature of about 170 MeV, the same as found at the SPS. 
The
chemical potential is going to be small; here a value of 10 MeV is used
as an upper limit. Strangeness and charge conservation require then
values of $\mu_s$ and $\mu_{I_3}$ of 2.5 and -0.2 MeV.

\begin{figure}[htb]

\vspace{-1.6cm}

%\hspace*{1.5cm}
\epsfig{file=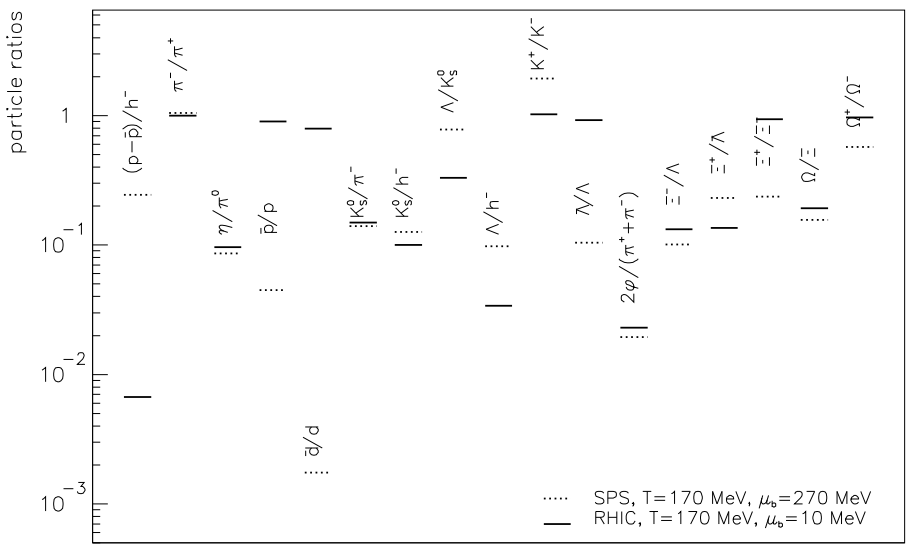,width=13.5cm}
\vspace{-1.9cm}

\caption{\small Hadron yield ratios from a thermal model (see text)}
\label{fig:yieldratios}

\vspace{-.8cm}
\end{figure}

The yield ratios as predicted for RHIC using these parameters are shown
in Figure~\ref{fig:yieldratios}. The predictions are 
shown together with a best fit from
the same model to data for central Pb + Pb collisions at the CERN SPS.
One notices drastic differences as compared to present data, in 
particular in the yield
ratios of antibaryons to baryons which are predicted at RHIC to be about
1 (see Fig.~\ref{fig:yieldratios}) due to the small chemical potential.

In order to predict absolute yields one has to estimate the volume per
unit rapidity at the time when hadronization is complete. Starting from an
initial temperature of $T_i$ = 500 MeV at a time $\tau_i$ = 0.2 fm/c and
using a transverse expansion in the plasma of $\beta_{t_p}$ = 0.16 this
volume is 3600 fm$^3$ at a freeze-out temperature of 170 MeV. A
selection of the predicted hadron yields are shown in
Table~\ref{tab:dndy}. The model was used to compute yields for
193 different hadron species in total. The typical systematic error of 
this prediction,
based on the freeze-out parameters, is judged as follows. The chemical
potential is bounded at a lower limit of 0. Yields for $\mu_b$ of 1 MeV
were found to not be noticeably different.  One finds small differences
in the absolute yields of baryons with respect to antibaryons. The
corresponding results are given in column 3 of Tab.~\ref{tab:dndy} where
they differ from the values for $\mu_B$ = 10 MeV. Further, present SPS data
put a lower limit of 160 MeV on the freeze-out temperature. Typical
results for this lower temperature are given in column 4 of
Tab.~\ref{tab:dndy}. Generally particle yields are reduced by about 10
\% (as indicated for pions in Tab.~\ref{tab:dndy}), for the heaviest particles
the reduction is 35 \% ($\Omega$, d).

\vspace{-.5cm}

\begin{table}[htb]
% -----------------------------------------------------
% adapted from TeX book, p. 241
\newlength{\digitwidth} \settowidth{\digitwidth}{\rm 0}
\catcode`?=\active \def?{\kern\digitwidth}
% -----------------------------------------------------
\caption{Hadron yields per unit of rapidity near mid-rapidity}
\label{tab:dndy}
\begin{tabular}{rrrr}
\hline
&&&\\
particle & T=170 & T=170 & T=160 \\
 & $\mu_b$=10 & $\mu_b$=1 & $\mu_b$=10\\
\hline
\(\rm \pi^- \approx \pi^+ \) &  630 &  & 555\\
\(\rm \pi^o \) &  700 &  & \\
\(\rm \eta \) & 68 & & \\
\(\rm \rho^o \) &  70 &  & \\
\(\rm \omega \)  &  56 &  & \\
\(\rm \eta' \)  &  6 &  & \\
\(\rm \phi \)  &  14 &  &  \\
\(\rm K^+ \approx K^- \approx K^0_s \)  & 97 & & \\
\hline
\(\rm p \)  &  62 & 59 & \\
\(\rm n \)  &  57 &  & \\
\(\rm \overline{p} \)  &  56 & 58 & \\
\(\rm p-\overline{p} \)  &  6 & 0.7 & \\
\(\rm \Lambda \)  & 32 & 30 & \\
\(\rm \overline{\Lambda} \)  & 29 & 30 & \\
\(\rm \Lambda(1405) \)  &  2 &  & \\
\(\rm \Xi^- \approx \Xi^+ \)  & 4 &  & 3\\ 
\(\rm \Omega^- \approx \Omega^+ \)  & 0.78 &  & 0.50\\
\(\rm d \)  &  0.23 &  & 0.14\\
\(\rm \overline{d} \)  &  0.18 &  & \\
\hline
\end{tabular}
\end{table}

%13
\newpage
\subsection{J. Rafelski: Strange Hadrons from QGP at RHIC} 
%****************************************************************
% qm99 proceedings for Friday session 
% outline updated 5/31/99
%Organizer: Miklos Gyulassy <gyulassy@nt3.phys.columbia.edu>
%13  section 2 subsection 3  11:25 Rafelski, J., Arizona: 
%\setcounter{section}{2}
%\setcounter{subsection}{2}
%****************************************************************
%\subsection{

%\hfill Jan Rafelski, Arizona}
%%%%%%%%%%%%%%%%%%%%%%%%%%%%%%%%%%%%%%%%%%%%%%%%%%%%%
(Strange) hadronic particle abundances and spectra  
have been obtained \cite{Raf1},
assuming a thermally equilibrated,
chemically non-equilibrated deconfined {\small QGP} source undergoing a 
sudden freeze-out at RHIC. The theoretical framework of this approach 
is presented in some detail in the contribution by {\bf Jean Letessier} 
\cite{Raf2}.

 { }  From the study of the SPS data we deduce the universality of
physical properties of hadron chemical  freeze-out 
\cite{Raf2,Raf3},
occurring practically at the same condition as the kinetic 
freeze-out. The differences in system dependent 
particle $m_\bot$-slopes are understood 
to be a consequence of differences in collective  flow in a deconfined 
{\small QGP} source. We believe that this remains the 
general situation at the 10 times higher  collision energies reached at 
RHIC. We expect that the {\small QGP} break-up temperature 
$T_f^{\small \rm SPS}\simeq 145\pm5$ MeV  will see a minor upward change to say
 $T_f^{\small \rm RHIC}\simeq 150\pm5$ MeV. Thus all we need 
to do in order to characterize particle production at RHIC is to adapt 
the picture of sudden QGP hadronization we have developed to the 
greater longitudinal flow, lower baryon density environment we expect at RHIC. 

In our understanding of the freeze-out, there is complete absence of a pure
hadronic gas, or even a mixed phase in the dense matter evolution. The
deconfined {\small QGP} state evaporates over  a few fm/c in time, during 
which time it is kept by a balance of dynamical processes  near to
the freeze-out temperature: the evaporation and work done against
the vacuum cool the surface down, which is internally heated by 
the explosive energy flows. Empirical evidence at SPS
suggests that the  transverse flow velocity imparted on emitted
hadrons does not surpass
$  v_\bot \simeq {c}/{\sqrt{3}} \equiv v_{\rm sound} $.
The other QGP properties imparted on hadronic particles 
within this sudden  freeze-out model are: 
\begin{eqnarray*}
 \lambda_s=1 &\Leftrightarrow& \langle s-\bar s\rangle =0\,; \\
\gamma_q\simeq 1.5\mbox{--}1.8&\Leftrightarrow& \mbox{
entropy\ enhancement\  in plasma\ and 
\ gluons converted to quarks;}\\
\gamma_s>1&\Leftrightarrow& \mbox{strangeness\ equilibrates\ with\ }
T>T_f^s>T_f;\\
{unknown:\ } \lambda_q&\Leftrightarrow& \mbox{baryon density;
baryon:energy stopping relation}. 
\end{eqnarray*}
 However, $\lambda_q$ must be smaller than seen at 
SPS in S--S ($\simeq 1.4$)  so the range
$\lambda_q\simeq 1.2\pm0.15$ which we base our predictions on, 
is nearly certainly correct. 
%%%%%%%%%%%%%%%%%%%%%%%%%%%%%%%%%%%%%%%%%%%%%%%%%%%%%%%%%%%
 
We thus expect to find, at RHIC,  a kinetically equilibrated, but chemically 
evolving,  QGP source, with  direct hadron emission 
from a flowing surface, just as we found in the case of SPS energy 
collisions.  We expect, based and  compared to the 
Pb--Pb 158$A$ GeV results:\\
\indent $\bullet$ Shape identity  of  RHIC $m_\bot$ and $y$
 spectra of antibaryons 
 $\bar p\,,\ \overline\Lambda\,,\ \overline\Xi\,,$ since 
in our approach there is no difference in 
their production mechanism;\\
\indent $\bullet$  The $m_\bot$-slopes of these antibaryons
 should be very similar to the result we
have from  Pb--Pb 158$A$ GeV
since only a slight increase in the freeze-out temperature 
occurs, and no increase in collective transverse flow is expected;\\
\indent $\bullet$ 
Major changes compared to SPS must be expected in rapidity 
spectra of mesons, baryons and
antibaryons. 

Work on an extension of our 
sudden hadronization  model to include the expected substantial
longitudinal flow at RHIC is underway. We report here preliminary 
results that  deserve immediate attention.
Strangeness is here an extremely interesting observable for:\\
\indent {\bf 1.}  Gluons make strangeness effectively in 
the early stage of the collision and thus 
at SPS strangeness is more central than baryon number.
It is possible that this will be even more accentuated at 
 RHIC leading to  `strangeness dominance' at central rapidity.\\
\indent {\bf 2.}  Central strangeness  production will
further be enhanced by  an increase in initial temperature 
compared to SPS. We anticipate at RHIC
stunning strange hadron abundance anomalies.

%%%%%%%%%%%%%%%%%%%%%%%%%%%%%%%%%%%%%%%%%%%%%%%%%%%%%%%%%%%
In  Table~\ref{table1}  we present  two  scenarios which 
differ by the  strangeness phase space occupancy excess: we take in 
both cases the chemical freeze-out temperature and velocity
 $T_f=150\,\mbox{MeV},$ 
$v_\bot=1/\sqrt{3}\simeq 0.58c$ and the 
chemical light quark conditions: 
$ \lambda_q=1.2\pm 0.15,\ \gamma_q=1.8,$
as well as QGP strangeness conserving $\lambda_s=1$, but, in first case, 
we take $\gamma_s=2 $ and in second $ \gamma_s=1.5\,. $
To  obtain the rapidity densities one of the 
key distributions needs to be known: we use a net baryon density
${d(B-\bar B)}/{dy}\simeq 70$, which if nearly flat in $\Delta y\simeq\pm 3$,
allows to fully account for all participating baryons in a zero impact parameter 
collision. Employing the ratios given as RHIC-1) in  Table~\ref{table1}  
we obtain the results shown in  Table~\ref{table2}, which
imply hyperon-dominance of the baryon yields. RHIC-2) scenario 
would further enhance  the hyperon-dominance.

%%%%%%%%   TABLE RHIC 
\begin{table}[tb]
\caption{\label{table1}Sample of hadron ratios expected at RHIC,
see text explanation of the model parameters}
\small
%\vspace*{-0.1cm}
\begin{center}
\begin{tabular*}{0.396\textwidth}{lcc}
\hline
%\begin{tabular}{|l|c|c|}
%\hline
 Ratios                                              &  RHIC-1) &            RHIC-2)   \\
\hline
${\Xi}/{\Lambda}$                                    &0.18 $\pm$ 0.02 &0.14 $\pm$ 0.02 \\
${\overline{\Xi}}/{\bar\Lambda}$                     &0.25 $\pm$ 0.03 &0.19 $\pm$ 0.03 \\
${\bar\Lambda}/{\Lambda}$                            &0.49 $\pm$ 0.15 &0.49 $\pm$ 0.15 \\
${\overline{\Xi}}/{\Xi}$                             &0.68 $\pm$ 0.15 &0.68 $\pm$ 0.15 \\
${\Omega}/{\Xi}$                                     &0.14 $\pm$ 0.03 &0.10 $\pm$ 0.03 \\
${\overline{\Omega}}/{\overline{\Xi}}$               &0.20 $\pm$ 0.03 &0.15 $\pm$ 0.03 \\
${\overline{\Omega}}/{\Omega}$                       &1.   &1.  \\
$(\Omega+\overline{\Omega})\over(\Xi+\bar{\Xi})$     &0.17 $\pm$ 0.01 &0.12 $\pm$ 0.01 \\
$(\Xi+\bar{\Xi})\over(\Lambda+\bar{\Lambda})$        &0.21 $\pm$ 0.02 &0.17 $\pm$ 0.02 \\
%${K^0_{\rm s}}/\phi$                                 &8.1 $\pm$ 0.02 &10.9 $\pm$ 0.03 \\
 ${K^+}/{K^-}$                                 & 1.35 $\pm$ 0.25 &1.34 $\pm$ 0.25  \\
$p/{\bar p}$                                         &2.9 $\pm$ 1.5 &2.9 $\pm$ 1.5 \\
${\bar\Lambda}/{\bar p}$                             &2.4 $\pm$ 0.3 &1.8 $\pm$ 0.3 \\
%${K^0_{\rm s}}$/B                                    &0.6 $\pm$ 0.25 &0.50 $\pm$ 0.25 \\
${h^-}$/B                                            &4.6 $-$2+10 &4.6 $-$2+10\\
\hline
\end{tabular*}
\end{center}
\vspace*{-.6cm} 
%\end{table}
%\begin{table}[b]
\caption{\label{table2}RHIC rapidity distribution of all (anti)baryons assuming 
flat net-baryon rapidity distribution}
\vspace*{-.2cm} 
\begin{center}
\begin{tabular*}{0.35\textwidth}{ccc}
%\begin{tabular}{|r|c|c|}
\hline
$ {dN}/{dy} $   &  Baryons           &   Antibaryons   \\
\hline
 protons        &    25\ \ ($p$)         &   8.5\ ($\bar p$)    \\
${\Lambda}$     &   42\ \ (${\Lambda}$)  &  20\ \ ($\overline{\Lambda}$)  \\ 
$\Xi$           &  7.5\ ($\Xi$)          &     5\ \,\ \ ($\overline\Xi$) \\ 
$\Omega$        &  1\ \,\ \ ($\Omega$)   &  1\ \,\ \ ($\overline\Omega$) \\ 
\hline
\end{tabular*}
\end{center}
\vspace*{-1.2cm} 
\end{table}

The key result we see in  this extrapolation of the SPS results to RHIC is that
 the (anti)baryon abundances (rapidity distributions) are strongly 
strangeness dominated, which as we have been arguing in past two decades 
is the characteristic behavior in QGP phase break-up.\\

%\end{document}

%14
\subsection{S.E. Vance:    Baryons, Junctions and Hyperons}

Recently, the baryon junction mechanism was 
proposed to explain the observed baryon 
stopping at the SPS\cite{khar_bj96,vance_hijb}.  
The baryon junction appears when writing the QCD gauge 
invariant operator of the baryon, being the 
vertex which links the three color flux (Wilson) lines flowing 
from the valence quarks.   Being a gluonic configuration,  
the junction can be easily transported into the mid-rapidity 
region in hadronic reactions.   There near mid-rapidity, as the strings 
connecting the junction to the valence quarks fragment, a baryon is 
produced being composed of three sea quarks.
This gluonic mechanism is also 
able to explain\cite{kop_hera} a striking, preliminary measured 
baryon asymmetry observed approximately 8 units 
of rapidity away from the proton's fragmentation region in $ep$ 
collisions at HERA.
 
The valence baryon junction mechanism was recently extended and  
a new mechanism for anti-hyperon production was proposed\cite{vance_hijbb}.   
Like the valence baryon junction mechanism, this junction-anti-junction 
loop $(J\bar{J})$ mechanism is also derived from the topological gluon 
structure of the baryon and originates in the context of Regge phenomenology.
The $(J\bar{J})$ mechanism was shown to strongly enhance the anti-hyperon 
yields in nuclear collisions and is needed 
to provide reasonable anti-hyperon to hyperon ratios.
Both of these mechanisms were implemented in HIJING/${\rm B\bar{B}}$,
a modified version of the HIJING event generator.

\begin{figure}[h]
\hspace{5cm}
\psfig{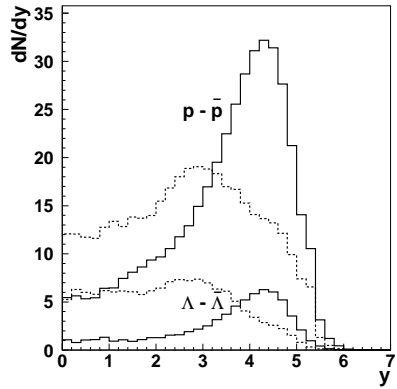}
\caption{Predictions for the initial valence proton rapidity 
distribution \ (upper two curves) and for the initial valence hyperon 
rapidity distribution (lower two curves) are given for 
Au+Au collisions at $E_{cm} = 200$ AGeV by HIJING (solid) 
and HIJING/${\rm B\bar{B}}$(dashed).}
\label{svfig1}
\end{figure}
 
\begin{figure}[h]
%\vspace{3cm}
\hspace{3.0cm}
\psfig{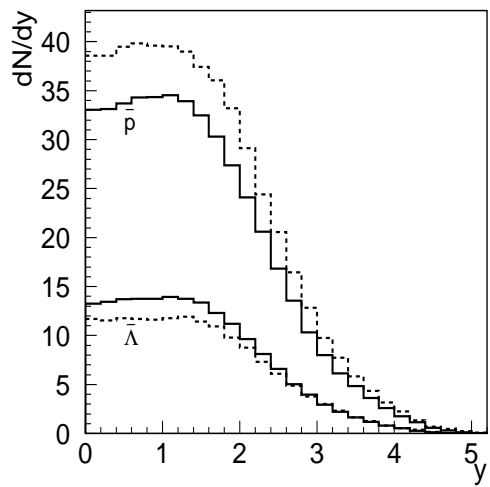}
\vspace{-20 pt}
\caption{Predictions for the initial $\bar{p}$ 
and $\bar{\Lambda}$ rapidity distributions are given for HIJING (solid lines) 
and for HIJING/${\rm B\bar{B}}$ (dashed curves) for 
Au+Au collisions at $E_{cm} = 200$ AGeV for $b = 0 \;{\rm fm }$.}
\label{svfig2}\end{figure}

The effects of these mechanisms at RHIC energies were explored using 
HIJING/${\rm B\bar{B}}$.  Estimates for the initial valence $p$ 
and $\Lambda$ rapidity distributions in $Au+Au$ collisions at 
RHIC energies ($\sqrt{s}$ = 200 GeV) for  $b = 0 \;{\rm fm }$ 
are shown in Figure \ref{svfig1}.
HIJING/${\rm B\bar{B}}$ predicts approximately twice the initial number 
of valence protons and five times the initial number of valence 
hyperons of HIJING at mid-rapidity leading to a prediction of twice 
the initial baryon density, $\rho(\tau_0) \approx 2 
\rho_0 \approx 0.3/{\rm fm}^3$.   Estimates for the total $\bar{p}$ 
and $\bar{\Lambda}$ have also been made (see Figure \ref{svfig2}), 
where it was found that at these energies, 
the yields of the anti-baryons are only sensitive to the relative 
string fragmentation probability of producing diquarks to quarks, 
and not to the small cross section for producing $J\bar{J}$ loops.   
While the $J\bar{J}$ loops do not significantly effect the absolute yields,
they are important in producing rapidity correlations between baryons
and anti-baryons, such as the $\Delta^{++} (uuu)$ 
and $\bar{\Omega}^+ (\bar{s}\bar{s}\bar{s})$, which are absent 
in present baryon pair production schemes.  

In the phenomenology associated with the baryon junction, many new 
states being gauge invariant combinations of junctions, anti-junction, 
quark and anti-quarks\cite{rossi_77} were predicted, 
e.g. a quarkless hybrid glueball $(J\bar{J}$ = $M^J_0)$.
The experimental search for these states has been inconclusive (presumably
due to their large decay widths), with only several broad resonances 
remaining as possible candidates.  We propose that the effect 
of large glueball states (e.g. $S^6_0=(J\bar{J}J\bar{J}J\bar{J})$) 
can be observed at RHIC by seeing a non-trivial rapidity correlation 
between two baryons (e.g. $pp$ rapidity correlations) in $NN$ collisions 
(see \cite{qm99}).   

%In summary, we propose to view the effect of the junction at RHIC through 
%the baryon stopping and nontrivial baryon-anti-baryon correlations 
%such as the $\Delta^{++} (uuu)$ and $\bar{\Omega}^+ (\bar{s}\bar{s}\bar{s})$.
%In addition, we propose that the effects of large mass hybrid junction 
%glueball states can be observed in novel two baryon rapidity correlations.

%\newpage

\section{ HBT AND COLLECTIVE FLOW CORRELATIONS}
%15
\subsection{S. Pratt:   HBT at  RHIC }
Prediction of QGP time delay too late for these
proceedings. [M.G.: Look for it at RHIC
in spite of heretical views 
in sec. 3.3. See D.H. Rischke, M. Gyulassy, NPA 608 (1996)
479]
%16
%\newpage
\subsection{D. Teaney:     Nutcracker Flow and HBT}
%##########################################################################

\begin{figure}[htb]
        \begin{center}
\centerline{
\psfig{file=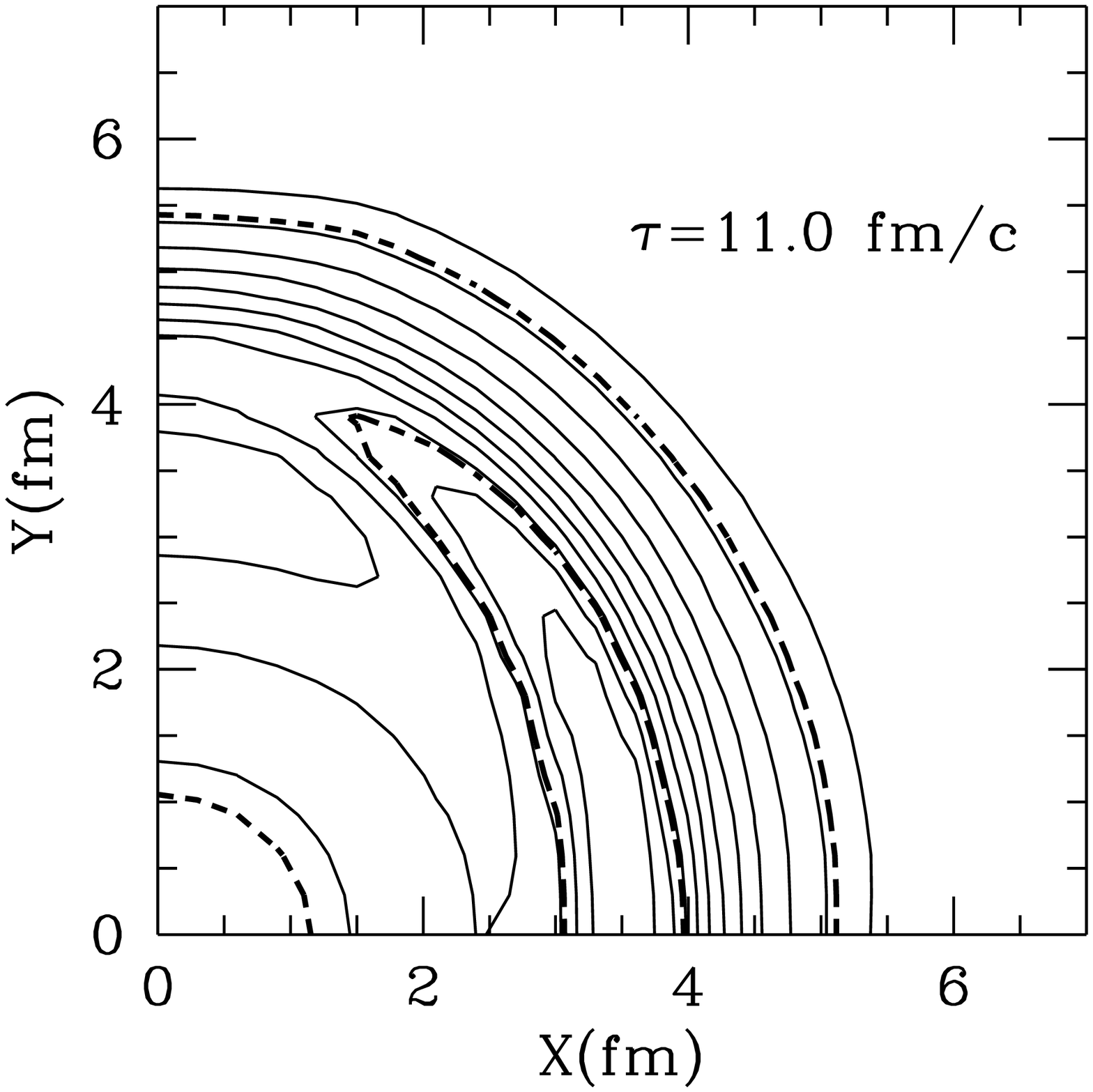,height=5.0cm,width=6.0cm}}
%                \vspace{  1.0cm}
\centerline{\psfig{file=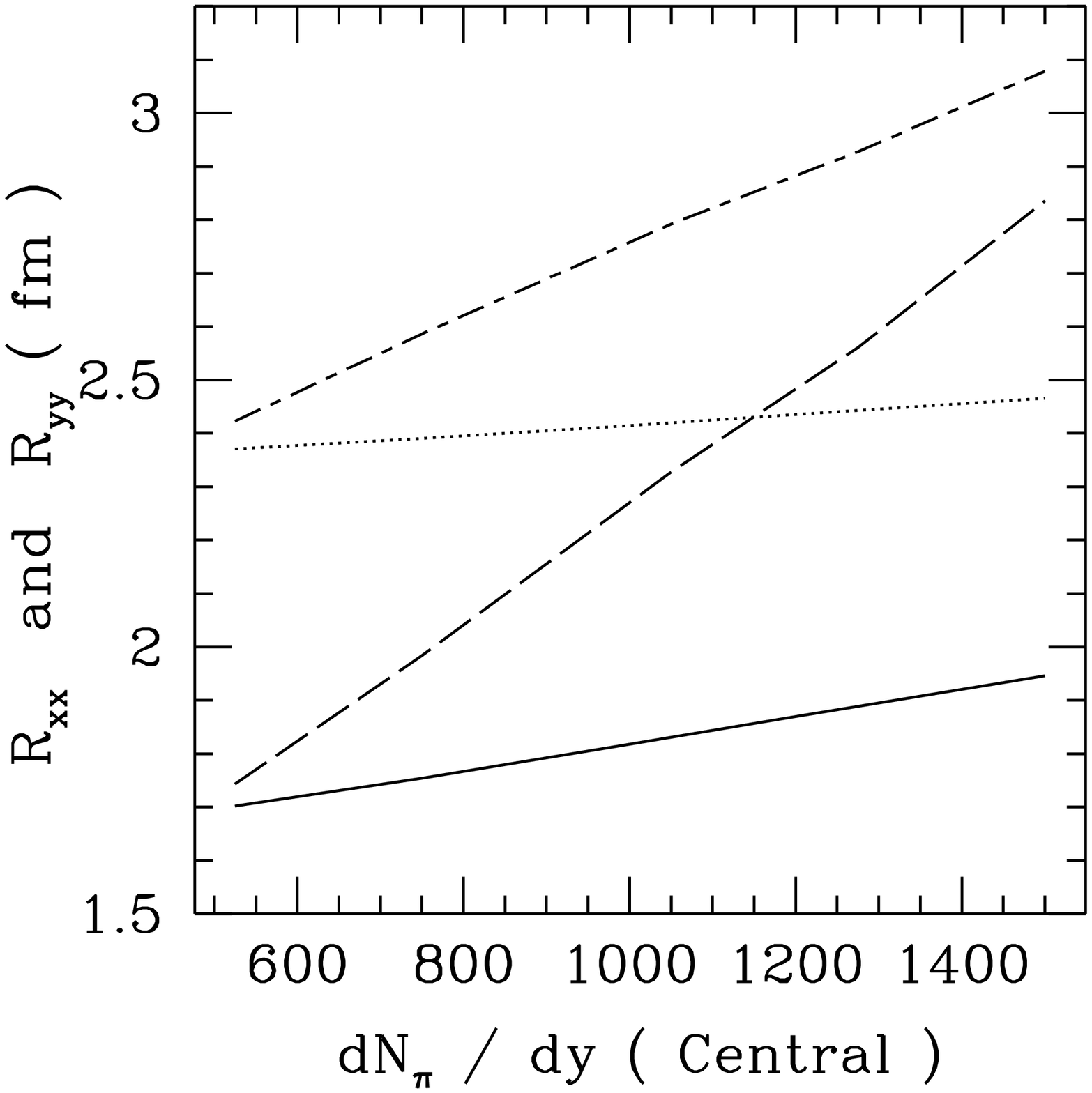,height=5.0cm,width=6.0cm}}
        \end{center}
%        \vspace{-2.5cm}
        \caption
        {
        \label{nut-fig}
   (a) Typical matter distribution in the transverse plane
        at mid rapidity, 
        calculated using boost invariant hydrodynamics with
        a bag model EoS, for a PbPb collision at b=8 fm at 
        RHIC( $dN_{\pi}/dy=850$ ) .
        The solid lines show contours of constant energy density.
        The inner and outer dashed lines show temperatures T=140 MeV
        and T=120 MeV respectively.
   (b) HBT radii, $R_{xx}$ and $R_{yy}$, as a function of the final 
     pion multiplicity.  For a resonance gas EoS $p=.2\epsilon$,  
     the solid lines and dotted lines show $R_{xx}$ and $R_{yy}$ respectively.
     For a bag model EoS, the dashed and dashed-dotted lines show 
     $R_{xx}$ and $R_{yy}$ respectively.
   }
     \vspace{-.7cm}
\end{figure}

Flow in non-central collisions at the SPS has been 
studied experimentally \cite{Na49} and theoretically
\cite{Ollitrault,Sorge-elliptic} and shows
sensitivity to the EoS.
However, high and low energy collisions have different
acceleration histories, due to the QCD phase transition.
In the mixed phase the EoS is soft \cite{HS-soft}. 
For AGS/SPS collision energies  the matter is produced
close to the ``softest point" and therefore 
the initial transverse acceleration is small. 
Using a bag model EoS and fixing the initial entropy to the
SPS multiplicity, we have simulated PbPb collisions 
at b=8fm with boost invariant hydrodynamics.
The matter expands preferentially in the impact parameter direction
(the x direction).
 Since the matter is produced close to the softest point, it retains 
its initial elliptic shape, accelerating only
in the late hadronic stages.
For  higher collision energies,
the hot QGP is formed well above the transition temperature  
 and the 
early pressure forces early acceleration.  
Unfortunately, the final radial flow velocity at RHIC is
quite similar to the SPS \cite{HydroUrqmd} as is the final 
elliptic flow \cite{TS-nut,Kolbp-nut}.
The early quark push has important consequences, however. 
The resulting velocity
has a long time, $\sim 10 fm/c$, to change the matter 
distribution before  freeze-out. 
The stiff QGP in the center , with $T>>T_c$ , pushes against the soft 
matter on the exterior, with $T\approx T_c$,   producing two shell-like 
structures which separate in the x direction leaving two holes in
the y direction.  
Since the final distorted distribution rather resembles a nut
and its shell,  we call this picturesque configuration the 
$nutshell$.  
In Fig.\ref{nut-fig}(a) we show the ``nutty" energy distribution 
in the transverse plane at zero rapidity. 
No shell like structures are seen
for an ideal gas EoS. The matter expands elliptically and the
final distribution resembles a gaussian.
%
%#############################################################################
%
%
%###################################################################
%

The early acceleration may be seen with HBT interferometry since
the early velocity increases the size of the system.
{} From the two pion correlation function, the following radii may
be extracted 
by carefully choosing the pion pair momenta \cite{Uli-review,TS-nut}. 
\begin{equation}
        R_{xx}^2 =  < x^2 > - < x >^2 \nonumber   
\end{equation}
\vspace{-0.3cm}
\begin{equation}
        R_{yy}^2 =  < y^2 > - < y >^2 \nonumber
\end{equation}
(The average momentum
of the pair was taken to be 500 MeV).
These radii are shown in Fig.\ref{nut-fig}(b) for
PbPb collisions at b=8 fm, as a function of the pion 
multiplicity  scaled by the number of participants 
to central collisions (not b=8 fm).
$R_{xx}$ and 
$R_{yy}$ are shown for a bag model EoS and for 
a simple resonance gas EoS.
For low energies near the left hand side of the plot, the two EoS show
approximately the same radii, roughly corresponding to the initial
elliptic shape of the matter distribution.  For a simple resonance gas EoS
the HBT Radii  
show little energy dependence, while
for  an EoS with the phase transition
the Radii increase steadily with 
beam energy.
The rapid increase of 
$R_{xx}$ can be
understood qualitatively. 
The nutshells separate at higher 
collision energies and leave a large homogeneous region in the 
center. The constancy of $R_{xx}$ for an ideal gas
EoS reflects the energy gradients in the final distribution. 
For an EoS with a phase
transition, the source function is ``box" like and the ``box" 
increases in size, while for an ideal gas EoS the source function is
more gaussian.

Other signatures of the nutcracker flow were presented in \cite{TS-nut} and
similar qualitative features were found by \cite{Kolbp-nut,Miklos-nut}.

Acknowledgments:  Work supported by U.S. DOE, under Contract 
No. DE-FG02-88ER4038.
%
%###################################################################

%%%%%%%%%%%%%%%%%%%%%%%%%%%%%%%%%%%%%%%%%%%%%%%%%%%%%%%%%%%%%%%%%%%%%%%%

%\newpage
%17
\subsection{B.R. Schlei:     The QGP Stall or Not}

In 1996, D.H. Rischke and M. Gyulassy \cite {rischke96} proposed to use the 
time-delay signal in Bose-Einstein correlations (BEC) as a signature for 
the possible formation of a quark-gluon plasma (QGP) in relativistic 
heavy-ion collisions.
In particular, they suggested to measure the ratio of the transversal
interferometry radii, $R_{out}/R_{side}$, which can be obtained by
fitting experimental BEC data with a parametrization introduced by
G. Bertsch \cite{bertsch88} et al. in 1988. The transverse radius 
parameter $R_{out}$ has compared to the transverse radius parameter 
$R_{side}$ an additional temporal dependence which should be sensitive to 
a prolonged lifetime of a fireball, in case a QGP was formed in a 
relativistic heavy-ion collision. This latter phenomenon is known under the 
term ``QGP stall''.
In the following, I shall explain why I believe that the time-delay signal 
in Bose-Einstein correlations {\it is not} a good signature for the possible 
formation of a QGP.

Let us consider two quite different equations of state (EOS) of nuclear
matter within a {\it true} relativistic hydrodynamic framework 
(i.e., HYLANDER-C) \cite{schlei97}. The first one, EOS-I, has a 
phase-transition to a QGP at $T_C$ = 200 $MeV$ ($\epsilon_C$ = 1.35 
$GeV/fm^3$) \cite{schlei99a}. The second EOS, EOS-II, is a purely hadronic EOS, 
which has been extracted from the transport code RQMD (cf., Ref. 
\cite{schlei98}) under the assumption of fully achieved thermalization. 
If one assumes for each EOS {\it different} initial conditions before the
hydrodynamical expansions, one can fit simultaneously hadronic single 
inclusive momentum spectra and BEC, which have been measured recently by 
the CERN/NA44 and CERN/NA49 Collaborations (cf., \cite{schlei99a,schlei98}), 
respectively. 
In particular, for the acceptance of the NA44 experiment a ratio 
$R_{out}/R_{side} \approx$ 1.15 was found \cite{schlei99a} while 
using both EOS, EOS-I and EOS-II. Little difference was seen in the BEC of 
identical pion pairs while considering the two different EOS.

\begin{figure}[t]
\epsfig{scale=1.0,file=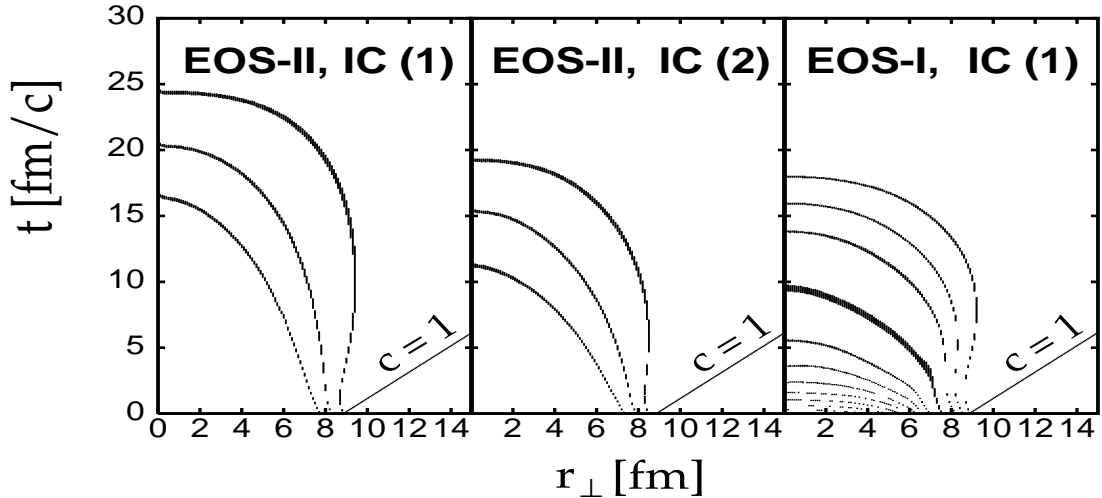}
\caption{Isotherms in steps of $\Delta T = 20MeV$. The outer contours equal
$T_f = 140MeV$.}
\label{schleifig}
\end{figure}

In the following, we shall assume for central Au+Au collisions at BNL/RHIC 
beam energies a set of fireball initial conditions, IC-I, which are similar
to those as described in Ref. \cite{schlei99c}. From these fireball initial 
conditions, IC-I, single inclusive momentum spectra have been calculated 
while using EOS-I and EOS-II in the hydrodynamic expansions. We note, that 
the rapidity spectra of both calculations differ in width and normalization 
significantly \cite{schlei99b}. 
Fig.~\ref{schleifig}. shows, that the isotherms of the 
transversely expanding fluids at 
longitudinal coordinate $z=0$ also differ significantly. Since there will be 
only one set of measured data, we shall fit the calculation using EOS-II to 
the single inclusive momentum spectra of the calculation using EOS-I. 
In doing so, we find new initial conditions, IC-II \cite{schlei99b}. 
But now the space-time picture of the evolving fireball at freeze-out is 
again very similar to the one using EOS-I with IC-I.

If one calculates BEC of identical pion pairs or identical kaon pairs,
respectively, one finds \cite{schlei99b} while comparing the calculations 
using EOS-I with IC-I and EOS-II with IC-II, respectively, {\it no} 
significant differences in the extracted ratios $R_{out}/R_{side}$ regardless 
of the pair kinematics under consideration. In particular, the assumption of 
the PHENIX detector acceptance \cite{schlei99b} leads to a ratio 
$R_{out}/R_{side} \approx$ 1.65 for both choices of EOS.

In summary, the larger ratio $R_{out}/R_{side}$ at RHIC beam energies
appears to be rather a consequence of the expected higher energy deposit
in the fireball during the heavy-ion collision, but it appears {\it not} to
be an indicator of the present or absent phase-transition to a QGP.
Of course, more theoretical analysis is necessary, but there is strong 
evidence, that BEC do not provide a good QGP signature, since we do not
understand the initial state of a heavy-ion collision well enough yet.

%18
\subsection{J.Y. Ollitrault: Elliptic Flow}
See J.Y. Ollitrault,
%Anisotropy as a signature of transverse collective flow,
Phys. Rev. {\bf D46}, 229 (1992).

%19
%\newpage
\subsection{D. Molnar:  Elliptic Flow via Inelastic Parton Cascade}
Transverse energy and elliptic flow are of recent theoretical interest
because they signal collectivity and thus may be sensitive
to a possible phase transition.
Also, they are relatively easy to measure
and therefore will be one of the first results at RHIC.

Until thermal and chemical equilibrium
are shown to be achieved in relativistic heavy ion collisions,
a transport theory approach is absolutely necessary.
Two forerunners \cite{Bin_Et,Bin_v2}
of the present study utilize the ZPC elastic parton cascade \cite{Bin_ZPC},
which has colorless gluons with a single elastic $2\to 2$ process.
However,
for the question of chemical equilibration
it is essential to include inelastic processes,
too.

If we insist on having on-shell partons,
the simplest set of particle number changing processes
consists of a $2\to 3$ and a $3\to 2$ process.
By adding these two new processes to ZPC,
we developed an on-shell inelastic parton cascade,
MPC \cite{QMtrans}.

To have a well-defined theoretical model that is independent of the code,
%avoid ${\rm Model\equiv Code}$,
we regard MPC as a tool for solving the Boltzmann transport equation
\begin{eqnarray}
p^\mu \partial_\mu f_1 &=&
  %2<->2
  2 \int\limits_2\! \int\limits_3\! \int\limits_4
    \left[
      \frac{(2\pi)^6}{g^2} f_3 f_4 
      - \frac{(2\pi)^6}{g^2} f_1 f_2 
    \right]
    \, W_{12\to 34}
    \, \delta(12 - 34)
   \nonumber\\
  &+&
  %2->3
  3 \int\limits_2 \int\limits_3 \int\limits_4 \int\limits_5
    \left[
      \frac{(2\pi)^6}{g^2} f_4 f_5
      - \frac{(2\pi)^9}{g^3} f_1 f_2 f_3
    \right]
    \, W_{123\to 45} \, \delta(123 - 45)
  \nonumber\\
  &+&
  %3->2   
  2 \int\limits_2 \int\limits_3 \int\limits_4 \int\limits_5
    \left[
      \frac{(2\pi)^9}{g^3} f_3 f_4 f_5
      - \frac{(2\pi)^6}{g^2} f_1 f_2 
    \right]
    \, W_{12\to 345} \, \delta(12 - 345) \ ,
\label{BTE_with_TR}
\end{eqnarray}
where $f$ is the Lorentz invariant phase space density,
the Lorentz invariant $W$-s are proportional 
to the corresponding $2\to 2$, $2\to 3$ and $3\to 2$ matrix elements,
$f_i \equiv f(p_i, x)$,
$\int\limits_i \equiv \int \frac{d^3 p_i}{E(p_i)}$,
and $g$ is the degeneracy of the partons.
Furthermore, 
combinations like $12 - 34$ are for $p_1 + p_2 - p_3 - p_4$,
etc.,
and the arguments of $W_{12\to 34}(p_1, p_2, p_3, p_4)$,
etc.,
are suppressed. 

For simplicity, we took our gluons to be massless.
The $2\rightarrow 2$ and $2\leftrightarrow 3$
differential cross sections were
taken to be uniform in phase space with theta function cutoffs
$\Theta((p_i+p_j)^2-\mu^2)$
for each possible pair of ingoing and each possible pair of outgoing particles
in order to prevent collinear collisions.
The leftover momentum dependence of the cross sections 
was fixed by imposing energy independent {\em total} crossections for $\mu^2 =
0$,
while the $3\to 2$ cross section was given by detailed balance.
The gluon degeneracy was $g=16$ (3 colors).

\begin{figure}[htb]
\begin{minipage}[b]{78mm}
\epsfig{scale=0.6,file=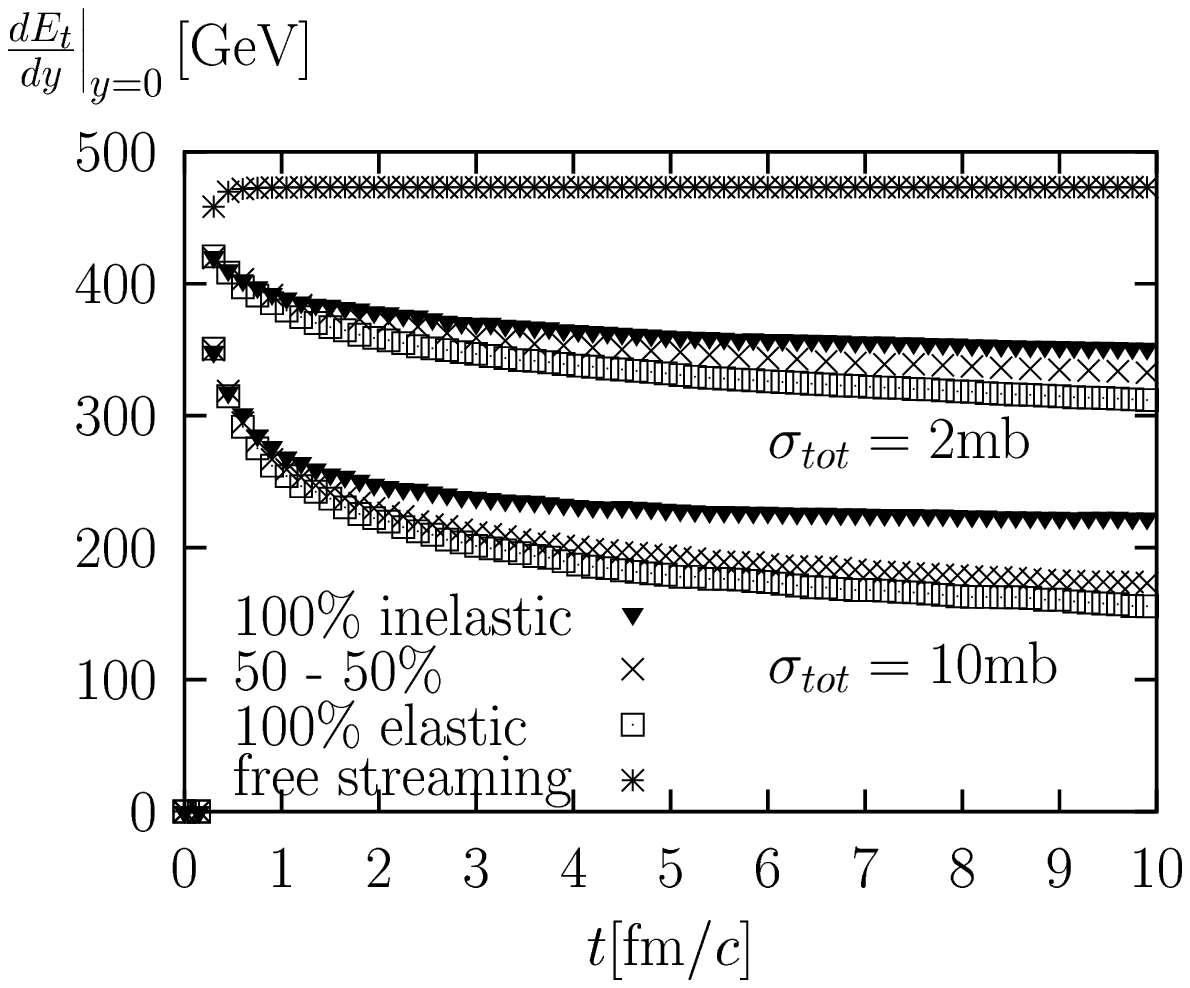}
\caption{$E_t$ evolution for a 1-dimensional Bjorken expansion}
\label{fig:Et}
\end{minipage}
\hspace{\fill}
\begin{minipage}[b]{78mm}
\epsfig{scale=0.6,file=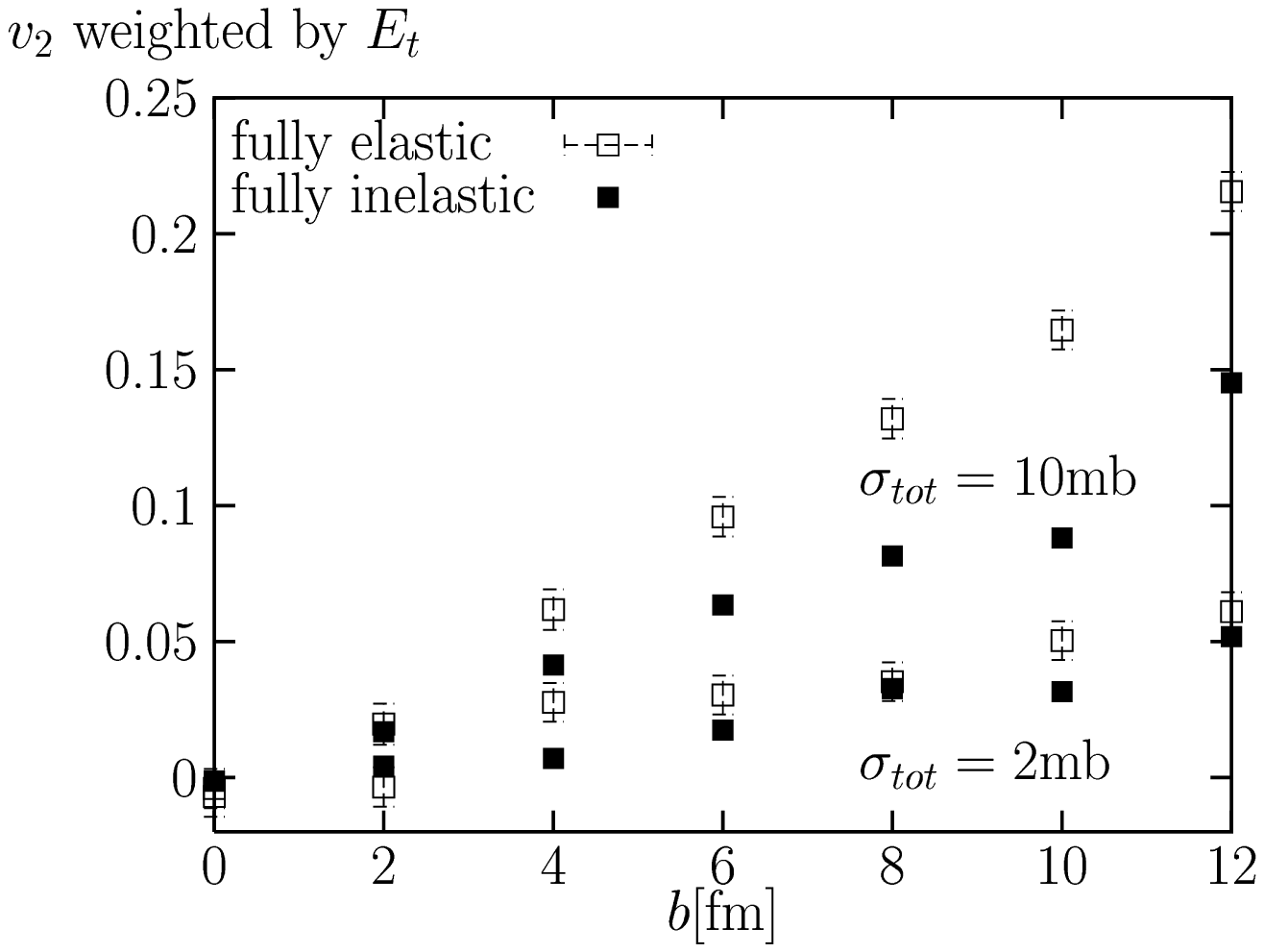}
\caption{Elliptic flow vs. centrality and inelasticity}
\label{fig:v2}
\end{minipage}
\end{figure}

Fig.~\ref{fig:Et}. shows the transverse energy loss in a one-dimensional
Bjorken expansion starting from a locally thermal distribution.
Keeping the total $2\to X$ cross section constant,
we studied how the energy loss changes with the ratio
of the elastic and inelastic part of the cross section.
We considered three cases:
fully elastic, 50\% inelastic, and fully inelastic.
The initial conditions were the same as in Ref. \cite{Bin_Et}:
$\tau_0 = 0.1 {\rm fm}/c$,
$T_0 = 500{\rm MeV}$,
$dN/d\eta = 400$,
$-5 \le \eta \le 5$,
$A_t = 100 {\rm fm}^2$, 
and we averaged over 20 events.

We can see that the transverse energy at midrapidity
is sensitive only to the total $2\to X$ cross section.
With increasing inelasticity there is somewhat less work done
because opening up the inelastic channels leads to parton production
which compensates for the decrease of work.
There also shows a tendency that until the cross section is dominantly
inelastic the energy loss is rather insensitive to the degree of inelasticity.

Fig.~\ref{fig:v2}. shows how the elliptic flow parameter $v_2$ changes
with centrality and inelasticity.
We took a similar initial condition as Ref. \cite{Bin_v2} but with
$dN/d\eta d^2 x_t = 3/{\rm fm}^2$,
$T_0 = 500{\rm MeV}$,
$R = 7{\rm fm}$,
$-5 \le \eta \le 5$.
We averaged over 20 events.

As we go from the fully elastic to the fully inelastic case
elliptic flow drops,
showing that the details of the microscopic dynamics {\em are} important.
This is also supported by another study \cite{Sorge_v2}
using the RQMD 3.0 event generator.

The 50\% reduction of $v_2$ is {\em in addition}
to the original 50\% drop relative to ideal hydrodynamics \cite{Ollitraultb}
given by dissipative corrections at the elastic level \cite{Bin_v2}.
The centrality dependence is very similar to that from hydrodynamics.

A study with more realistic gluon cross sections and diffuse initial geometry
is underway.

\newpage

\section{ JETS AND PENETRATING PROBES AT RHIC}
%20
%\newpage
\subsection{R. Rapp: Low Mass Dileptons}
%%%%%%%%%% espcrc1.tex %%%%%%%%%%
%\documentclass[12pt,twoside]{article}
%\usepackage{fleqn,espcrc1}

% change this to the following line for use with LaTeX2.09
%\documentstyle[12pt,twoside,fleqn,../espcrc1,epsfig]{article}

% if you want to include PostScript figures
%\usepackage{graphicx}
% if you have landscape tables
%\usepackage[figuresright]{rotating}

% put your own definitions here:
\newcommand{\cZ}{\cal{Z}}
\newcommand{\eg}{{\it e.g.}}
\newcommand{\ie}{{\it i.e.}}
%   ...
%\newcommand{\ttbs}{\char'134}
%\newcommand{\AmS}{{\protect\the\textfont2
%  A\kern-.1667em\lower.5ex\hbox{M}\kern-.125emS}}

% add words to TeX's hyphenation exception list
%\hyphenation{author another created financial paper re-commend-ed Post-Script}

% declarations for front matter
%\title{Low-Mass Dileptons at RHIC}

%\author{Ralf Rapp\thanks{supported by the A.-v-.Humboldt Foundation as a
%Feodor-Lynen fellow, and by US-DOE grant DE-FG02-88ER40388.}} 
%\vspace{0.3cm}
%

%\begin{document}

% typeset front matter
%\maketitle

%\begin{abstract}
%A prediction is given for the low-mass dilepton signal from a thermal
%fireball in central in 200~AGeV Au+Au collisions at RHIC, focusing
%on the region of  $\rho$ and $\omega$ mesons and their potential medium 
%modifications.
%\end{abstract}

%%%%%%%%%%%%%%%%%%%%%%%%%%%%%%%%%%%%%%%%%%%%%%%%%%%%%%%%%%%%%%%%%%%%%%
%\section{Introduction}
%%%%%%%%%%%%%%%%%%%%%%%%%%%%%%%%%%%%%%%%%%%%%%%%%%%%%%%%%%%%%%%%%%%%%%
Low-mass dilepton spectra in (ultra-) relativistic heavy-ion collisions
are expected to provide information on the in-medium properties of the 
light vector mesons in connection with signals from chiral symmetry
restoration. At RHIC energies, the major experimental challenge is  
a large (combinatorial) background which limits the sensitivity
to well-defined resonance structures.
In this contribution~\cite{qm99web} we will thus address the 
question to what extent $\rho$ and $\omega$ spectral functions are 
affected in meson-dominated matter (following paragraph) 
and how this reflects itself in space-time integrated dilepton spectra
to be measured at RHIC (last paragraph).   

\begin{figure}[!htb]
\begin{minipage}{7.7cm}
\psfig{file=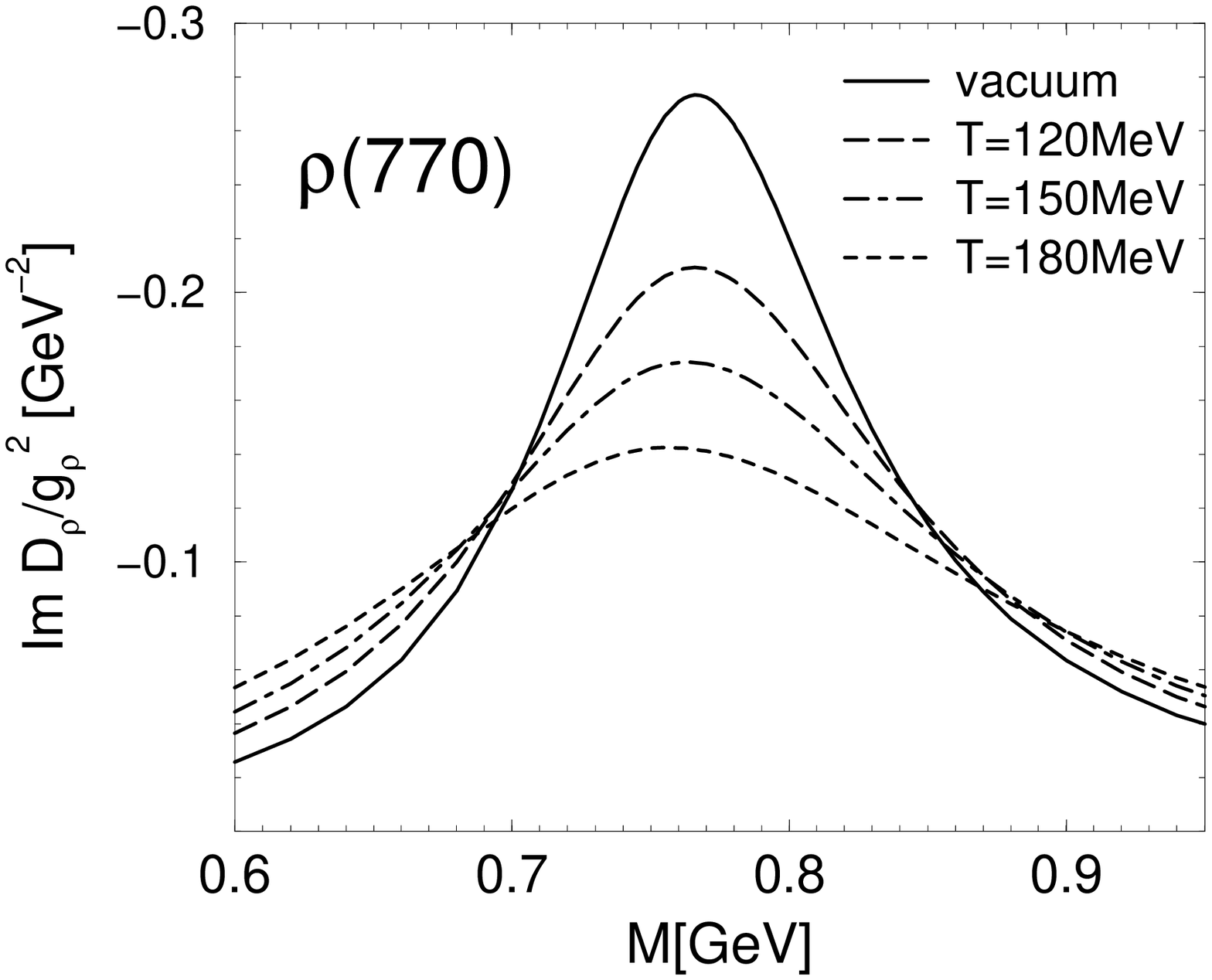,width=7.7cm}
\end{minipage}
\hspace{\fill}
\begin{minipage}{7.7cm}
\psfig{file=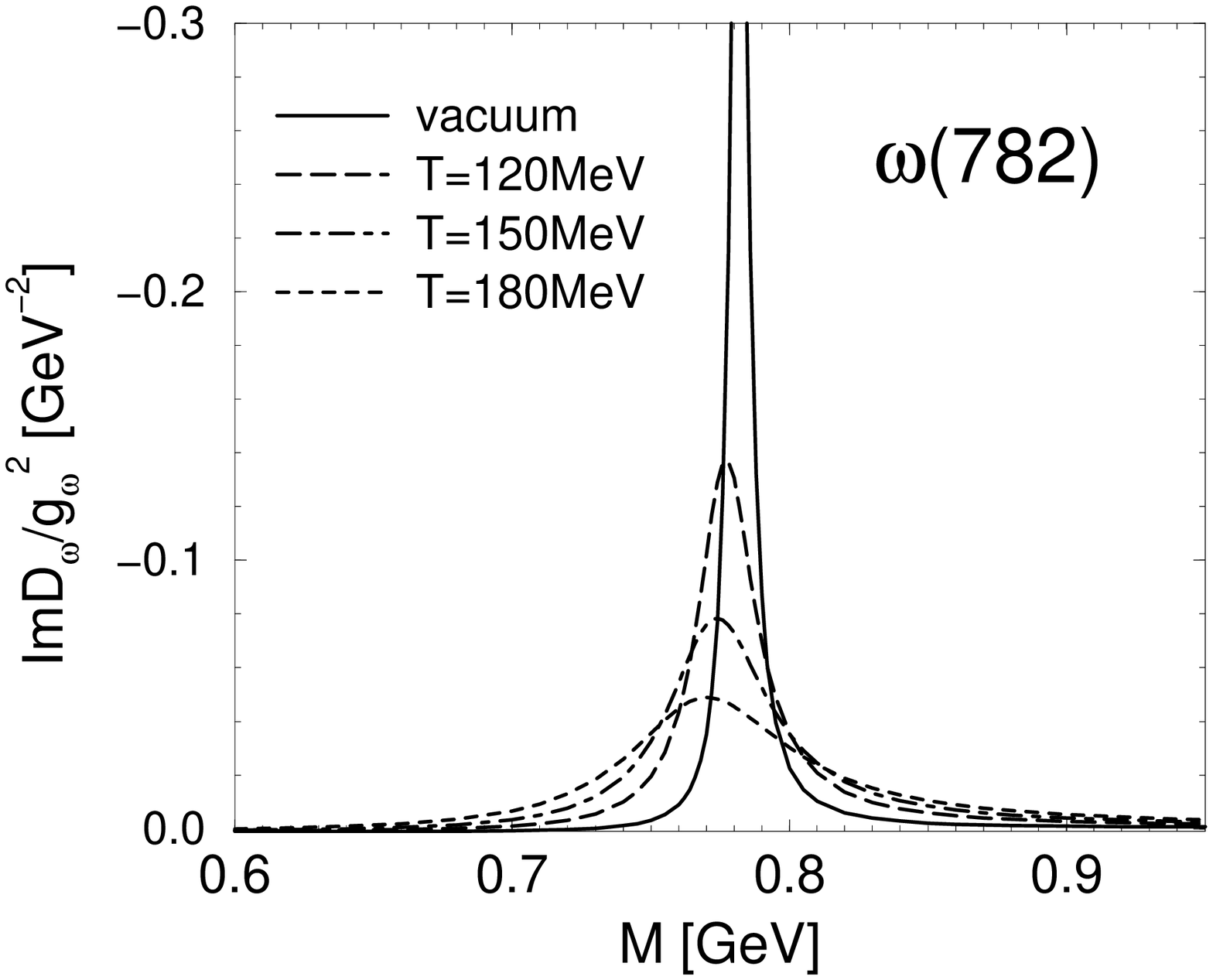,width=7.7cm}
\end{minipage}
%\vspace{-1.3cm}
\vspace{-1.1cm}
\caption{$\rho$- (left panel) and $\omega$- (right panel)
spectral functions (weighted by the corresponding VDM coupling constant)
at fixed 3-momentum in a thermal meson gas.} 
%at temperatures appropriate for the hadronic phase at RHIC.}
\label{fig_Avec}
\end{figure}

%%%%%%%%%%%%%%%%%%%%%%%%%%%%%%%%%%%%%%%%%%%%%%%%%%%%%%%%%%%%%%%%%%%%%%
%\section{$\rho$ and $\omega$ Mesons at Finite Temperature}
%%%%%%%%%%%%%%%%%%%%%%%%%%%%%%%%%%%%%%%%%%%%%%%%%%%%%%%%%%%%%%%%%%%%%%
Thermal dilepton radiation from a hot and dense medium is governed 
by the thermal expectation value of the electromagnetic current-current 
correlator, $\Pi_{em}$. Up to invariant masses of $\sim$~1~GeV its 
hadronic part can be accurately saturated with light vector mesons
(Vector Dominance Model). The pertinent
thermal dilepton production rate can then be expressed through the 
vector meson spectral functions, ${\rm Im} D_V$, as 
\begin{equation}
\frac{dN_{ll}}{d^4xd^4q}=-\frac{\alpha^2}{\pi^3} \ \frac{f^B(q_0;T)}{M^2}
 \ {\rm Im} \Pi_{em}(q;T) \quad , \quad
{\rm Im} \Pi_{em}(q;T)= \sum\limits_{V=\rho,\omega,\phi} 
\frac{m_V^4}{g_V^2} \ {\rm Im} D_V(q;T) \ , 
\label{rate}
\end{equation}
where $f^B$ denotes a thermal Bose distribution function and $M$ is 
the invariant mass of the lepton pair. 
Medium effects in the $\rho$ propagator are accounted for  
through modifications in the $\rho\to \pi\pi$ width
(most notably from a Bose enhancement of the two-pion states 
characterizing 'stimulated emission') and by 
collisions with surrounding thermal $\pi$, $K$ and $\rho$ mesons
via $s$-channel resonance formation ($\omega$, $a_1$, $K_1$, $f_1$, etc.),
see, \eg, Refs.~\cite{RG99}. At a temperature $T$=150~MeV the total 
thermal broadening of ${\rm Im} D_\rho$ amounts to about 80~MeV with
no significant shift in mass, cf.~left panel of Fig.~\ref{fig_Avec}. 

\newpage
Along similar lines, the $\omega$ meson decay width (which in free 
space consists of a combination of $\rho\pi$ and direct $3\pi$ decays)
has been evaluated including both pion- and rho Bose factors as well 
as the in-medium $\rho$ spectral function. 
Also, an in-medium selfenergy from $\pi\omega$ scattering through 
the $b_1(1235)$ resonance~\cite{b1} has been inferred from the 
70\% branching ratio of the free $b_1\to \omega\pi$ decay.
As a result, the finite temperature $\omega$ spectral function 
exhibits a broadening of several times its vacuum width, cf.~right
panel of Fig.~\ref{fig_Avec}.

\begin{figure}[htb]
\vspace{-0.7cm}
\begin{center}
\psfig{file=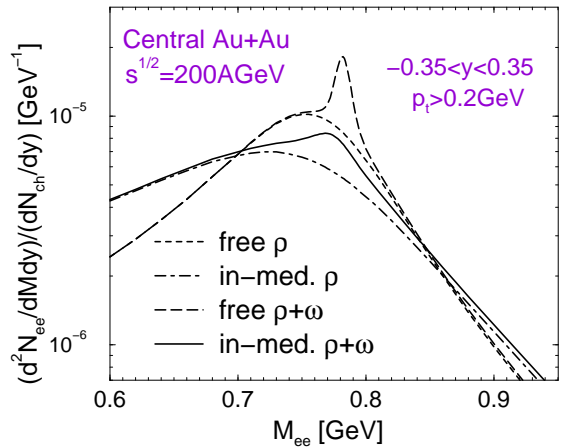,width=8cm}
\end{center}
\vspace{-1.7cm}
\caption{Dilepton spectra from a thermal fireball through
$\rho,\omega\to e^+e^-$ decays in the hadronic phase including a schematic
acceptance for the PHENIX detector.}
\label{fig_dlspec}
\end{figure}

%%%%%%%%%%%%%%%%%%%%%%%%%%%%%%%%%%%%%%%%%%%%%%%%%%%%%%%%%%%%%%%%%%%%%%
%\section{Low-Mass Dilepton Spectra at RHIC}
%%%%%%%%%%%%%%%%%%%%%%%%%%%%%%%%%%%%%%%%%%%%%%%%%%%%%%%%%%%%%%%%%%%%%%
Dilepton spectra from $\rho$ and $\omega$ decays are readily calculated
by integrating the thermal rate, Eq.~(\ref{rate}), over the hadronic
space-time volume of a central Au+Au collision at RHIC. For that we assumed 
an adiabatic expansion (including a mixed phase at $T_c$=180~MeV) 
with an entropy over baryon density
of $s/n_B=220$ together with a charged particle multiplicity of
$dN_{ch}/dy\simeq$~1100 around midrapidity. The 3-volume expansion has been 
modelled in accordance with hydrodynamic simulations.
The final  3-momentum integrated dilepton invariant mass  
spectra from $\rho,\omega\to e^+e^-$ decays are
displayed in Fig.~\ref{fig_dlspec}. 
The broad $\rho$ structure can most likely
not be disentangled from the large background expected in PHENIX. On the
other hand,  the thermal broadening of the $\omega$ (leading to a net 
$\sim$~40~MeV wide signal) may be 'just right', \ie, narrow enough to be 
possibly detectable above the background and the $\rho$ contribution, 
but substantially wider than the free $\omega$, which allows for a 
discrimination from (free) $\omega$ decays after freeze-out and thus 
enables important insights into its finite temperature behavior.

%\end{document}
\newpage
%21
\subsection{I. Vitev:     Jet Quenching in Thin Plasmas}
\newcommand{\vk}{{\vec k}}
\newcommand{\vb}{{\vec b}}
\newcommand{\vp}{{\vec p}}
\newcommand{\vq}{{\vec q}}
\newcommand{\vQ}{{\vec Q}}
\newcommand{\vx}{{\vec x}}
\newcommand{\tr}{{{\rm Tr}}}

One of the predicted  new observables  in  nuclear collisions at RHIC energies 
$(\surd s \sim 200$ AGeV) is jet quenching due to gluon bremsstrahlung. We 
study the detailed gluon probability distribution in the case of few 
($n_s=1,2,3$) scattering centers which leads to the suppression of the high 
$p_\perp$ tails of the hadronic spectra \cite{GLV,HTTP}.

The Landau-Pomeranchuk-Migdal (LPM) effect \cite{LPM}, first studied for 
the case of QED, plays an important role in the non-abelian gauge theories
as well, leading to nontrivial results \cite{QUENCH}. Unlike in 
electrodynamics, where the energy of the final electron and photon can be 
directly measured, the energy of a single parton is not measurable.
Therefore, in order to test the non-abelian energy loss of jets \cite{QUENCH},
the detailed form of the angular distribution, $dN_g/dyd^2 \vk_\perp$,
of induced gluon radiation must be known.  The angular integrated
energy loss, $dE/dx$, is theoretically interesting but insufficient to
address the experimentally accessible question of how much are
moderate $p_\perp \sim 3-10$ GeV pions expected to be suppressed relative 
to their known spectrum in elementary nucleon-nucleon reactions. 

We use the Gyulassy-Wang (GW) model of locally thermally equilibrated plasma
\cite{QUENCH}. In the case of one scattering center there are three relevant
diagrams \cite{HTTP} that give the gluon conditional probability distribution:
\begin{eqnarray}
\frac{dN^{(1)}_g}{dyd^2 \vk_\perp} &=& C_R {\alpha_s \over \pi^2} 
\left (\vec{H}^2 + R\,(\vec{B}^2_1 +\rho(\vb_1)\; \vec{C}^2_1  )
- R\left ( \vec{H} \cdot \vec{B}_1 \cos (t_{10}\omega_0) 
\right) \right.  \nonumber \\ 
&& -  R\,\rho (\vb_1/2 ) \left. \left(\vec{H}\cdot\vec{C}_1 
\cos (t_{10} \omega_{10}) - 2\vec{C}_1\cdot\vec{B}_1
\cos (t_{10}\omega_1) \right) \right ) \,  ,\; \label{dn10}
\end{eqnarray} 
where we have rearranged the terms following \cite{GUNION} and introduced 
the following notation:
\begin{eqnarray}
&&\vec{H}={\vk_\perp \over k^2_\perp }\; , \quad
\vec{C}_{1}={(\vk - \vq_1 )_\perp 
\over (k - q_1 )^2_\perp } \; , \quad
\vec{B}_1= \vec{H} - \vec{C}_1 \; ; \nonumber \\[1.5ex]
&&\omega_0\equiv k_\perp^2/2\omega=1/t_f\;  , \quad \omega_1\equiv 
(k-q_1)_\perp^2/2\omega \; , \quad \omega_{10} \equiv \omega_{1}- 
\omega_{0}\; . 
\end{eqnarray}
The phase factors $\omega_0,\, \omega_1, \, \omega_{10}$ have the
formation physics built in. The normalized transverse density profile
$\rho(\vb_1)\equiv T(\vb_1)/T(0)$ eliminates rescatterings of the emitted
gluon at large angles and keeps consistency with the geometry of the process.
$R \equiv C_A/C_R$ is 1 or 9/4 for gluon or quark jets respectively.

Similar formulas cam be derived for the case of two and three scattering 
centers \cite{GLV}. In all cases a couple of simple analytical limits can be
obtained.
\begin{eqnarray}
&& \lim_{k_\perp \rightarrow \; 0} {dN^{(n_s)}_g \over dyd^2\vk_\perp}=
{dN^{(0)}_g \over dyd^2\vk_\perp}\;, \quad
\lim_{k_\perp \rightarrow \; \infty} {dN^{(n_s)}_g \over dyd^2\vk_\perp}=
{dN^{(0)}_g \over dyd^2\vk_\perp}\,(1+R)^{\,n_s}\; .   
\end{eqnarray}
Therefore we can make an ansatz, allowing us to consider multiple 
scattering extrapolations:
 \begin{eqnarray}
{\Delta{dN^{(n_s)}_g \over dyd^2\vk_\perp}}&\approx&
{dN^{(0)}_g \over dyd^2\vk_\perp}{\left( (1+R)^{N_s}-1\right)}
f \left( \frac{\lambda \mu^2}{\omega},\frac{k^{\;2}_\perp}{\mu^2}         
\right)\; .      
\end{eqnarray}      
{}From Fig.24. one can see that the quantitative agreement between the scaled 
result for one scattering center and the exact and much more complicated 
$n_s=2$ result is within 5\% everywhere in the kinematically allowed
domain. 

\begin{figure}
\begin{center}
\vspace*{8.25cm}
\includegraphics{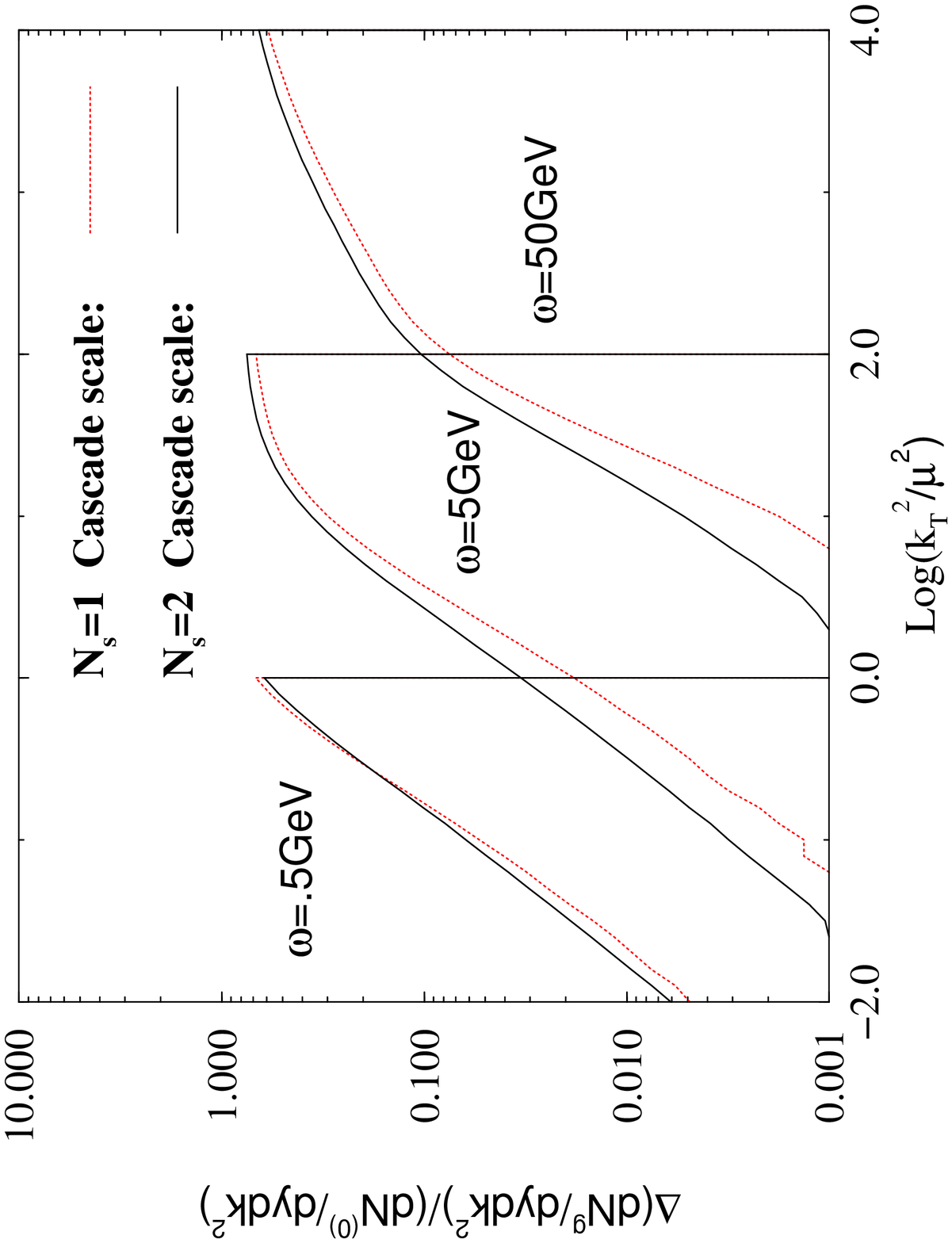}
\includegraphics{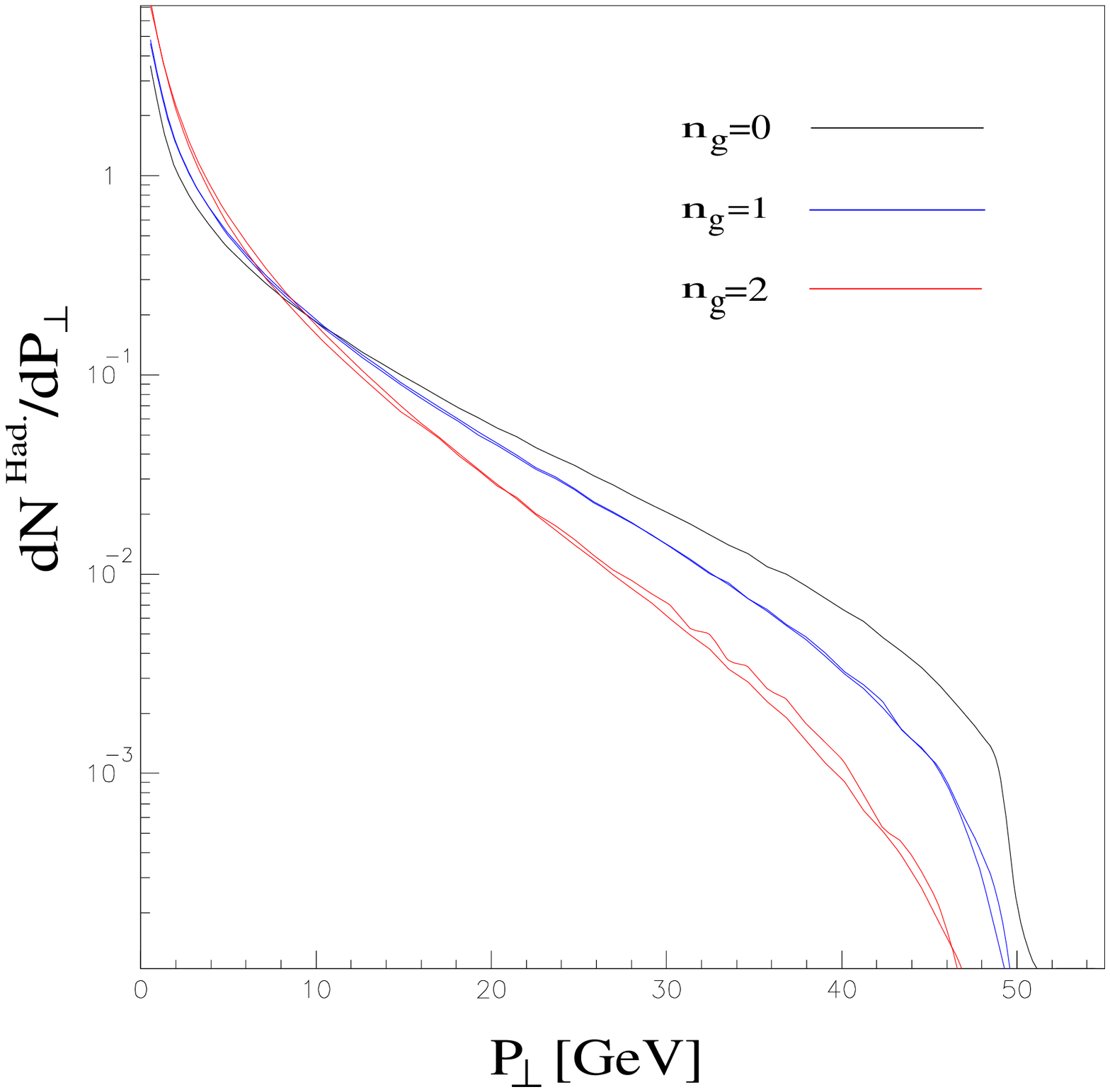}
\vskip -90pt
\end{center}
\begin{minipage}[t]{14.7cm}
\vskip -27pt
\begin{tabular}{p{7.3cm}p{0.1cm}p{7.0cm}}
\caption{ Scaling of the gluon probability
distribution with the number of scattering centers. $E_{jet}=50\; GeV$}
&$\,$ &
\caption{ High $p_\perp$ tail suppression of the hadronic 
distribution for partonic event $E_{C.M.}=20\; GeV$ (JETSET).}
\end{tabular}
\end{minipage}
\end{figure}

%\vskip -7truemm

Jet quenching occurs even in free space where there are no final state
interactions, leading to characteristic double-logarithmic probability
distribution \cite{GLV,GUNION}. In practice, the large logarithm implies t
hat multiple  gluon emission must also be considered. This leads to a 
Sudakov form factor for the jet and a probabilistic Alterelli-Parisi 
parton shower \cite{FIELD}. 

The final state multigluon shower can be calculated
most readily by one of the many Monte-Carlo event generators \cite{GENERAT} 
that encode empirical parton to hadron fragmentation functions and thus 
allow the detailed study of the effect of parton showering on the final 
hadron distributions. We have used the Lund JETSET \cite{GENERAT} 
string fragmentation routine to hadronize a high energy partonic event 
accompanied by gluon bremsstrahlung.  Fig.25. clearly shows the quenching 
effect of the medium induced and modified radiation.

Considering the significant role that ``hard'' physics will play at RHIC 
energies as compared to SPS we expect this effect to be roughly
on the order of two.

% Local Variables: 
% mode: latex
% TeX-master: "y"
% End: 

%22
\newpage
\subsection{A.M. Snigirev:       Monojet Rate at RHIC}
%23
Hard jet production is considered to be an effective probe for formation of
super-dense matter -- quark- gluon plasma (QGP) in future heavy ion collider
experiments at RHIC and LHC. High $p_T$ parton pair (dijet) from a single hard
scattering is produced at the initial stage of the collision process
(typically, at $\sim 0.01$ fm/c).  It then propagates through the QGP formed due
to mini-jet production at larger time scales ($\sim 0.1$ fm/c), and interacts
strongly with the comoving constituents in the medium. The various aspects of
hard parton passage through the dense matter are discussed
intensively~\cite{appel,blaizot,heinz,ryskin,gyul90,thoma91,gyul94,gyulqm95,baier,zakharov,lokhtin1,lokhtin2}.
In particular, the strong acoplanarity of dijet transverse
momentum~\cite{appel,blaizot,heinz}, the dijet quenching (a suppression of high
$p_T$ jet pairs)~\cite{gyul90} and a monojet-to-dijet ratio
enhancement~\cite{gyulqm95} were originally proposed as possible signals of
dense matter formation in ultrarelativistic ion collisions.

In the simple QCD picture for a single hard parton-parton scattering without
initial state gluon radiation (i.e. when jets from dijet pair escape from
primary hard scattering vertex back-to-back in azimuthal plane with equal
absolute transverse momentum values, $p_{T1} = p_{T2}$) a monojet is created
only if one of the two hard partonic jets loses so much energy due to multiple
scattering in the dense matter that effectively we can detect only one single
jet in the final state. The monojet rate is obtained by integrating the dijet
rate over the transverse momentum $p_{T2}$ of the second (unobserved) jet with
the condition that $p_{T2}$ be smaller than the threshold value $p_{cut}$ (or
the threshold jet energy $E_T = p_{cut}$~\footnote{Due to fluctuations of the
  transverse energy flux arising from a huge multiplicity of secondary
  particles in the event, the "true" jet recognition in ultrarelativistic heavy
  ion collisions is possible beginning only from some energy
  threshold~\cite{kruglov}.}). Then rate of dijets $R^{dijet}$ with $p_{T1},
p_{T2} > p_{cut}$ and monojets $R^{mono}$ with $p_{T1} > p_{cut}$ ($p_{T2} <
p_{cut}$) in central $AA$ collisions is calculated as integral over all
possible jet transverse momenta $p_{T1}$, $p_{T2}$ and longitudinal rapidities
$y_1$, $y_2$.

At first in the framework of the simple model~\cite{lokhtin3} we demonstrate
that monojet-to-dijet ratio can be related to mean the acoplanarity measured in
the units of the jet threshold energy, namely
\begin{equation}
%~~~~~~~~~~~~~~~~~~~~~~~~~~~~~~~~~~~~~~~~~~~
\frac{R^{mono}}{R^{dijet}} \propto \frac{<|K_T|>}{E_T}.   
\end{equation}

%\parindent 0mm
%\vspace{6mm}
\begin{figure}[thbp]
\centerline{\psfig{file=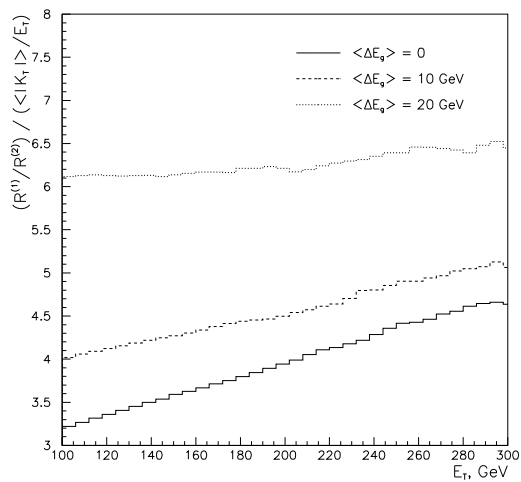,height=3.0in}}
\caption{(Monojet/dijet)/(mean acoplanarity/$E_T$) 
ratio as a function of jet energy threshold $E_T$. }
\end{figure}
%\vspace{-20mm}
%\begin{small}
%Figure 1: 
%\end{small}
%
%\hspace{\fill}
%
%\vspace{-80mm}
%\hspace{90mm}
%\begin{minipage}{70mm}
  The results of physics simulation have been obtained in the three scenarios
  for jet quenching due to collisional energy losses of jet partons in
  mid-rapidity region $y=0$~\cite{lokhtin1,lokhtin3}: $(i)$ no jet quenching,
  $(ii)$ jet quenching in a perfect longitudinally expanding QGP (the average
  collisional energy losses of a hard gluon $<\Delta E_{g}> \simeq 10$ GeV,
  $<\Delta E_{q}> = 4/9 \cdot <\Delta E_{g}>$), $(iii)$ jet quenching in a
  maximally viscous quark-gluon fluid, resulting in $<\Delta E_{g}> \simeq 20$
  GeV. Initial state gluon radiation has been taken into account with the
  PYTHIA Monte-Carlo model~\cite{pythia} at c.m.s. energy $\sqrt{s} = 5.5 A$
  TeV.
%\end{minipage}

%\vskip 0.5 cm   

Thus we conclude that rescattering of hard partons in medium results in weaker
$E_T$-dependence of ratio $R^{mono}/R^{dijet}$ to $<|K_T|>/E_T$. With growth of
energy losses the ratio we are interested in has a tendency to be constant,
what would be interpreted as the signal of super-dense matter formation.

%\end{document}
\newpage
\subsection{D.K. Srivastava:  Direct Photons and Dileptons at RHIC}
% D. K. Srivastava
% Variable Energy Cyclotron Centre, 1/AF Bidhan Nagar, Calcutta 700 064
%
The radiation of single photons from quark matter has recently been
estimated to the order of two loops~\cite{aurenche98}. This provides
a large bremsstrahlung contribution as well as a new mechanism of
quark-annihilation with scattering. Together these lead to a much large
radiation of photons than the estimates based on one-loop 
calculations~\cite{kapusta91}.
These rates along with the radiations from the hadronic matter
due to the hadronic reactions have been used to estimate the
yield of photons from $Pb+Pb$ collisions at RHIC~\cite{sri99}
with the assumption that a thermally and chemically equilibrated
quark gluon plasma is formed at $\tau_0=0.5$ fm/$c$ and having
$T_0=$ 310 MeV, corresponding to $dN_\pi/dy=$ 1734 
(see Fig.~\ref{shrhydro}).

However, it is rather unlikely that the QGP would be created in a
chemically equilibrated form. The radiation of photons from
an equilibrating plasma~\cite{sri97} due to Compton and annihilation 
processes (alone) and also from the pre-equilibrium stage
(i.e., pre-thermally equilibrated quark matter) within a parton
cascade model~\cite{sri98} are given in Fig.~\ref{shrall}.  

\begin{figure}[htb]
\begin{minipage}[t]{48mm}
%\epsfxsize=3.5in
%\centerline{
\psfig{file=gyshrif1.epsi,height=2.1in,width=1.5in}
%}
\end{minipage}
\hspace{0.04\linewidth}
\begin{minipage}[b]{100mm}
\caption{Radiation of photons from  central collision of lead nuclei
at RHIC energies from the hadronic matter (in the mixed phase
and the hadronic phase) and the quark matter (in the QGP phase
and the mixed phase). The contribution of the quark matter
while using the rates obtained by Kapusta et al. and
Aurenche et al. and those from hard QCD processes are shown 
separately.}
\label{shrhydro}
\end{minipage}

\begin{minipage}[t]{ 80mm}
%\epsfxsize=3.5in
%\epsfbox{Fig2.ps}
\psfig{file=gyshrif2.epsi,height=2.1in,width=1.5in}
\end{minipage}
%\hspace{0.04\linewidth}
\hspace{-1.in}
\begin{minipage}[b]{100mm}
\caption{Radiation of photons from  central collision of gold nuclei
at RHIC energies from an equilibrating plasma
where the initial conditions are taken from a self-screened parton
cascade model. Only Compton and annihilation processes are
included. The pre-equilibrium contribution within a parton cascade
model is also shown.}
\label{shrall}
\end{minipage}

\end{figure}
\vspace{-2mm}

Correlated charm decay presents the largest source of dileptons at
RHIC energies, though the spectrum of such dileptons will depend
sensitively on the extent of thermalization of the charm quarks
and D-mesons~\cite{gavin96}. In case it is absent the upper limit of dilepton 
production can be seen in Fig.~\ref{shrsource}, while the radiation of
dileptons due to annihilation of quarks (alone) from a
chemically equilibrating plasma~\cite{sri97} is shown in Fig.~\ref{shrdil}.       

\begin{figure}[htb]
\begin{minipage}[t]{80mm}
%\epsfxsize=3.5in
%\epsfbox{Fig3.ps}
\psfig{file=gyshrif3.epsi,height=2.5in,width=2.in}
\end{minipage}
%\hspace{0.04\linewidth}
\hspace{-0.8in}
\begin{minipage}[b]{100mm}
\caption{Sources of high mass dileptons. A thermalized and
chemically equilibrated QGP is assumed. The correlated charm 
and bottom decay is estimated assuming no energy-loss for
the heavy quarks and no thermalization of D mesons.
}
\label{shrsource}
\end{minipage}

\begin{minipage}[t]{80mm}
%\epsfxsize=3.5in
%\epsfbox{Fig4.ps}
\psfig{file=gyshrif4.epsi,height=2.5in,width=2.0in}
\end{minipage}
%\hspace{0.04\linewidth}
\hspace{-0.8in}
\begin{minipage}[b]{100mm}
\caption{Radiation of large mass dileptons from a chemically
equilibrating quark gluon plasma.}
\label{shrdil}
\end{minipage}

\end{figure}

%\end{document}

%24
\subsection{Z. Lin:          Open Charm and Drell Yan at RHIC}
%25
%\subsection{4:35 Lin, Z. Open Charm and Drell-Yan at RHIC}
%\subsection{Lin, Z. Open Charm, Drell-Yan and Thermal Dileptons at RHIC}

One of the signatures for the formation of the quark-gluon 
plasma in heavy ion collisions at RHIC is the thermal dileptons
emitted from such a matter. However, dileptons from both 
final heavy meson decays and initial Drell-Yan 
processes can contribute significantly to the continuum background, thus 
making it difficult to observe the thermal dileptons from the quark-gluon 
plasma.  Quantitative studies of dilepton production from these processes
are therefore essential. 

Although production of charm quarks in proton-proton interactions
can be reasonably well described by 
perturbative QCD, their fragmentation to charm mesons is non-perturbative.
In order to calculate the final open charm observables, one needs to know the 
intrinsic transverse momentum $k_t$ of the partons inside a proton, 
the K-factor which takes into account higher order contributions, 
and the parameter $\epsilon$ when the Peterson charm fragmentation function is 
used.
{} From the fit to E791 data on the angular distribution between the 
charm and anti-charm mesons, $k_t$ is found to be about $1$ GeV.  
In fitting the charm meson $p_\perp$ spectra from E706 and 
the $p_\perp^2$ spectra from E769, $K_c \simeq 3$ and 
$\epsilon \simeq 0.01$ are obtained.
For Drell-Yan processes, we simply take the leading-order (LO) $q\bar q$ 
annihilation cross sections with a K-factor, $K_{DY}=1.7$, determined
by fitting to E772 data. 
Thermal dileptons from the LO $q\bar q$ annihilation 
from the quark-gluon plasma is calculated in the parton cascade model ZPC, 
where the initial gluon and quark distributions are taken 
from the HIJING model. 

\begin{figure}[ht]
%\centerline{\psfig[width=3.5in,height=5.in,angle=270]{Lin/fig1-lin.eps}}
\centerline{\psfig{figure=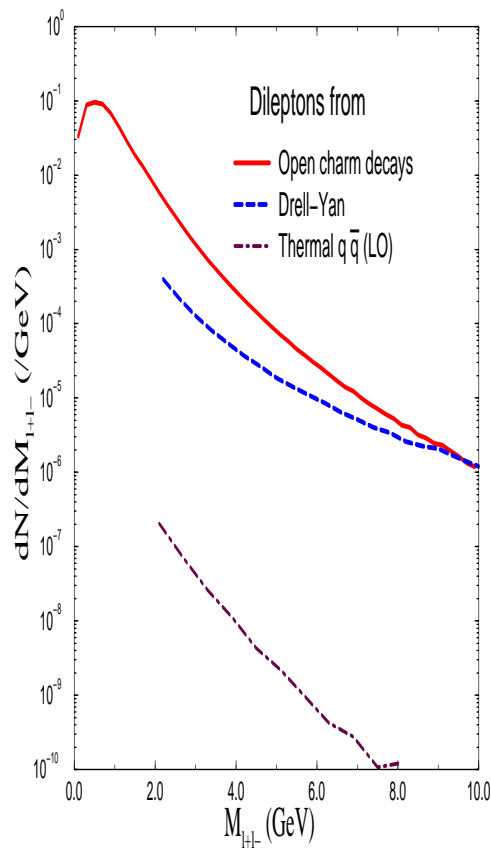,height=2.5in,width=4.5in,angle=270}}
\caption{Dilepton spectra from central Au+Au collisions at RHIC.} 
\label{linfig}
\end{figure}

In Fig.~\ref{linfig}, we show the dilepton invariant-mass spectra 
for central Au+Au collisions at RHIC.  
The solid, dashed and dot-dashed curves represent dileptons from 
correlated open charm decays, Drell-Yan processes, and LO thermal $q\bar q$ 
annihilation, respectively. The dilepton yield from open charm decays is
lower than earlier estimates where the charm fragmentation function
is taken to be $\delta(1-z)$, which is
equivalent to taking $\epsilon = 0$ in the Peterson formula.
For dileptons with mass below 10 GeV, the open charm contribution 
is higher than that from Drell-Yan processes, which is much larger
than the thermal dilepton yield from LO $q\bar q$ annihilation
from the quark-gluon plasma. 

Our preliminary results thus indicate that to detect the
thermal dileptons from the quark-gluon plasma does not seem feasible.
However, we have underestimated the thermal dilepton yield.
First, only $2\rightarrow 2$ partonic processes are included in ZPC, 
so the total number of partons does not increase as the system evolves.  
Secondly, we have not included thermal dilepton production from 
gluonic processes 
such as $gq \rightarrow q \gamma^*$ and $gg \rightarrow q \bar q \gamma^*$. 
Although cross sections for these two processes are suppressed 
by $\alpha_s$ and $\alpha_s^2$, 
respectively, they could have large effects on the thermal dilepton
yield because gluons greatly outnumber quarks 
in the early partonic stage of heavy ion collisions at RHIC.

Also, results on open charms are only based on information taken 
from $pp$ collisions. In heavy ion collisions, charm mesons are expected 
to undergo final-state hadronic interactions. 
We have recently studied the hadronic scatterings between charm mesons and 
hadrons such as pion, rho, and nucleon and have found that 
the charm meson spectra are significantly affected by these rescatterings.
The increase of the invariant mass of charm meson pairs due to 
scatterings in hadronic matter leads to an increase of
the invariant mass of dileptons from their decays, and this has been shown
to provide a possible explanation for the observed large dimuon enhancement in 
the intermediate-mass-region (IMR) of $1.5-2.5$ GeV in the NA50 experiment 
at SPS. Such final-state hadronic rescattering effects are different
from those due to initial-state interactions, e.g., 
an increase of the parton intrinsic $k_t$ in the nuclei effectively 
boosts the charm pair in the transverse direction, thus causing negligible
change of the pair mass spectra and the resulting IMR dileptons
from their decays.

At RHIC energies and beyond, partonic interactions of charm quarks
in the quark-gluon plasma are also possible, such as the energy loss of 
fast quarks. Even a moderate energy loss of $0.5$ GeV/fm has 
been found to suppress by an order of magnitude both the yield of 
high $p_\perp$ charm mesons and that of large invariant-mass dileptons 
from their decays. 

For a more reliable evaluation of the dilepton spectra in heavy ion
collisions at RHIC, one needs to further study the effects due to 
final-state interactions of both charm quarks and charm mesons as well as 
thermal dilepton production from gluonic processes in the quark-gluon plasma.

%----------------------------References---------------------------------

%\end{document}            
\newpage
\subsection{R. Thews:     $B_c$ Mesons at RHIC}
The $B_c$ meson is the bound state of $b\bar{c}$ (or $\bar{b}c$)
whose recent detection is the first step toward completion of
 the spectroscopy of heavy quark mesonic states.
The  $b$-$c$ states have properties that conveniently fill the gap  
between the $J/\psi$ and the  $\Upsilon$ states. Thus it is 
probable that at RHIC the $B_c$ mesons will serve 
as a probe of deconfined  matter \cite{Thews1}. We find that
significant differences arise for $B_c$ formation in deconfined and
confined matter \cite{Thews2}.  Our initial calculations suggest that:

(a) The rates of normal hadronic production mechanisms at RHIC energies
are {\underbar{not}} sufficient to produce a detectable number of
$B_c$ mesons.  

(b) If a region of deconfined quarks and gluons is formed,
the production (and survival) rate can be enhanced by several orders
of magnitude. 
 
(c) The observation of $B_c$ mesons at RHIC would signal a source of
deconfined charmed quarks, and the rate of $B_c$ production will
be a measure of the initial density and temperature of that source.

We note that the study of the $b$-$c$ sector has the advantage of 
a long history of
potential model analysis in the $b\bar{b}$ and $c\bar{c}$ sectors.  
These studies have provided robust predictions
 for the mass and lifetime of
the $B_c$ states \cite{Thews4}, and the recent measurements 
by CDF are consistent with those calculations \cite{Thews5}. 

First, we estimate the  production rate of
$B_c$ which one would expect if it results just from a
superposition of the initial nucleon-nucleon collisions  at RHIC. 
For heavy quark production, pQCD calculations for $p$-$p$ interactions
fit present accelerator data and bracket the RHIC energy range.
Hard Probes Collaboration estimates indicate about 10 $c\bar{c}$
pairs and 0.05 $b\bar{b}$ pairs per central collision \cite{Thews6}.   
$J/\psi$ and $\Upsilon$ production involves the use of some model, 
such as the Hard Probes color singlet fits \cite{Thews6}, which predict
bound state
fractions of order somewhat less than the one percent level.
A similar analysis for $B_c$ production involves substantially smaller
bound state fractions, since the $b$ and $\bar{c}$ must be produced
in the same hard interaction, a process of order $\alpha_s^4$.  
At RHIC energies, typical values are 
$3-10 \times 10^{-5}$ \cite{Thews8}, with the uncertainty from 
the scale choice in the pQCD calculations. 
  
To get predictions for RHIC, we have looked at two scenarios for
the luminosity.  a)  The ``first year'' case assumes a luminosity of
20 $\mu b^{-1}$ with no trigger.  b)  The ``design'' luminosity
assumes 65 Hz event rate with a 10\% centrality trigger in
Phenix, and uses $10^7$ sec/year.
The event rate predictions listed in the Table include 
both the weak branching fraction of the $B_c$ plus
the dimuon decay fraction for $J/\psi$.  Results are also
shown for $J/\psi$ and $\Upsilon$ production and detection via 
$\mu^+\mu^-$ and the underlying heavy quark production 
which may be useful to make contact with other estimates.  One sees
easily that in this scenario there is no hope of seeing $B_c$'s 
at RHIC. 

Now the principal reason for this work - could deconfinement
change the $B_c$ production rate at RHIC?  We have investigated the following
scenario:  For those events in which a $b\bar{b}$ pair are produced, 
the small $B_c$ formation fraction could be avoided if one  
utilized the 10 $c\bar{c}$ pairs already produced by independent
nucleon-nucleon collisions in the same event.  This can occur
if and only if there is a region of deconfinement which allows 
a spatial overlap of the $b$ and $c$ quarks.  
Also, one would expect additional $c\bar{c}$
production in the deconfined phase during its lifetime, as a result of the
approach to chemical equilibration.   The large binding energy
of $B_c$ (840 Mev) would favor their early ``freezing out"
and they will tend to survive as the temperature drops 
to the phase transition
value.  The same effect for the $B$ mesons and indeed for the $B_s$ will
not be so competitive, since these states are not bound at the 
initial high temperatures (or equivalently they are ionized 
at a relatively high rate by thermal gluons).

To do a quantitative estimate, we first utilized the quarkonium
break-up cross section based on the operator product expansion
\cite{Thews9}, to calculate the dissociation rate of bound states
due to thermal gluon collisions.
We then used detailed balance for the formation rates, and 
calculated the equilibrium fraction of $B_c/b$-quarks (see Ref.\cite
{Thews2} for details).  At T = 160 GeV, this fraction 
drops to as low as a few percent, but it is at least a factor of
100 above the no-deconfinement scenario.   To get a more
realistic limit, we repeated the calculation, using only the 
initially-produced $c$ quarks.  This actually produces a much larger
bound state fraction, since the $c$ quark density only decreases
with a $T^3$ volume factor rather than the $\exp(-m_c/T)$ of
chemical equilibrium.  The last three rows in the Table show
the corresponding $B_c$ numbers at RHIC in this scenario.  They
depend quite strongly on the initial temperature, which determines
the final charm density through the assumed isentropic expansion.

We are in the process of refining these preliminary numbers.
The kinetic equations will be followed numerically, using exact
time evolution of the deconfined region and the effect of 
approach to chemical equilibrium for the charm quarks.
It appears that the sensitivity to the parameters of the deconfined
state will remain, making the observation of any $B_c$'s at RHIC both
a ``smoking gun" signal of deconfinement and a probe of the 
initial temperature of the system and the initial density of
deconfined charm.
%%%%%%%%%%%%%%%%%%%%%%%%%%%%%%
%\begin{table}[ht]
%\caption{RHIC yields for heavy quark systems.}
\begin{center}
\begin{tabular}{|l|cc|} \hline\hline
Observable&First Year&Design Luminosity\\
\hline
$c\bar c$-pairs & $2.8\,10^8$ & $6.5\,10^9$ \\
$b\bar b$-pairs & $1.2\,10^6$ & $3.2\,10^7$ \\
$J/\Psi\to\mu^+\mu^-$ & $1.6\, 10^5$ & $3.9\,10^6$ \\
$\Upsilon(1s)\to\mu^+\mu^-$ &140 & 3800 \\
\hline\hline
$B_c\stackrel{2.5\%}{\to} J/\psi\it{l}\nu\stackrel{6\%}{\to}\mu^+\mu^-\it{l}\nu$&&\\
\hline\hline
(No Deconfined Phase) & 0.05--0.18 & 1.5--4.9 \\
\hline
(QGP+$c\bar{c}$ in Chemical Equil.)& 18 & 490 \\
\hline
(Only initial $c\bar{c}$ at $T_o$ = 500 MeV) & 130 & 3530 \\
\hline
(Only initial $c\bar{c}$ at $T_o$ = 400 MeV) & 235 & 6420 \\
\hline
(Only initial $c\bar{c}$ at $T_o$ = 300 MeV) & 475 & 12900 \\
\hline
\hline
\end{tabular}
\end{center}
\vspace{-0.4cm}
%\end{table}
%%%%%%%%%%%%%%%%%%%%%%%%%%%%%%%%%%%%%%%%%%%%%%%%%%%%%%%%%%%%%%%%%%%%%%%%%%

%26
\subsection{R. Vogt:       Predictions for J/Psi Suppression}
%\input{Vogt/vogt.tex}
%vogt qm99
The predictions are based on the model for $J/\psi$ absorption by nucleons,
comovers, and quark-gluon plasma discussed in the recent review \cite{rv}.
As in Ref.~\cite{rv}, we take a nucleon absorption of 4.8 mb for color octet 
$c \overline c$ pairs and assume that the final-state charmonia can be broken
up by mesonic comovers with an 0.67 mb cross section. The comover density is
assumed to be proportional to the participant density in Au+Au collisions,
$n_{\rm AuAu}$.

\begin{figure}[htb]
\setlength{\epsfxsize=0.75\textwidth}
\setlength{\epsfysize=0.30\textheight}
\centerline{\epsffile{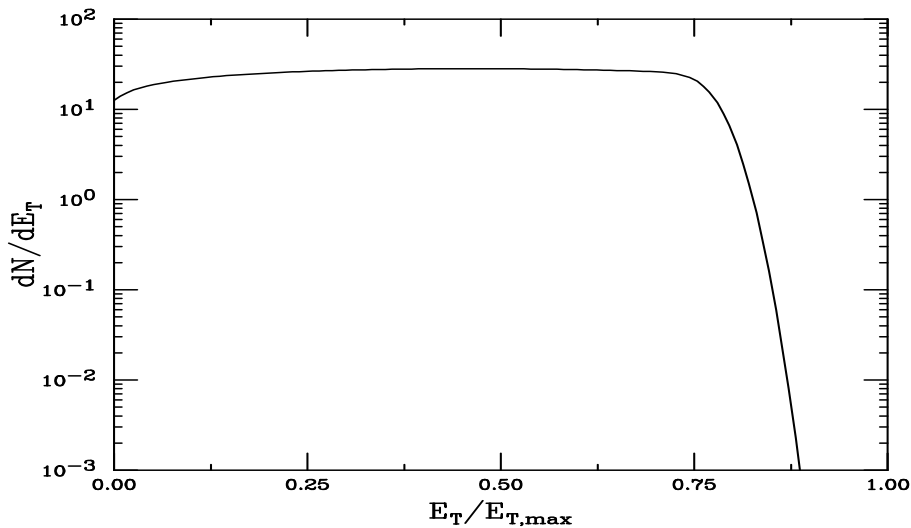}}
\caption[]{Lepton pair $E_T$ distribution for RHIC. }
\label{etdist}
\end{figure}

The scale of the suppression is set by the transverse energy distribution of
lepton pairs, shown in Fig.~\ref{etdist}.  In this calculation, the average
transverse energy at a given impact parameter is the sum of hard and soft
components and is approximately proportional to the number of collisions.
Nuclear shadowing is included in the estimate of the hard component, see
Ref.~\cite{ekkv} for details.  The distribution is plotted as a function of
$E_T/E_{T, {\rm max}}$ to be detector independent.  Note that $E_{T, {\rm 
max}} \sim 1600$ GeV for STAR and 700 GeV for PHENIX with an uncertainty of
10-20\% depending on the shadowing model.

\begin{figure}[htb]
\setlength{\epsfxsize=0.95\textwidth}
\setlength{\epsfysize=0.35\textheight}
\centerline{\epsffile{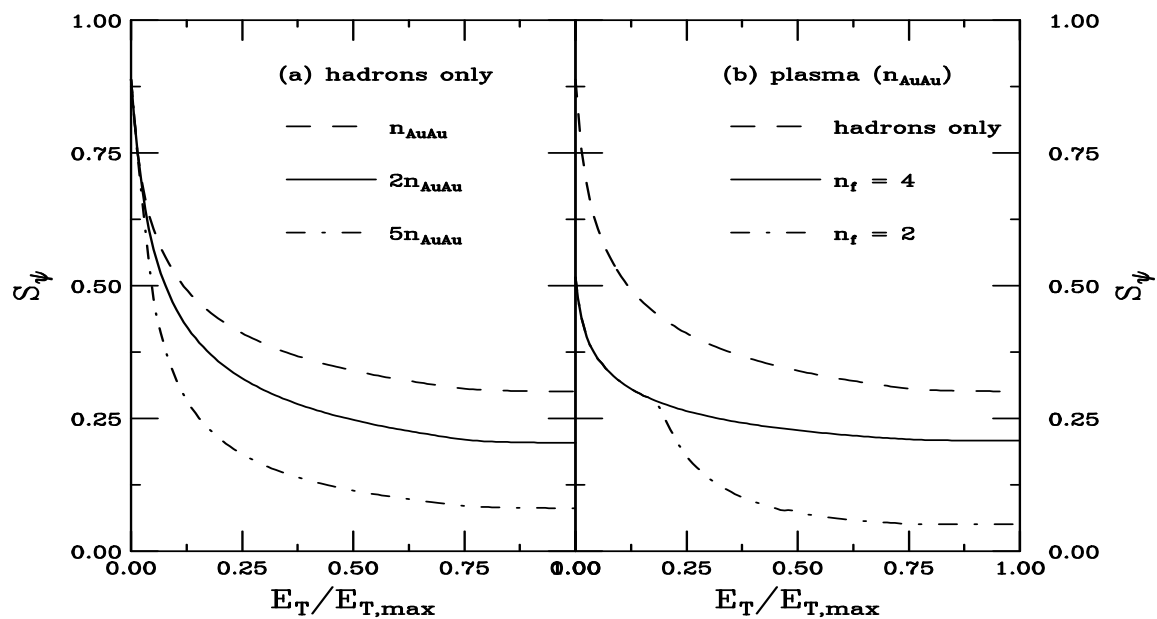}}
\caption[]{(a) $J/\psi$ suppression by hadrons only for comover densities
proportional to $n_{\rm AuAu}$ (dashed),  $2n_{\rm AuAu}$ (solid), and 
$5n_{\rm AuAu}$ (dot dashed). (b) $J/\psi$ suppression when quark-gluon plasma
is produced for a plasma with 2 (dot dashed) and 4 (solid) quark flavors.  
The dashed curve is the same as in (a).
}
\label{rat}
\end{figure}

The resulting $J/\psi$ survival probabilities are shown in
Fig.~\ref{rat}. Figure~\ref{rat}(a) is a completely hadronic scenario where the
comover density is proportional to: $n_{\rm AuAu}$, as for NA38 S+U; 
$2n_{\rm AuAu}$, as for NA50 Pb+Pb; and $5n_{\rm AuAu}$, beyond reasonable
expectations for hadronic matter.  In Fig.~\ref{rat}(b), plasma production
is assumed in addition to comover absorption with density $n_{\rm AuAu}$.
Plasmas with
2 and 4 quark flavors are studied.  The energy density as a function
of $E_T$ is determined from the hard and soft components to the average $E_T$
\cite{ekkv}.  When $n_f = 4$, only the $\chi_c$ and $\psi'$ are suppressed,
but for $n_f = 2$, direct $J/\psi$ suppression occurs when $E_T/E_{T, {\rm
max}} \sim 0.2$.  Note that plasma is created in even the lowest $E_T$'s
produced at RHIC so that $\chi_c$ and $\psi'$ suppression begin immediately in
both cases.

The $J/\psi$ and Drell-Yan cross sections were calculated in perturbative QCD
with and without nuclear shadowing.  In the lepton pair mass range $4 < m < 9$
GeV, $B\sigma_{J/\psi}/\sigma_{\rm DY} \approx 190-280$ depending on
the shadowing parameterization. The Drell-Yan cross section is calculated at
next-to-leading order.  The $J/\psi$ cross section has been calculated both in
the color evaporation model and in the non-relativistic QCD approach.  The two
methods yield nearly identical results at RHIC when no nuclear shadowing is
included.  See Ref.~\cite{ekkv} for details of the cross
section calculations and \cite{qm99site} for more numbers.

\newpage

\section{ EXOTIC POSSIBILITIES AT RHIC}
%27
\subsection{J. Schaffner:   \protect{$U_A(1)$} Restoration  Signatures}
%28
%\input{Schaf/schaf.tex}
%***********************************************************************
% qm99 proceedings for friday session 
% outline updated 5/31/99
%Organizer: Miklos Gyulassy <gyulassy@nt3.phys.columbia.edu>
%
% 27
% J''urgen Schaffner-Bielich
%

\medskip

Strange Signals from the Chiral Phase Transition? \cite{Schaf}
\medskip

%\footnote{see http://www.qm99.to.infn.it/rhic\_pred/schaffner/schaffner.html}
Recently, there have been strong indications from lattice calculations, that
the chiral $U_A(1)$ symmetry is partially restored just above the critical
temperature \cite{Chen98}. It was found that
the screening mass splitting between the $\delta$ meson (the $a_0$, a scalar-isovector
meson) and the pion is reduced to about 5\% or less. That means that the mass 
gap has changed by at least an order of magnitude. We are now going to discuss possible 
signals for RHIC associated with that effect.
We explore the consequences with an SU(3)$\times$SU(3) chiral Lagrangian.
If the chiral SU(3) symmetry is effectively restored, then the masses of the
pion and the sigma meson and the masses of the eta and the $\delta$ meson are 
the same separately.
If the chiral $U_A(1)$ symmetry is effectively restored, then this mass gap
vanishes and all four meson masses are the same:
$  m_\pi = m_\sigma \approx m_\delta = m_\eta$ for $c\approx 0$
which gives two new observable effects for RHIC \cite{Schaffi99_ua1}.
\medskip

{\em Prediction 1:} the number of $\eta$s is enhanced by up to an factor four.
The $a_0(980)$ meson has a width of $\Gamma=50-100$ MeV. Its mass decreases
with temperature, as seen on the lattice. Already below $T_c$ the decay
$a_0\to \eta + \pi$ must be blocked by phase space (see figure). Also the matrix elements
${\cal M}(a_0\to\eta_{ns}+\pi) = 4\lambda\sigma\approx0$ in the chiral SU(3) phase and
${\cal M}(a_0\to\eta_s+\pi) = 2c\approx0$ in the chiral $U_A(1)$ phase. Hence,
the inelastic channels are closed. The elastic channels are still large as
they are proportional to the coupling constant $\lambda$ only. The $a_0$ can
not decay to two pions which is forbidden by isospin. Also the decay to two
kaons is heavily suppressed as the $a_0$ is actually lighter than one kaon
alone. If the expansion from the chiral phase at $T_c$ until the freeze-out
temperature is fast enough (faster than the lifetime of the $a_0$ which is
about 2--4 fm), then the numbers of $a^{+-0}_0$ is three times the numbers of
the pions due to isospin counting. As the $a_0$ decays mainly to $\eta$s and
pions, the numbers of the $\eta$s can be increased by a factor up to four.
\medskip

{\em Prediction 2:} The scalar $\kappa$ ($K^*_0$) appears in the $K\pi$ mass
spectrum. 
Recent studies of the $\pi K$ scattering data show the existence of a broad
scalar, strange resonance, the $\kappa(900)$ \cite{Ishida97}.
The mass of this particle couples also to the $U_A(1)$ anomaly, i.e.\ the mass 
splitting to the kaon is proportional to $\Delta m_K\propto c\cdot \sigma$.
The matrix element for the strong decay of the $\kappa$ to a pion and a kaon
decreases in the chiral $U_A(1)$ phase from values around $\Gamma\approx .8$
GeV to $\Gamma\leq .2$ GeV. Hence, the barely visible broad resonance gets a
much smaller width in the chiral $U_A(1)$ phase. As the mass of the $\kappa$
approaches that of the kaon, the strong decay $\kappa\to K+\pi$ is blocked by
phase space already below $T_c$. Now, if the system freezes out around $T_c$,
there appears a cusp structure in the $K\pi$ invariant mass spectrum between
$m_K=m_\pi = .64$ GeV and $m_\kappa=.9$ GeV due to the opening of the decay
channel of the $\kappa$. The cusp should be pronounced as the width of the
$\kappa$ has substantially decreased. 
\medskip

{\em Prediction 3:} strange clusters of two baryons are formed at RHIC.
The status quo about the baryon-baryon interaction is, that there is some
(very limited) knowledge about the $N\Lambda$ ($N\Sigma$, $N\Xi$, $\Lambda\Lambda$) 
interactions from $\Lambda$ ($\Sigma$, $\Xi$, $\Lambda\Lambda$) hypernuclei
and $N\Lambda$, $N\Sigma$ scattering and $\Sigma^-$ atoms. There is absolutely 
no experimental information about the $\Lambda\Sigma$, $\Lambda\Xi$,
$\Sigma\Sigma$, $\Sigma\Xi$, $\Xi\Xi$ interaction at all. As hyperons are
produced abundantly in relativistic heavy-ion collisions, this opens a new
domain of strong interaction physics for RHIC. 
General SU(3) flavor symmetry predicts that there are (quasi) bound states in
e.g.\ $\Sigma^+p$, $\Sigma^-\Sigma^-$, $\Sigma^-\Xi^-$. The new Nijmegen
soft-core potential fitted to scattering data indeed predicts bound states of
$\Sigma\Sigma$, $\Sigma\Xi$, and $\Xi\Xi$ \cite{Stoks99a}. These exotica are
detectable by their weak two-body (nonmesonic) decay to final states like
proton, $\Lambda$, and $\Xi^-$. 
Detectable candidates are e.g.\ $(\Sigma^+p)_b\to p+p$,
$(\Xi^0p)_b\to\Lambda+p$, $(\Xi^0\Lambda)_b\to\Xi^-+p$ or $\Lambda+\Lambda$, and
$(\Xi^-\Xi^0)_b\to \Lambda+\Xi^-$. The branching ratio is about $(5-30)$\%
depending on the channel and the binding energy \cite{Schaffi99_memos}. 
The decay length
is about $c\tau=1-5$ cm, the more bound the candidate, the shorter the
lifetime. The production rates have been estimated to be around 0.02 to 0.001
per single event using RQMD2.4 and wavefunction coalescence
\cite{Schaffi99_memos}.

% Local Variables: 
% mode: latex
% TeX-master: "y"
% End: 

%\newpage
\subsection{D.H. Rischke:     Parity Violation and Color Superconductors}
%\title{Parity violation through color superconductivity}

%\author{R.D.\ Pisarski\address{Physics Dept., 
%Brookhaven National Laboratory, \\
%Upton, NY 11973--5000, U.S.A.} and D.H.\ Rischke\address{RIKEN-BNL Research
%Center, Brookhaven National Laboratory, \\
%Upton, NY 11973--5000, U.S.A.}}

%\begin{document}

% typeset front matter
%\maketitle

%\begin{abstract}
%We give a pedagogical discussion of how color superconductivity 
%can produce parity violation in cold quark matter at very high densities.
%\end{abstract}
%\newpage
%\section{Parity violation through color superconductivity}

We give a pedagogical discussion of how, for massless quarks
at very high densities, the formation of a spin-zero color superconducting
condensate spontaneously breaks both the axial $U(1)$ symmetry and parity
\cite{pisris:pr}.
This observation is implicit in the seminal work of Bailin and Love, 
is noted by Alford, Rajagopal, and Wilczek, and is
explicitly discussed by Evans, Hsu, and Schwetz \cite{pisris:bl}.

For simplicity, consider 
two degenerate flavors of quarks, and assume that a quark-quark condensate
forms in the color-antitriplet channel 
\cite{pisris:pr,pisris:bl,pisris:2,pisris:br}.
For massless quarks, 
two of the four possible condensates with total spin $J=0$ are 
\cite{pisris:pr}
\begin{equation}
\langle \phi^a_1\rangle  = 
\epsilon^{abc} \, \epsilon_{fg}\, \langle \,{q^b_f}^T \, C \,
\gamma_5 \, q^c_g \, \rangle \,\,\,\, {\rm and} \,\,\,\,
\langle \phi^a_2 \rangle = 
\epsilon^{abc} \, \epsilon_{fg}\, \langle \,{q^b_f}^T \, C \,
{\bf 1} \, q^c_g \, \rangle \,\,,
\end{equation}
where $a,b,c=1,2,3$ are $SU(3)_c$ color indices, $f,g=1,2$ are
$SU(2)_f$ flavor indices, and $C$ is the charge conjugation matrix.
$\phi^a_{1,2}$ are antitriplets under $SU(3)_c$ gauge transformations and
singlets under $SU(2)_f$ rotations \cite{pisris:pr}. The condensate $\phi^a_1$
is even under parity, $J^P=0^+$, while $\phi^a_2$ is odd,
$J^P=0^-$. There are two other condensates \cite{pisris:pr}, but 
they do not change our qualitative arguments about parity violation,
and so we omit them.

In the limit where mass and instanton-induced terms can be neglected,
the effective Lagrangian for color superconductivity is
\begin{equation} \label{pisris:eq1}
{\cal L}_0 = \left| \partial_\mu \phi_1 \right|^2 + 
\left| \partial_\mu \phi_2 \right|^2 +
\lambda \left( \left| \phi_1 \right|^2 + \left| \phi_2 \right|^2
- |v|^2 \right)^2  \,\, ,
\end{equation}
where $|\phi|^2 \equiv \sum_a (\phi^a)^* \phi^a$.
When mass and instanton effects are neglected,
the Lagrangian is symmetric under axial $U(1)$ transformations, which
rotate $\phi^a_1$ and $\phi^a_2$ into each other. 

\newpage
Therefore, there is
only one quartic coupling, $\lambda$. The Lagrangian (\ref{pisris:eq1})
generates nonzero
vacuum expectation values for the $\phi^a$'s, which can be written as
\begin{equation}
\langle \phi^a_1 \rangle = v^a\, \cos \theta \,\,\,\, , \,\,\,\,
\langle \phi^a_2 \rangle = v^a\, \sin \theta \,\,.
\end{equation}
Condensation picks out a given direction in color space for $v^a$, and
a given value for $\theta$. $v^a \neq 0$ breaks the $SU(3)_c$ color 
symmetry, which produces color
superconductivity. 
$\theta \neq 0$ breaks the axial $U(1)$ symmetry. Further,
whenever $\theta \neq 0$, there is
a nonzero $J^P=0^-$ condensate $\langle \phi^a_2 \rangle$;
this represents the spontaneous breaking of parity
(relative to the external vacuum).
\smallskip

This breaking of parity is actually familiar from the spontaneous breaking of
chiral symmetry. Consider two flavors of {\em massless\/}
quarks; the effective potential is $O(4)$-symmetric, involving the
$J^P=0^+$ $\sigma$- and $J^P=0^-$ $\pi$-meson fields. For massless quarks, it
is as likely for a parity-odd pion condensate to form as it is for a 
parity-even $\sigma$-meson condensate. This does not happen in nature,
because nonzero quark masses break chiral symmetry explicitly,
and thus favor a $0^+$ condensate.
\smallskip

Similarly, it is important to add to the effective Lagrangian 
(\ref{pisris:eq1}) terms which explicitly
break the axial $U(1)$ symmetry:
\begin{equation}
{\cal L}' = 
- c \left( \left| \phi_1 \right|^2 - \left| \phi_2 \right|^2 \right)
+ m^2 \left| \phi_2 \right|^2 \,\, .
\end{equation}
As shown by Berges and Rajagopal \cite{pisris:br}, 
the first term is due to instantons, with $c$ proportional to the 
instanton density. 
Instantons are attractive in the $J^P=0^+$ channel, and repulsive in 
the $J^P=0^-$ channel, so $c$ is positive. 
\smallskip

In the second term, each power of the current quark mass $m_q$ is accompanied
by one power of $\phi^a_2$. Since $\phi^a_2$ itself is not gauge invariant,
the simplest gauge-invariant term is $m_q^2 |\phi_2|^2$
\cite{pisris:pr}, so $m \sim m_q$.
Thus, the pseudo-Goldstone boson for the axial $U(1)$ symmetry is
extremely light, $m \sim 10$ MeV, taking $m_q$ to be 
the up or down quark mass and assuming the constant of proportionality between
$m$ and $m_q$ to be of order 1. 
This is in contrast to the explicit breaking of chiral symmetry,
where the corresponding term is linear in the quark mass. 
The pseudo-Goldstone bosons are the pions which are relatively heavy,
$m_\pi \simeq 140 \, {\rm MeV} \sim \sqrt{m_q}$.
\smallskip

Both instanton and mass terms act to favor 
the formation of the $0^+$ condensate $\phi_1$ 
over that of the $0^-$ condensate $\phi_2$. Consider, however, the
limit of very high densities. When the quark 
chemical potential $\mu \rightarrow \infty$, the instanton density and so 
$c$ vanish like $\sim \mu^{-29/3}$ (for two flavors). 
The real question is whether at some density the current
quark mass is negligible compared to the scale of the condensate. If this
happens, we reach an
``instanton-free'' region in which quarks are effectively massless, 
${\cal L}'$ can be neglected, and parity is spontaneously broken.
\smallskip

Because mass terms are always present, the true thermodynamic ground state is
always the parity-even $0^+$ condensate, i.e., $\theta=0$. There is, however,
a finite probability for the system to condense in a parity-odd
state, i.e., $\theta \neq 0$. The size and lifetime of this state
is set by the mass of the pseudo-Goldstone bosons. 
For chiral symmetry breaking, the characteristic
scale is $1/m_\pi \sim 1.4$ fm. This is 
small compared to the time and length scales of a heavy-ion collision, so
that parity-odd fluctuations average to zero. On the other
hand, the region in space-time over which a parity-odd 
color superconducting condensate forms
is large, $1/m \sim 20$ fm.
If the collision time is shorter than this time
scale, there is a finite probability that the system decays in
a parity-odd state. 
We therefore propose to trigger on phase-space regions where nuclear
matter is cool and dense, in order to observe
the formation of parity-odd color-superconducting
condensates on an event-by-event basis.
A possible global parity-odd observable was discussed in 
\cite{pisris:kpt}.

\newpage
%29

\subsection{D. Kharzeev:    CP Violation in Au+Au?}

It has been proposed by R. Pisarski, M. Tytgat and myself 
\cite{pap,pap1} 
that the discrete symmetries of strong interactions -- 
parity $\cal{P}$ and $\cal{CP}$ -- can be spontaneously violated 
in the vicinity of the deconfining phase transition. This 
would lead to a variety of dramatic effects which can be observed 
experimentally. What follows below is a brief and  elementary introduction 
to the ideas of \cite{pap,pap1}. I refer the reader  
to these papers for all details and further references.

$\cal{P}$ and $\cal{CP}$ violation in strong interactions was never observed. 
However, in QCD, these symmetries cannot be taken for granted, and  
$\cal{P}$ and $\cal{CP}$ invariance should be regarded as a property 
of the ground state of the theory -- the vacuum. When QCD vacuum is 
excited in a high energy nuclear collision, 
the properties of this excited vacuum state under 
$\cal{P}$ and $\cal{CP}$ transformations are in general not pre--determined, 
and will be defined  by its structure. With these considerations in mind, 
let us recall briefly what is known at present about the 
$\cal{P}$ and $\cal{CP}$ symmetries of QCD.

Classical equations of motion of QCD are known to possess topologically 
non--trivial solutions -- four-dimensional 
configurations of the gluon field which are 
characterized by different values of the ``winding number" $n$. 
In this situation, the 
true vacuum of the theory should be represented as a linear superposition 
of states with different $n$; this is analogous to the structure of the 
ground state in a crystal (Bloch wave function):  
\begin{eqnarray}
|\theta \rangle = \sum_{n} e^{i \theta n} |n \rangle, \label{theta}
\end{eqnarray}
where $\theta$ is called ``$\theta$ angle", and $n$ in terms of the 
gluon fields is defined 
as $n = \int d^4 x \  Q(x)$, with the topological charge density 
$Q(x) = g^2/(32 \pi^2) tr(G_{\mu \nu} \widetilde{G}^{\mu \nu})$.  
Once an expectation value of a local observable is computed via 
the path integral, 
one has to include configurations of different winding numbers $n$ 
with the weight given in  
(\ref{theta}). This procedure is equivalent to adding to the QCD Lagrangian 
a term  
\begin{eqnarray}
{\cal{L}}_{\theta} = \theta \ Q = \theta  \  {g^2 \over 32 \pi^2}\ tr(G_{\mu \nu} 
\widetilde{G}^{\mu \nu}) \label{tterm}.
\end{eqnarray}
Since in terms of
the color electric, $\vec{E}$, and color magnetic, $\vec{B}$,
fields, $G_{\mu \nu} \widetilde{G}^{\mu \nu} \sim \vec{E} \cdot \vec{B}$, it is easy to see 
that the ``$\theta$-term" (\ref{tterm}) explicitly violates $\cal{P}$ and $\cal{CP}$ invariances 
if $\theta \neq 0$. $\cal{P}$ and $\cal{CP}$ conservation in QCD thus is not guaranteed 
{\it a priori}; however current 
experimental constraints on $\theta$ are very stringent, $\theta < 10^{-9}$, and this 
constitutes so-called ``strong $\cal{CP}$ problem".

In terms of an effective theory of Goldstone bosons, the $\theta$-term (\ref{tterm}) 
manifests itself in the structure of the potential for the chiral $U(N_f)$ matrix.   
By a global chiral rotation,
a constant $U$ field can be rotated into a diagonal matrix.  With
$U_{i j} = exp(i \phi_i) \;\delta^{i j}$, the effective potential
can be written down as
\begin{eqnarray}
V(\phi_i) = f_\pi^2 \left(
- c \sum_{i} m_i \; cos(\phi_i)
+ \frac{a}{2} (\sum_i \phi_i - \theta )^2 \right) \; ,
\label{e3}
\end{eqnarray}
where the last term  
is proportional to the topological susceptibility,
$a \sim \int d^4 x \, \langle Q(x) Q(0)\rangle$ -- the correlation function of topological charge 
at zero momentum, and $m_i$ are current quark masses. At zero temperature, the second term in 
(\ref{e3}) dominates, and the ground state is trivial, with $\langle \phi_i \rangle =0$.

Based upon an analysis in the limit
of a large number of colors, we suggested \cite{pap}  
that near the phase transition, $a$ becomes much smaller than
its value at zero temperature.  
In a mean field type of analysis, with $T_d$ the temperature
of the deconfining transition, and $t = (T_d -T)/T_d$ the
reduced temperature, we found that 
$c(T) \sim 1/t^{1/2}$ and
$a(T) \sim t$, so that the relevant ratio, $a/c$, scales
as $a(T)/c(T) \sim t^{3/2}$. Once $a/c$ gets small near $T_d$, the first term in (\ref{e3}) 
becomes comparable in magnitude to the second, and (\ref{e3}) 
admits non--trivial metastable vacuum solutions with $\sum_i \langle \phi_i \rangle \neq 0$. 
It is clear from 
(\ref{e3}) that non-zero values of $\langle \phi_1 + \phi_2 + \phi_3 \rangle$ 
act like having a system 
with non-zero $\theta -$ angle, $\theta \neq 0$. These solutions therefore correspond to 
metastable domains of $\cal{P}-$ and $\cal{CP}-$odd vacua. Once the system is heated 
above the deconfinement phase transition and cools down, it can be trapped in one 
of these metastable $\cal{P}-$ and $\cal{CP}-$odd states. 

An obvious question
is how this breaking of the discrete symmetry by a metastable bubble
could be measured in nuclear collisions.
As the bubble is odd under $\cal P$ and $\cal CP$,
the pions produced by its decay must also be in a state which is
odd under these symmetries. 
 For the collisions of nuclei with equal atomic
number, as the initial state is even under $\cal P$, the observation
of a $\cal P$-odd final state must be due to parity violation.
In \cite{pap} we
proposed measuring, on an event by event basis, 
a global variable which is odd under $\cal P$: 
\begin{eqnarray}
{\bf J} = 
\sum_{\pi^+, \pi^-}
(\hat{p}_{+} \times \hat{p}_{-})\cdot \hat{z} \; ,
\label{eqj}
\end{eqnarray}
where the sum includes all $\pi^+ \pi^-$ pairs in a given event, and 
$\hat{z}$ is a fixed vector of unit norm. Various choices for this vector, 
as well as a general classification of  $\cal{P}-$ odd observables, can be 
found in \cite{pap1}. 
Since the effective potential (\ref{e3}) is symmetric, 
$\bf{J}$ on the average should vanish when summed over many events, and 
the distribution in $\bf{J}$ should be symmetric 
with respect to zero. $\cal{P}-$ odd effects therefore will manifest themselves in 
a non-zero width of this distribution.     

It is easy to understand why $\cal P$-odd bubbles induce 
non-zero values of $\bf{J}$ in a given event. 
In terms of the underlying gluonic fields, the $\cal P$-odd bubbles
arise from fluctuations in the topological charge density, $G_{\mu \nu}
\widetilde{G}^{\mu \nu}$.  Consider the propagation of
a quark anti-quark pair through a region in which 
$G_{\mu \nu} \widetilde{G}^{\mu \nu} \sim \vec{E} \cdot \vec{B} \neq 0$. 
If $\vec{E}$ and $\vec{B}$ both lie along the $\hat{z}$ direction,
then a quark is bent one way, the anti-quark the other,
so that $(\vec{p}_q \times \vec{p}_{\overline{q}}) \cdot \hat{z} \neq 0$,
where $\vec{p}_q$ and $\vec{p}_{\overline{q}}$ are the three-momenta
of the quark and anti-quark, respectively. An estimate of the effect 
can be done by considering the topologically 
non--trivial solutions directly in terms of collective pion fields \cite{pap1}; 
we find that the $\cal{P}-$ odd asymmetries can be relatively
large, at least $\sim 10^{-3}$. $\cal{P}$ and $\cal{CP}$ 
violation in nuclear collisions is therefore possible, and should be searched for.

\newpage
%30
\section{M. Gyulassy: Concluding Remarks}

We have seen considerable variation in section one on the
predictions of global observables 
(Bass, Bleicher, Cassing, Drescher, Eskola, Wang).
Much of the factor of two uncertainty in the initial
entropy and transverse energy are inherent our ignorance of the initial
conditions and especially its  soft component.
However, as emphasized by Eskola and Wang a large part of that uncertainty
is also due to interesting controllable physics. That component,
nuclear shadowing and anti-shadowing of the structure functions,
can in fact be inferred from systematic $p+A$ studies at RHIC (Wang, Lin)
or in $e+A$ when electron beams can eventually
be aimed through RHIC (Kovchegov).
The $p+A$ experimental studies are ${\bf mandatory}$ prerequisites
not only to reduce the uncertainties in the initial conditions
at RHIC, but also from the fundamental goal of simply
understanding the nuclear
wavefunction on the light cone. At first, light ion data, e.g. $Si+Au$,
will be essential to get a feeling for the size of those effects, 
but eventually a full complement of $pp$ and $pA$
data will be needed as at the AGS and RHIC to untangle the physics.

The A,B dependence of asymmetric systems as well as multiplicity
dependence is also key to understanding the deviations from phase
space saturation. Fireball models as shown in section 2
suggest that hadro-chemical equilibration is approached 
in strangeness to a surprising degree, and most microscopic
models cannot reproduce the observation without invoking
novel concepts such as baryon junctions (Vance) 
and higher Casimir flux tubes (Sorge).
However, the linear variation of $\langle K\rangle/\langle \pi\rangle$
with multiplicity in NA49 for example suggests that nonequilibrium
effects play an important role. Deviations from flavor equilibration
 predicted by Stachel 
will also be important to look for at RHIC. Those deviations
can provide us a handle to learn about the  physics of dense matter.
They may be due to phase space coalescence (Zimanyi),
 small relevant transport cross sections,
or interesting medium mass modifications near the chiral symmetry
boundary (Rapp, Schaffner). Just as the interesting physics are deviation from OSCAR certified
\cite{OSCAR}
transport theories, the deviations from fireball fits must be
carefully scrutinized as emphasized by Rafelski. 

I cannot over emphasize enough the importance
of gaining control over the initial
conditions. As Schlei and Dumitru pointed out, the data can be fit with 
almost any transport or hydro model is we are allowed the
freedom to dial in arbitrary initial condition.
The goal of the RHIC program is however
not to tabulate arbitrary initial conditions but to understand
in detail the energy and A dependence of the initial as well as the final
conditions. Only with knowledge of the initial conditions
can we interpret  collective observables
such as directed, transverse, and azimuthal asymmetric flow
and such as possible time delay via HBT as evidence for new physics. 

One collective probe  emphasized by Dumitru
 from the hydrodynamic point of view
is $dE_\perp/dy$ that decrease in general if work is done by the system.
However, given the present uncertainties in the absolute height of 
"Mount RHIC", I suggest that the following
simpler relative measure of longitudinal collectivity:
$$R(\tau)=dE_T/d\eta /dN_{ch}/d\eta$$ 
Since Bjorken's work, it is
known that an ideal $p=\epsilon/3$ equations of state leads to 
$R(\infty)/R(\tau_0)\sim 1/2$ due to pdV work associated with longitudinal
Bjorken expansion. 
Covariant transport theory as discussed by Zhang and 
Molnar predicts less  but still significant longitudinal work
$R(\infty)/R(\tau_0)\sim 3/4$, , due
to finite size dissipation. 

One of the most interesting observations at the
SPS in my opinion was  in fact ${\bf NO}$ evidence for any
longitudinal work. Both $E_\perp$ and $N_{ch}$ are observed to scale almost
perfectly with the number of wounded nucleons.  Models such as RQMD and UrQMD
provide a possible answer to the missing work puzzle.  
Heavy resonances and strings in
such models mimic the mixed phase in QCD where the speed of sound is
anomalously small. The mixed phase is notoriously lazy, as Hagedorn taught us! 
Excitation energy is wasted making heavy resonances instead
of converting into collective motion.
At RHIC, much higher initial
energies are expected to be produced via mini-jets than at SPS,
well into the plasma phase.
There the speed of sound is expected to approach $1/\surd 3$ again, the freshly
produced  plasma can push its neighboring cells
 more efficiently  down the beam pipe with a  resulting in decreasing $R$.  

It is
important to distinguish this type of collectivity from that observed
associated with radial and elliptic flow and discussed in section 3. Those
may  also
be sensitive to interesting variations in the equation of state. However, the
longitudinal work occurs mostly during the early times when the longitudinal
gradients $\sim 1/\tau$ are the largest. Transverse collectivity may develop
over longer times because the transverse gradients are $\sim 1/R_A$.  The
system must spend its first three fm deep in the plasma phase to get
longitudinal work going. At the SPS lazy mixed
phase produces  no significant
longitudinal work.  Note that the hadron systematics in
section 2 indicate that AA at those energies
 cross the transition region but the simplest signal of longitudinal
collectivity was not seen.  So my prediction is that we will start to see
$R$ decrease for the first time at RHIC as a function of centrality and A.

\begin{figure}[htb]
\centerline{
\psfig{figure=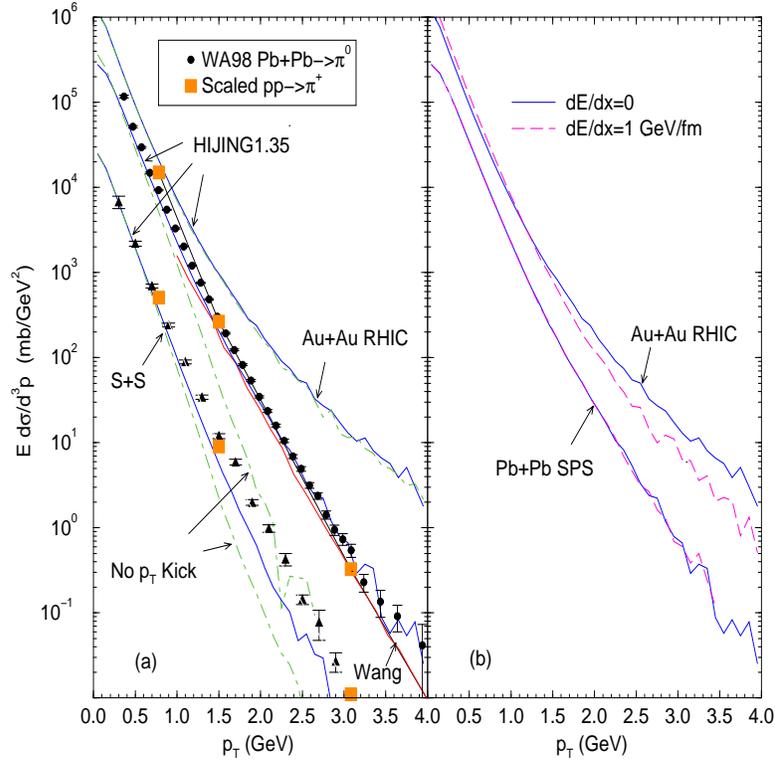,width=4in,height=4.0in,angle=-90.}}
\caption{ Opening of the high \protect{$p_\perp$}
window at RHIC
from \protect{\cite{gylev}}.
a) The WA80 $S+S$ data  (triangles) and 
the WA98 $Pb+Pb\rightarrow \pi^0$ data 
 (dots) are compared to HIJING1.35
 with soft $p_\perp$ kicks (full lines)
and without $p_T$ kicks (dot-dashed curves). The later  scale
with the wounded projectile number
times \protect{$\sigma_{AA}$} times
the invariant distribution calculated for \protect{$pp$}.
The parton model curve from X.N.Wang
is labeled by 'Wang'. 
The filled squares show \protect{$pp\rightarrow \pi^+$} data
scaled by the (Glauber) number of binary collisions 
times \protect{$\sigma_{AA}$} for both $SS$ and $PbPb$.
 b) Significant jet quenching predicted  for RHIC energies 
\protect{\cite{MGXW92}},
is masked by soft physics at SPS energies in the HIJING model. 
}
\label{fig1mg}
\end{figure}

One of the most exciting new frontiers at RHIC is high $p_\perp$
nuclear physics (see Eskola, Wang, Vitev, Snigirev, Srivastava). 
Unlike at SPS, where much theoretical ambiguity in pQCD
results from uncertainties associated with intrinsic $k_\perp$
smearing as discussed by Wang, 
at RHIC collider energies the pQCD power law tail
is predicted to stick out clearly. This provides a calculable, reliable
base or calibration point from which deviation
can be used to extract the physics of dense matter.
In Fig. \ref{fig1mg} from \cite{gylev}
the contrast between the now known SPS and the predicted RHIC
domain is evident. At the SPS HIJING accidently fits the data,
but variations of uncontrolled soft hadronization assumptions
leads to extreme discrepancies. At RHIC on the other hand,
that soft component has much smaller influence on the high $p_\perp> 3$ GeV
domain. Therefore it becomes possible to look for new physics
such as the non-linear energy loss of gluons predicted by BDMSP.
The magnitude of predicted jet 
quenching phenomena is illustrated in the figure.

Heavy flavor physics at RHIC will also be interesting as emphasized
in section 4. The open questions left by NA50 can be clarified
at RHIC energies which is well above the charm thresholds
and the high energy densities occur already in much smaller A systems.
The anomalous $J/\psi$ deficit has been predicted to occur by Satz already
in $Cu+Cu$, and brand X comover alternative scenarios will be much
more easily ruled out as emphasized by Vogt. The topic of open charm
is also rich with speculation and possibilities as discussed by Lin and Thews.

Finally, I comment on section 5. The predictions for
truly exotic and novel dynamics associated with
chiral restoration, P violation at high baryon densities and especially
possible CP violating domains add great excitement
to exploratory experiments at RHIC.
 While the predictions are based on very bold and even further
extrapolations from known physics than those discussed in the previous
sections, those directions associated with subtle many-body quantum
phenomena  should also be pursued vigorously because the potential payoff 
is so high.

\end{document}